\documentclass[12pt,epsfig]{article}

\usepackage{multirow}
\usepackage[dvips]{graphics}
\usepackage{rotating}
\usepackage{epsfig}
\def\lesssim{\mathrel{\hbox{\rlap{\hbox{\lower5pt\hbox{$\sim$}}}\hbox{$<$}}}}
\def\gtrsim{\mathrel{\hbox{\rlap{\hbox{\lower5pt\hbox{$\sim$}}}\hbox{$>$}}}}

\newcommand{\ntrl}[1]{\chi^0_#1}
\newcommand{\chpm}[1]{\chi^\pm_#1}

\def\squark{\tilde{q}}
\def\squarkc{\tilde{q}^*}
\def\gluino{\tilde{g}}

\def\hpm{H^\pm}

\def\tbar{\bar{t}}       %

\def\mone{M_1}
\def\mtwo{M_2}
\def\mthree{M_3}

\newcommand{\mntrl}[1]{m_{\chi^0_#1}}      
\newcommand{\mchpm}[1]{m_{\chi^\pm_#1}}

\def\mhpm{m_{H^\pm}}

\def\ptmiss{\not\!\!{p_T}}

\def\beq{\begin{equation}}   %
\def\eeq{\end{equation}}   %

\begin{document}

\begin{flushright}
   {\bf HRI-P-08-11-003 \\
   HRI-RECAPP-08-014}
\end{flushright}

\vskip 30pt

\begin{center}
{\large \bf Probing non-universal gaugino masses via Higgs boson production under SUSY cascades at the LHC: A detailed study}\\
\vskip 20pt
{Priyotosh Bandyopadhyay\footnote{priyotosh@mri.ernet.in},
}  \\
\vskip 20pt
{ Regional Centre for Accelerator-based Particle Physics  \\
Harish-Chandra Research Institute  \\
Chhatnag Road, Jhunsi, Allahabad, India 211019 }\\

\end{center}

\vskip 65pt

\abstract{Cascade decays of Supersymmetric (SUSY) particles are likely to be
prolific sources of Higgs bosons at the Large Hadron Collider (LHC).
In this work, we explore, with the help of detailed simulation, the 
role of non-universal gaugino masses in the production of the Higgs 
bosons under SUSY cascades. The analysis is carried out by choosing 
an appropriate set of benchmark points with non-universal gaugino 
masses in the relevant SUSY parameter space and then contrasting the
resulting observations with the corresponding cases having universal
relationship among the same. It is shown that even of data at an early
 phase of the LHC-run  with 10 fb$^{-1}$ one would be able to see,
under favourable situations, the imprint of non-universal gaugino
masses by reconstructing various Higgs boson resonances and comparing 
their rates. With increased accumulated luminosities, the indications 
would naturally become distinct over a larger region of the parameter 
space.}

\newpage

\section{Introduction}

One of the principal goals of the Large Hadron Collider (LHC) is to unravel 
the mechanism of electroweak symmetry breaking (EWSB). 
Supersymmtery (SUSY) offers
a unique kind of technical solution to the hierarchy problem that
cripples the Higgs sector of the Standard Model (SM) of particle physics thus
attaching a special significance to the idea of electroweak symmetry breaking
via the Higgs mechanism. It also goes one step forward in a much satisfying 
and assuring direction by suggesting how the EWSB can be triggered 
dynamically within a supersymmetric framework, a feature, not present
in the SM. However, all these come at some price. The Higgs sector of a
SUSY scenario is necessarily an extended one with at least two Higgs doublets
in contrast to only one such present in the SM.

The Minimal Supersymmetric extension of the Standard Model (the MSSM) thus
contains 5 physical Higgs bosons on EWSB, viz., two  CP-even
neutral scalars ($h$ and $H$), one CP-odd neutral Higgs ($A$) and a pair of mutually
conjugate charged Higgs bosons ($H^\pm$). It is thus clear that in such a 
scenario, the phenomenology of Higgs bosons assumes a special significance 
in terms of its richness and resulting complications \cite{Djouadi:2005gi},
 \cite{AguilarSaavedra:2005pw}, \cite{Haber:1984rc}, \cite{HHG}, \cite{Martin}. Naturally, with the 
LHC soon starting its operation, the hunt for such scalars would be a priority.

As has been pointed out in earlier works \cite{Datta:2001qs}, \cite{Datta:2003iz} the phenomenology of
the MSSM Higgs bosons becomes extremely involved due to interactions that are 
present among various SUSY particles and the former. Thus, Higgs bosons are
expected to be looked for at the LHC in all probable channels to extract 
information about the underlying framework through various consistency checks.
This is also important in the sense that the SM-like production processes 
(with reference to which the current experimental constraints on the MSSM 
Higgs sector are derived) are known to be only sensitive to specific ranges
of some SUSY parameters. For example, for most of the SM-like production
processes like $gg \to h, H, A$, the associated production with heavy quarks
like $gg,q\bar{q} \to h,H,A + t\bar{t}$ or $b \bar{b}$ for the neutral
Higgs bosons and $t \to H^+ b$ or associated production with top quarks like
$gg/q\bar{q} \to H^+bt$ for the charged Higgs bosons, the productions processes
are only enhanced for extreme values of $\tan\beta$ i.e., the ratio of the
vacuum expectation values of the two Higgs doublets which are responsible
for the breaking of the electroweak symmetry in the MSSM. Another example
of limitations of the SM-like production processes is when an otherwise viable
signal of a Higgs boson produced in association with superparticles fails
due to presence of CP-violation \cite{Bandyopadhyay:2007cp}. Hence, other characteristic interactions 
of the MSSM Higgs bosons, if present, must be exploited for the purpose.

In particular, there exist nontrivial interaction-vertices among the 
electroweak gauginos/higgsinos (the charginos and the neutralinos) and 
the MSSM Higgs bosons. This implies that decays of 
charginos/neutralinos may lead to Higgs bosons. On the other hand, 
the major source of charginos
and neutralinos at the LHC would be the cascades of squarks and gluino which
are expected to be produced copiously there due to their strong interactions. 
It has already been demonstrated in earlier works \cite{Datta:2001qs,Datta:2003iz} that
the overall suppression due to different branching fractions along a cascade
leading to Higgs bosons 
could be more than compensated by the strong production cross section.
Hence, the strong production of squarks and gluino may turn out to be a
major source of Higgs bosons at the LHC. The compositions and the masses of
the charginos and the neutralinos play a big role in the process. These in
turn are determined by the values of the $U(1)$ and the $SU(2)$ gaugino
masses ($\mone$ and $\mtwo$ respectively, which breaks SUSY softly) 
and the value of $\mu$, the
SUSY-conserving higgsino parameter appearing in the superpotential. On the
other hand, the soft-SUSY breaking $SU(3)$ gaugino mass ($\mthree$) 
determines the mass of the gluino which in turn controls the strong 
production rates. Thus, it is clear that a correlation existing among these
masses would play a crucial role in the phenomenology of Higgs bosons under
SUSY cascades.

Canonical SUSY studies at colliders work in a paradigm that assumes
high scale universality among different gaugino masses. Although highly
economic and thus predictive (and hence, popular as well), there is no
deep reason as to why such a universality should be robust. In contrary,
it is now known that such universality is a result of a trivial form of
the so-called gauge kinetic function from which the common gaugino mass 
arises at a high scale as SUSY breaks in the hidden sector. In particular,
this happens when the gauge kinetic function involves a combination of
hidden sector fields which is singlet under the underlying gauge group
of the SUSY Grand Unified Theory (SUSY-GUT). However, contributions from
the non-singlet
higher GUT representations or from linear combinations of the singlet and
possible non-singlet representations may effectively induce non-universality
in the soft masses for the gauginos at the high scale itself [\cite{Gl}-\cite{Cremmer}]. Such a triggering of non-universality at the high scale
would distort the spectrum of the gaugino masses at the weak scale with 
respect to the one obtained with universality condition being intact at the
high scale. This leads to a modification in the compositions of
the charginos and neutralinos and alterations in masses for all of them 
including the gluino. As indicated in the last paragraph, this is bound to
have a significant impact on the resulting phenomenology at colliders
[\cite{Bt1}-\cite{Huitu:2007vw}] including SUSY-Higgs searches at the LHC.

Production of Higgs bosons under SUSY cascades has been discussed in detail
in the MSSM framework \cite{Datta:2001qs,Datta:2003iz}. In these works, the assumption
of high-scale 
universality between $\mone$ and $\mtwo$ was retained while that with respect
to $\mthree$ had been deliberately relaxed. As expected, these retained the
imprints of the high-scale universality of gaugino masses in the masses and
compositions of the charginos  and the neutralinos at the lowest order of
perturbation theory
(since they are determined by $\mone$, $\mtwo$ and $\mu$ only). However,
the above choice allowed for a different production-rate for the gluinos in the
first place and their possible different branching ratio to the charginos and
neutralinos when compared to the case where universality was kept intact over
all three gaugino mass parameters. 
In one of our recent works \cite{Bandyopadhyay:2008fp} we relaxed even the
universality relation between $\mone$ and $\mtwo$ at the weak scale.
As discussed in the last paragraph, this has a more direct effect on the
allowed pattern and strengths of the electroweak cascades involving 
charginos/neutrlainos and the Higgs bosons while production rate for the gluinos
may remain unaffected. As would be obvious, the most economic way to trigger
phenomenologically illustrative non-universality involving all the 
three gaugino masses is to tweak
$\mtwo$ while keeping $\mone$ and $\mthree$ fixed so that two rather
important ratios, $\mone/\mtwo$ and $\mthree/\mtwo$ get modified in a single
stroke. Nonetheless,
in an actual analysis, absolute values of all the involved 
masses and couplings (of the charginos and neutralinos) matter and hence,
for a general study, all the gaugino mass parameters along with the value 
of $\mu$ should be varied as independently as possible.

In this paper we continue to work within a framework of such a 
maximally relaxed version of 
non-universality as was also adopted in one of our recent works \cite{Bandyopadhyay:2008fp}.
We also assume that $R$-parity (defined as $R=(-)^{3B+L+2J}$, where $B,L,J$
stand for the baryon number, the lepton number and the spin of the particle concerned with
the convention that $R=+1$ for SM particles and $R=-1$ for their SUSY partners)
to be conserved so that the SUSY cascades ultimately end up in jets/leptons
and the stable lightest SUSY particle (LSP) (the lightest neutralino in the present study) which escapes detection and thus, is the source 
of missing transverse momentum (energy) in a SUSY event.

In the present work we systematically extend 
our earlier study \cite{Bandyopadhyay:2008fp} in the
following directions:
\begin{itemize}
\item identification of a set of `benchmark' points in the relevant SUSY
parameter space suitable for demonstrating the role of gaugino mass 
non-universality in the production of Higgs bosons under SUSY cascades,
\item detailed simulations of the Higgs-signals in these benchmark scenarios 
      with the help of an event-generator including the ones for the
      heavier neutral Higgs bosons ($H$ and $A$),
\item estimating the viability of different modes by performing a
      somewhat detailed simulations of the backgrounds to which contributions
      come from both SM and SUSY and
\item Establishing the robustness of our main results under varied situations.
\end{itemize}

The paper is organised as follows. In section 2 we  outline the production
mechanism of Higgs bosons under SUSY cascades. In section 3 we discuss our choice of  the
`benchmark' points in the relevant SUSY parameter space at which we later
carry out our analyses. Production rates for various Higgs bosons under SUSY
cascades are described in section 4. In section 5 we outline the signal and
describe the backgrounds in detail. The prospects of identifying neutral Higgs
bosons are discussed in detail in section 6 while in section 7 we deal with
the case of the charged Higgs bosons under different circumstances. In section 8 we present a discussion on squarks and the gluino have an inverted spectrum as compared to what assumed till section 7. In section 9 we discuss the robustness of the results with $\tan{\beta}$
and heavier squark and gluino masses. In section 10 we conclude.

\section{Higgs production in SUSY cascade}

As indicated in the Introduction, Higgs production under SUSY cascades
would be predominantly fed by copious productions of strongly interacting 
particles like squarks and gluinos. The heavier of these two SUSY particles
would always undergo a two-body decay (via strong interaction) 
to the lighter mate while for 
the latter there are only electroweak decay modes available in a $R$-parity
conserving scenario. Generic SUSY cascades ending up with Higgs bosons would
 thus look like:

\beq
p p \to  \gluino \gluino, \squark \squark, \squark \squarkc, 
\squark \gluino \to \chi^{\pm}_{i} , \chi^{0}_{j} + X 
\eeq
followed by
\beq
\chi^{\pm}_{i} , \chi^{0}_{j} \to \chi^{\pm}_{k} , \chi^{0}_1 + h,H,A,\hpm +X
\eeq

\noindent
In general, it is possible that, phase-space permitting, several different
charginos and/or neutralinos could decay into a specific Higgs boson. There are two
broad categories depending upon a generic pattern of the mass-spectra
for the charginos and the neutralinos. The first one is where the 
mass-splittings are such that only the heavier
of the charginos or neutralinos could decay to one of their lighter
counterparts plus a Higgs boson and the second possibility is where even the
lighter ones have enough splittings among them to accommodate a Higgs bosons
in their decays. The first possibility is known in the literature as the
`big cascade' (involving heavier gauginos and hence triggering a longer
chain of cascade decays)  and schematically given by
\beq
\chi^{\pm}_2, \chi^{0}_{3},\chi^{0}_4 \to \chi^{pm}_1,\chi^{0}_2, \chi^{0}_1 + h,H,A,\hpm
\eeq
while the second one is dubbed as `little cascade' in 
\cite{Datta:2001qs,Datta:2003iz} (while being known for quite some time 
 [\cite{hcascade}-\cite{oldcascade}]\footnote{The contrast between the `little' and `big'
cascades was drawn 
for the first time in \cite{Datta:2001qs,Datta:2003iz}.} and looks like
\beq
\chi^{\pm}_{1}, \chi^{0}_{2} \to  \chi^{0}_{1} + h,H,A,\hpm
\eeq
We will stick to this terminology throughout this paper.

Obviously, thus, the rates for Higgs bosons under SUSY cascades at the LHC
would crucially depend upon several different branching fractions of the 
involved decays. These branching fractions are determined by various 
SUSY masses and 
couplings involved in the problem. As hinted before, the most important roles
are played by the couplings of the Higgs bosons with the charginos and neutralinos.
It is to be noted \cite{Datta:2001qs,HHG} that the Higgs bosons 
couple favourably to charginos and neutralinos when the latter have 
significant components of gauginos and higgsinos while for the 
gauge bosons the couplings are maximal when the
charginos and neutralinos are dominated by higgsinos. Thus, it is clear
that the compositions of charginos and neutralinos would play a crucial role
in our analysis. Hence, the input parameters that turn out to be  instrumental
 are $\mone, \mtwo$ and $\mu$. For $\mu \gg \mtwo$ it is the so-called 
`gaugino region'. Here, the lighter neutralinos ($\ntrl1,\ntrl2$) and the 
lighter chargino ($\chpm1$) is gaugino-dominated with 
$\mntrl1 \simeq min(\mone,\mtwo)$ while 
$\mntrl2,\chpm1 \simeq max(\mone,\mtwo)$. Here, the heavier chargino and the 
 two heavier neutralinos are higgsino-like thus rendering $\mntrl3,\mntrl4,\mchpm2 \simeq \mu$.
The reverse is true when $\mu \ll \mtwo$. This is the so-called `higgsino
region' where the lighter gauginos (as grouped above) are almost degenerate and
their masses go as $\mu$. Note that in such `pure' regions the masses and
the contents (gaugino and/or higgsino) have direct correspondences.
In contrast, when one enters the mixed region, i.e., when 
$\mone, \mtwo \simeq  \mu$, the charginos and the neutralinos tend to get maximally
mixed in their gaugino and higgsino contents while their masses 
show no clear patterns although ultimately restricted by the values of $\mone, \mtwo$
and $\mu$. Thus, one could directly observe from the above discussion that
 the `big cascades' are favoured in the `pure' regions where
the heavier and the lighter set of charginos and neutralinos have different
contents (gaugino vis-a-vis higgsino). 
The `little' cascades only refer to the lighter set
and they are favoured in a `mixed' region where the 
lighter charginos
and neutralinos have appreciable gaugino and higgsino components.


\section{The benchmark points}

In \cite{Bandyopadhyay:2008fp} we discussed how the production rates for the neutral and the
charged Higgs bosons under SUSY cascades depend on the soft SUSY breaking
gaugino masses $M_1$, $M_2$, $M_3$ and the SUSY conserving higgsino mass
parameter $\mu$ in particular reference to possible nonuniversal patterns that
could exist among the three gaugino masses. There we summarised how the
relative rates for the lightest neutral Higgs boson and the charged
Higgs boson could be indicative of high-scale universality (or not) of the
gaugino masses in different regions of the $M_2$-$\mu$ parameter
space which govern the content of the charginos and the neutralinos
in a crucial way. Hence, establishing the robustness of such an
observation requires 
identification of regions of SUSY parameter space which could collectively
represent the host of situations that could be relevant for the problem in 
hand. With an understanding of the dynamics and the kinematics involved in
Higgs production under SUSY cascades (as discussed in section 2) we delineated 
such representative regions. These are called the `benchmark points' in the 
SUSY parameter space each of which is a set of relevant SUSY parameters. 
Further studies are then carried out in these benchmark points. In the
following we briefly discuss on how we converge on the benchmark points studied
in this work.

As stressed in the previous sections and also indicated in the last
paragraph, the benchmark points we settle for must encompass a 
phenomenologically significant region in the $M_2$-$\mu$ plane. This
would bring out varied spectra for the charginos and the neutralinos
with maximal spreads in their contents, i.e., the admixtures of
gaugino and higgsino in them whose role cannot be stressed enough
for the phenomena we are looking for. Thus, some representative sets
of values for $M_2$ and $\mu$ could well be enough for characterizing
the benchmark points for our present study. In Table 2 we present
four such benchmark sets indicated as BP1, BP2, BP3 and BP4. The
relative values of $M_2$ and $\mu$ are so taken that BP1 results in
higgsino-like lighter chargino and neutralinos while their heavier
siblings are gaugino like. The case is just the reverse for the set
BP2. For both BP3 and BP4 the charginos and neutralinos are heavy
mixtures of gauginos and higgsinos: in BP3 with all the masses being
a little heavier when compared to those in BP4. It is to be noted that
the benchmark points extend over a significant region in the
$M_2$-$\mu$ parameter space with their absolute values lying between
$350 \, {\mathrm {GeV}} <M_2< 700$ GeV for $M_2$ and
$150 \, {\mathrm {GeV}} < \mu < 700$ GeV for $\mu$.

Next we come to the important issue of fixing the masses for the strongly 
interacting SUSY particles, i.e., the squarks and the gluino. Since the
production rates of these excitations crucially govern subsequent rates for
the production of the Higgs bosons under SUSY cascades, masses of the former
controls the rates for the latter. For the purpose of demonstration we
fix masses for both squarks and the gluino a little below 1 TeV for which
the basic strong-production rates are moderately high and which are well 
within the direct reach of the LHC. In particular, we first work with
$m_{\tilde{g}} > m_{\tilde{q}}$ choosing $m_{\tilde{g}}=900$ GeV (which in
turn is indicative of $M_3$) and 
$m_{\tilde{q}}=800$ GeV. Subsequently, we also study the case with the squark 
and gluino masses flipped and justify how it could be relevant.

We then take up the case of sleptons. Although, as we would see later in this
work, we do not consider the leptonic final states in a major way (except for the $\tau$s in 
specific cases), the sleptons might
play a naturally crucial role in the SUSY cascades through the branching 
fractions of the charginos and the neutralinos which would compete
with the branchings of the latter to Higgs bosons, i.e., the main source
of Higgs bosons under SUSY cascades. In some cases, stau sleptons might contribute
to the SUSY background as would discussed later in an appropriate context.
The present study could 
have been rendered simpler with a somewhat massive slepton which decouples 
from the problem. We, instead, keep the sleptons light enough (which finds 
motivation in many SUSY scenarios) and fix their mass to 400 GeV in order to
keep them involved in the problem. In section 9, we demonstrate the 
robustness of our 
findings with an appropriately chosen higher values for the masses of the
squarks, the sleptons and the gluino.

The direct productions of the Higgs bosons in a SUSY framework are known to be
favoured for extreme values of $\tan\beta$ and thus the prospects of the LHC
in finding the Higgs bosons is rather limited for the intermediate values
of $\tan\beta$. In contrast, it is now known \cite{Datta:2001qs,Datta:2003iz} that the
production rates of Higgs bosons under SUSY cascades are more or less 
independent of $\tan\beta$ and thus could effectively probe into the regime of
intermediate $\tan\beta$. Motivated by this we fix $\tan\beta$ at an 
intermediate value of 10 for the main part of our study. In section 9, we 
resort to 
$\tan\beta=5(50)$ for demonstrating the robustness of our results in low(high)
$\tan\beta$ regime.

The benchmark values for the set of Higgs masses are fixed by requiring
$m_{H^\pm}=180$ GeV and 250 GeV, for two demonstrative cases. These particular
values are in turn motivated by the fact that they ensure two different final
states that arise dominantly from the decays of the charged Higgs boson thus
leading to a $\tau$ enriched final state in the first case while for the latter
the final state is rich in bottom jets. The corresponding Higgs spectra are
given in Table 3.

In Table 1 we collect all other input parameters that are relevant for our
analysis. The parton distribution function we use is
CTEQ6L \cite{CTEQ, CTEQ1} which is one from the recent times. We have made a conservative choice of
the renormalization/factorization scale of $\sqrt{\hat s}$. 
In conformity with recent Tevatron estimates, the mass of the top
quark is set to 172 GeV \cite{Group:2008nq}.
%

\begin{table}[hbtp]
\begin{center}
\begin{tabular}{||c|c|c||} \hline\hline
PDF & Scale& $m_t$\\
 && in GeV\\

\hline
CTEQ6LII&$\sqrt{\hat{s}}$& 172\\
\hline
\end{tabular}
\label{tab1}
\end{center}

\caption{Common inputs}
\end{table}
\begin{table}[hbtp]
\begin{center}
\begin{tabular}{||c|c|c||} \hline\hline
Benchmark&$M_2$&$\mu$\\
Point&(in GeV) &(in GeV)\\
\hline
BP1&600&150\\
\hline
BP2&350&700\\
\hline
BP3&700&550\\
\hline
BP4&350&400\\
\hline
\end{tabular}
\label{tab1}
\end{center}
\caption{Benchmark points in the ($M_2-\mu$) plane.}
\end{table}

After mentioning $m_{H^{\pm}}$ and $\tan{\beta}$ which effectively determine the other Higgs masses. The corresponding Higgs spectrum is given by,

\begin{table}[hbtp]
\begin{center}
\begin{tabular}{||c|c|c|c||} \hline\hline
$m_A$&$m_h$&$m_{H^{\pm}}$&$m_H$\\
(in GeV) &(in GeV)&(in GeV) &(in GeV)\\
\hline
162 & 109 & 180 & 164 \\
\hline
238 & 109 & 250 & 239 \\
\hline
\end{tabular}
\label{tab1}
\end{center}
\caption{The Higgs mass spectra, $m_A$'s are so chosen as to ensure $m_{H^{\pm}}$= 180 and 250 GeV. }
\end{table}

\begin{table}[hbtp]
\begin{center}
\begin{tabular}{||c|c|c|c|c|c|c||} \hline\hline
Benchmark&$m_{\chi^{\pm}_1}$ &$m_{\chi^{\pm}_2}$&$m_{\chi^0_1} $&$m_{\chi^0_2}$&$m_{\chi^0_3}$&$\mntrl4$\\
Point&(in GeV) &(in GeV)&(in GeV) &(in GeV)&(in GeV) &(in GeV)\\
\hline\hline
BP1(U)&145.0&612.0&135.3 &155.2&308.2&611.8\\
\hline
BP2(U)&341.7&713.2&173.6&341.9&703.3&712.8\\
\hline
BP3(U)&529.6&724.5&345.0&533.5&552.9&724.4\\
\hline
BP4(U)&311.3&445.5&171.0&312.9&404.8&445.8\\
\hline
BP1(NU)&145.0&612.0&84.3 &156.6&160.6&611.7\\
\hline
BP2(NU)&341.7&713.2&99.0&341.9&703.4&712.5\\
\hline
BP3(NU)&529.6&724.5&98.6&530.4&553.2&724.2\\
\hline
BP4(NU)&311.3&445.5&97.7&312.2&405.0&445.5\\
\hline\hline
\end{tabular}
\label{}
\end{center}
\caption{The gaugino mass spectrum for the universal (U) and the non-universal (NU) scenarios corresponding to the benchmark points in Table 2. For the universal case $M_1$ is taken to be $M_2/2$ while for non-universal case $M_1=100$ GeV is set.}

\end{table}
\section{Production rates}
Once we settel for the set of benchmark points, the immediate requirement in
 the understanding of the production rates of different Higgs bosons under SUSY
cascade at each of these points in both universal and non-universal scenarios. In Tables 5 and 6 we present these rates. From these tables we can easily see 
(as expected from our discussion), that the behaviour for the lightest neutral
 Higgs and charged Higgs could be very different and some time opposite in some
 benchmark points in the sense that one production is larger than other and
 vice-versa. Again this behaviour can be flipped when one goes from the 
universal to the non-universal scenario and vice-versa for that particular
 benchmark point. We will go through each of the scenarios (benchmark points 
in both universal and non-universal scenarios) and try to see and
 explain its results\footnote{As expected from our earlier discussions and our
 observation in \cite{Bandyopadhyay:2008fp}, we can easily notice in Table 5 
and 6, the very characteristic relative changes in $\sigma^{cascade}_{h}\, \&
 \, \sigma^{cascade}_{H^{\pm}}$ in going from a universal situation to a
 non-universal one. This includes flipping of the relative strengths of 
$\sigma_{h} \, \& \, \sigma_{H^{\pm}}$ under these two broad situations 
(i.e., compare the ration of the rates in columns 2 \& 3 with that for columns
 6 \& 7) reflecting the ``cross-over''. Of course, the individual rates are
 also to be kept track of since they determine the observability of
 individual signals as would from our subsequent discussion.}.

\begin{table}[hbtp]
\begin{center}
\begin{tabular}{||c|c|c|c|c|c|c|c|c|c|c||} \hline\hline
&\multicolumn{4}{|c|}{ Universal}&\multicolumn{4}{|c||}{Non-universal}\\\hline
Benchmark &\multicolumn{4}{|c|}{Effective
cross-section}&\multicolumn{4}{|c||}{Effective
cross-section}\\
Points &\multicolumn{4}{|c|}{(in fb)}&\multicolumn{4}{|c||}{(in fb)}
\\\hline\hline

&$\sigma_{h}$&$\sigma_{H^+}$&$\sigma_A$&$\sigma_H$&$\sigma_{h}$&$\sigma_{H^\pm}$&$\sigma_A$&\multicolumn{1}{|c||}{$\sigma_{H}$}\\
\hline
BP1&765.3&312.8&167.2&243.7&220.0&304.1&160.1&\multicolumn{1}{|c||}{136.4}\\\hline
BP2&657.2&1.7&4.8&303.6&350.0&1198.7&137.7&\multicolumn{1}{|c||}{488.3}\\\hline
BP3&290.4&124.0&54.0&76.4&231.4&375.7&105.4&\multicolumn{1}{|c||}{100.8}\\\hline
BP4&948.0&14.5&4.7&5.2&582.5&694.0&79.2&\multicolumn{1}{|c||}{296.4}\\\hline
\end{tabular}
\label{tab1}
\caption{Effective production rates for the $h$, $H^{\pm}$, $A$ and $H$ for $m_H^{\pm}=180$ GeV with other parameters as given in Table 1.}
\end{center}
\end{table}
\begin{table}[hbtp]
\begin{center}
\begin{tabular}{||c|c|c|c|c|c|c|c|c|c|c||} \hline\hline
&\multicolumn{4}{|c|}{ Universal}&\multicolumn{4}{|c||}{Non-universal}\\\hline
Benchmark &\multicolumn{4}{|c|}{Effective
cross-section}&\multicolumn{4}{|c||}{Effective
cross-section}\\
Points &\multicolumn{4}{|c|}{(in fb)}&\multicolumn{4}{|c||}{(in fb)}
\\\hline\hline
&$\sigma_{h}$&$\sigma_{H^+}$&$\sigma_A$&$\sigma_H$&$\sigma_{h}$&$\sigma_{H^\pm}$&$\sigma_A$&\multicolumn{1}{|c||}{$\sigma_{H}$}\\

\hline
BP1&741.1&262.5&138.7&128.3&228.8&257.4&136.7&\multicolumn{1}{|c||}{127.3}\\\hline
BP2&892.8&1.2&0.7&0.6&656.9&1.3&6.9&\multicolumn{1}{|c||}{254.4}\\\hline
BP3&227&0.2&0.0&0.1&226.8&268.8&105.4&\multicolumn{1}{|c||}{100.8}\\\hline
BP4&930.4&6.1&0.3&3.4&831.6&15.5&4.4&\multicolumn{1}{|c||}{4.6}\\\hline
\end{tabular}
\label{tab1}
\caption{Effective production rates for the $h$, $H^{\pm}$, $A$ and $H$ for $m_H^{\pm}=250$ GeV with other parameters as given in Table 1.}
\end{center}
\end{table}
 
Let us first consider the case for  $m_{H^{\pm}}=180$ GeV (Table 5). If we now compare scenarios, BP1(U) and BP1(NU), we can see that the production rates of the lightest Higgs drops down while going from universal to non-universal scenario, whereas the production rates of charged Higgs remain similar. This can be understood if we look at the gaugino mass spectrum of BP1 (both U \& NU). With $\mu=150$ GeV, the masses for the lighter gauginos are mostly driven by $\mu$ 
and thus tend to be rather degenerate (see Table 4, for both universal and non-universal scenarios) for the `little cascade's to open up. Hence the entire
 cascade-contribution comes from the `big cascade' alone as explained earlier in section 2 and \cite{Bandyopadhyay:2008fp}. For the universal case, $\mntrl3\sim M_2/2$. Thus it can provide the necessary mass-splitting between the neutralinos for $\ntrl3\to h \ntrl1$ to open up. This is responsible for having
 larger production rates for the lightest Higgs boson compared to the charged Higgs boson as, $\ntrl3\to H^{\pm} \chi^{\mp}_1$ is yet to open up. In the 
corresponding non-universal case (with $\mu>M_1$) it turns out that $\mntrl3\sim \mu$ and it is very easy to see from the BP1 case in Table 4 that $\mntrl3$ decays either to $h$ or to $H^{\pm}$ is kinematically forbidden.

In BP2, $\mu\gg \mtwo$. Thus, the masses of the lighter gauginos are determind by the values of $M_1$ and $M_2$ chosen for the study. Hence, a larger 
splitting can be obtained between the low-lying gauginos. This may potentially open up the `little cascade', which is clear from Table 4. For the universal 
case, $\mntrl2, \sim \mtwo$ and $\mntrl1 \sim \mone\sim \mtwo/2$. This opens up $\ntrl2\to h \ntrl1$ but forbids $\ntrl2\to H^{\pm} \chi^{\mp}_1$ and $\chi^{\pm}_1 \to H^{\pm}\ntrl1$ for our choice of $M_2$ and $\mu$)\footnote{In restrospect, the features in the variation of cross-sections ( the 'cross-over's) which were exploited to settel for the benchmark points depend very much on the fact that the `little cascades' are kinematically favoured in this part of the parameter space. For example, if $m_{\ntrl2}<2m_{h}$ (with $m_{\ntrl2}\simeq{2\ntrl1}$), then the `little cascade's would not open up and the features would be absent.}. Thus, rates for $h$ would be larger than that for $H^{\pm}$ in the universal case. As discussed in earlier in section 3, for the non-universal scenario
 we fix $M_1$=100 GeV thus having $\mntrl1 \sim \mone\sim 100$ GeV. This would then open up $\chi^{\pm}_1 \to H^{\pm}\ntrl1$ leading to an increment in the production rates of the charged Higgs. For the same reason, $\ntrl2\to A/H \ntrl1$ also opens up in the neutral counterpart, resulting in an effective
 suppression in the production rates of the lightest neutral Higgs boson.

With $M_2$ and $\mu$ becoming closer, BP3 becomes a mixed scenario unlike the
 previous two cases. Here, for the universal case  $\mntrl3,\mntrl2,m_{\chi^{\pm}_1}\sim{|\mu|}$,  $\mntrl4, m_{\chi^{\pm}_2}\sim{\mtwo}$ and  $\mntrl1 \sim \mone\sim \mtwo/2$. The decays, $\chi^0_{2,3}\to h \ntrl1$ are responsible for $h$-production. As for the charged
 Higgs boson, $\chi^{\pm}_1 \to H^{\pm}\ntrl1$  barely opens up. As before, in the non-universal scenario, $\mntrl1 \sim \mone\sim 100$ GeV.
 Thus, the rates for the charged Higgs boson increase thanks to larger available  phase space. In the neutral sector, the  production rates
for $A$, $H$  also increase because of a large available phase space thus effectively decreasing the rates for the light Higgs boson.\\

BP4 represents another mixed region with $\mntrl4,\mntrl3,m_{\chi^{\pm}_2} \sim |\mu|$, $\mntrl2, m_{\chi^{\pm}_1}\sim \mtwo$ and  $\mntrl1 \sim \mone\sim \mtwo/2$ for the universal case. Thus, for the lightest neutral Higgs boson, $\chi^0_{2,3}\to h \ntrl1$  channels are open but  $\chi^0_{2}\to A/H \ntrl1$ are kinematically forbidden. For the charged Higgs boson, none other than  $\chi^{\pm}_2 \to H^{\pm}\ntrl1$ is open. Thus, for the universal case,  the rate for the lightest Higgs boson is much larger than that for the cahrged Higgs boson. But in
 the non-universal case, as $\mntrl1 \sim \mone\sim 100$ GeV, the decays $\chi^{\pm}_{1,2} \to H^{\pm}\ntrl1$ open up resulting in a large
 rate for $H^{\pm}$ in the final state. On the other hand, in the neutral sector,  $\chi^0_{2}\to A/H \ntrl1$ are now kinematically allowed,
 leading to a suppression in the production of lightest Higgs boson.\\

For a heavier Higgs spectrum with $m_{H^{\pm}}=250$ GeV, in all of these four scenarios, the production rates for $H^{\pm}$, $A$ and $H$ experience phase-space suppression that affects the entire SUSY cascade leading to the Higgs bosons. In particular, for the non-universal cases under BP2 and BP4 the decays, $\chi^{\pm}_1 \to H^{\pm}\ntrl1$ do not open up thus keeping the production rates for charged Higgs low.

\section{Signal and Backgrounds}

Typically, the cascades occurring from the strong $2\to2$ processes like, $\tilde{g}\tilde{g}$, $\tilde{q_i}\tilde{q_j}$ and $\tilde{g}\tilde{q_i}$ would involve more number  of jets and large  missing energy in the final state. Higgs bosons produced in such cascades are to be identified  through their decay products as long as the former decay visibly, which is what we are interested in this work. Let us first discuss the case for the Higgs bosons. A generic signal for the Higgs bosons would be

$$n_{jets} \; (with \; at \; least \; two \;b-jets)+\not{p_T}$$
where  the invariant mass of the pair of $b$-jets has a reconstructed peak indicative of the mass of the neutral Higgs boson. Again, generically, such multijet signals at the LHC would have prohibitively large Standard Model QCD background. On top of that, backgrounds coming from different SUSY processes would be common and could be significant since cascades open up multiple alternate ways in which a particular final state can be obtained.

One, thus, has to study the appropriate backgrounds for this final state  extremely carefully and choose the signal-characteristics accordingly in order to optimize the signal to background ratio. This means, to specify the signal completely, one has to tailor the kinematic cuts appropriately by a detail understanding og the impacts of these cuts on both signal and the backgrounds.

\subsection{The Standard Model backgrounds}
 Let us first discuss the possible SM backgrounds. As common to many signals of physics beyond the standard model at the LHC, the signal discussed earlier above would also have significant background from $t\bar{t}$ production because of its large cross-section. Also, other
 QCD processes involving light quarks  like ($b\bar{b},c\bar{c}$ etc.) will contribute to the SM background. However, it is rather difficult to produce reliable QCD samples for these processes because of the requirement of extreme kinematical fluctuations necessary for this. However, the cuts which are effective to suppress $t\bar{t}$ are also expected to be effective in suppressing  these light quark contributions \cite{Datta:2003iz, Huitu:2008sa}.     

In practice, $t\bar{t}$ background can be effectively suppressed by exploiting some standard kinematic features of the final state it leads to vis-a-vis the ones for the signal. The differential distributions in the kinematic variables like the jet multiplicity, the missing transverse momentum, the transverse momentum of the hardest jet and the so-called effective mass could be characteristically different for the signal when compared to the $t\bar{t}$ background.

 In the following,  we study all these distributions in a little more detail to optimize the kinematic cuts for our subsequent analysis. To make our analysis a little more realistic we go one step beyond the bare parton level analysis. We simulated events (for both Standard Model and SUSY) with an event generator like  PYTHIA-6.4.13 \cite{Sjostrand:2006za}. Jets are constructed using PYCELL, the toy-calorimeter algorithm  default to PYTHIA with appropriate basic cuts. To add to further robustness, we incorporate the effects due to initial and final state radiations (ISR and FSR) through appropriate PYTHIA switches. For other standard inputs used in this analysis we refer to Table 1.
No explicit detector simulation is done. However, in some cases, some realistic detector effects are implicitly taken care of. These are indicated as and when incorporated. 

\begin{itemize}
\item {\bf Jet multiplicity distribution:} As pointed out earlier in the beginning of this section, signals under SUSY cascades could in general involve larger number of  jets. Obviously, when compared to final states arising from $t\bar{t}$, which can only contain a limited number of jets (at the parton level), jet multiplicity could emerge to be an effective discriminator. 

In Figure 1, we compare the  jet-multiplicity distributions for  the $t\bar{t}$ initiated final state and the ones arising from the SUSY cascades. Each plot contains two overlapping distributions. The red (grey) one stands for $t\bar{t}$ while the blue (black) one represents SUSY cascades. The left panel illustrates the case for the universal scenarios and the right one is for the corresponding non-universal ones. The four rows correspond to the benchmark scenarios, BP1, BP2, BP3 and BP4 as described in Table 2. The plots are generated with large enough samples of events (of same but arbitrary size) to ensure their reliability. The generic observation from Figure 1 is that for SUSY cascades the jet-multiplicity peaks at a higher value compared to the $t\bar{t}$ case. Note that,  for the $t\bar{t}$ case, it is  the same distribution for all the eight plots . It is thus clear  that a choice of  jet-multiplicity  $n_{jets}\geq5$ would optimize the signal at this level, against the $t\bar{t}$ background.

\item {\bf Missing-$p_T$ ($\not{p_T}$) distribution:} Figure 2 illustrates the  $\not{p_{T}}$ distributions for all the four scenarios as described in the case of jet-multiplicity. The  blue (dotted) lines represent the $t\bar{t}$ case while the red (solid) ones are for the SUSY cascades. Again, for the normalization we follow the same convention as for the case of jet-multiplicity (Figure 1). The plots show that $t\bar{t}$, which is the same for all the eight plots, peaks at around 50 GeV while the corresponding SUSY distributions are flatter in nature with peaks smeared at around 300 GeV. The locations of the peaks reflect the choice of input SUSY masses for squarks and gluinos (see section 3). The flatness of the plots is expected to be typical for events from cascade. This is because a little too many mass differences could be involved under a given cascade resulting in different amount of missing energies. As is obvious from the plots, requiring  $\not{p_T}>150$ GeV would efficiently remove the background from $t\bar{t}$.

\item {\bf Hardest jet $p_T$ distribution:} With the same convention as in the previous case,  Figure 3 demonstrates the $p_T$ distributions of the hardest-jet in the events for both $t\bar{t}$ and SUSY cascades. Clearly, requiring   $p^{(hardest \, jet)}_{T}\geq 300$ GeV would remove most of the $t\bar{t}$ backgrounds.
\item {\bf Effective mass:} The effective mass for a given final state is defined as the scalar sum of the transverse momenta of the jets and leptons present and the missing transverse energy, i.e., 
$$M_{eff}=\sum{p^{j_{i}}_{T}} +\sum{p^{\ell}_T} + \not{p_{T}}.$$
 
The peak in this distribution could be a rough indicator of the total mass of the particles produced at the top of the cascade. In Figure 4 we plot these distributions, the conventions being the same as before. Expectedly, a cut of $M_{eff}\geq 1200$ GeV would help bringing down the $t\bar{t}$ contribution to an insignificant level.

\end{itemize}

\noindent
While, with the help of all these cuts $t\bar{t}$ background can be reduced to a minimum, it is true that these quantities are not entirely independent of each other. Thus, there may remain  some avoidable redundancy in these cuts for as the backgrounds concerned which affecting the signal non-trivially. Nevertheless, being on a conservative side we retain the above-mentioned cuts.

Also, there are possible all-jet SM events (from $Z+n_{jets}$ and  $W^{\pm}+n_{jets}$ etc.). But, as backgrounds, they are subdominant since they carry only very little missing $p_T$. In Table 7 we list all the basic cuts that we impose to minimize the SM backgrounds. 

We enter the next phase of our analysis with events that satisfy the kinematic criteria listed in Table 7, thus having the $t\bar{t}$ background already under control. The acceptance of $t\bar t$ events under the basic cuts is also indicated in Table 7.
\begin{table}[hbtp]
\begin{center}
\begin{tabular}{||c|c||} \hline\hline
Basic Cuts & Number of events(Acceptance)\\
\hline
$n_{jet}\ge 5$ & 731913(0.18)\\
\hline
$\not{p_T}\ge 150$ GeV&80902(0.11)\\
\hline
$M_{eff}\ge 1200$ GeV &11306(0.14)\\
\hline
$p^{hardest-jet}_T\ge 300$ GeV&10771(0.95) \\
\hline
\hline
\end{tabular}
\label{tab1}
\end{center}
\caption{List of basic cuts imposed and the corresponding acceptances of $t\bar{t}$ events. A total of $4\times{10^6}$ $t\bar{t}$  events were generated corresponding to an integrated luminosity of $\mathcal{L}\sim 10$ fb$^{-1}$.}
\end{table}

\subsection{The SUSY background} We have already indicated the origin of a generic  SUSY background in the beginning of the present section.  To be specific in the present context, SUSY background to the signal of a particular Higgs boson search is constituted of those events which are similar to the signal events but do not contain that Higgs down the cascade. For example, while looking for the lightest neutral Higgs boson, SUSY background would be comprised of everything except those originating from the lightest Higgs boson. This means that while we look for a specific Higgs via hunting a peak in a suitable kinematic distribution (\emph{viz}., some invariant mass, say) events that contribute to the background would form a continuum. As we will see subsequently in this work, the SUSY background can be serious and thus calls for dedicated treatments. It may also need a case by case analysis in which some special kinematic cuts are to devised as would be discussed in sections 6 and  7.

The issues are similar in the search for the  charged Higgs under SUSY cascades. However, as we will see later, the signal for the charged Higgs boson under SUSY cascades would naturally be different from that for the neutral Higgs cases. A detailed study of the corresponding backgrounds is also discussed in sections 6 and 7. 


\begin{figure}[hbtp]
\begin{center}
\vspace*{-2.2cm}
{\hspace*{-1in}{\epsfig{file=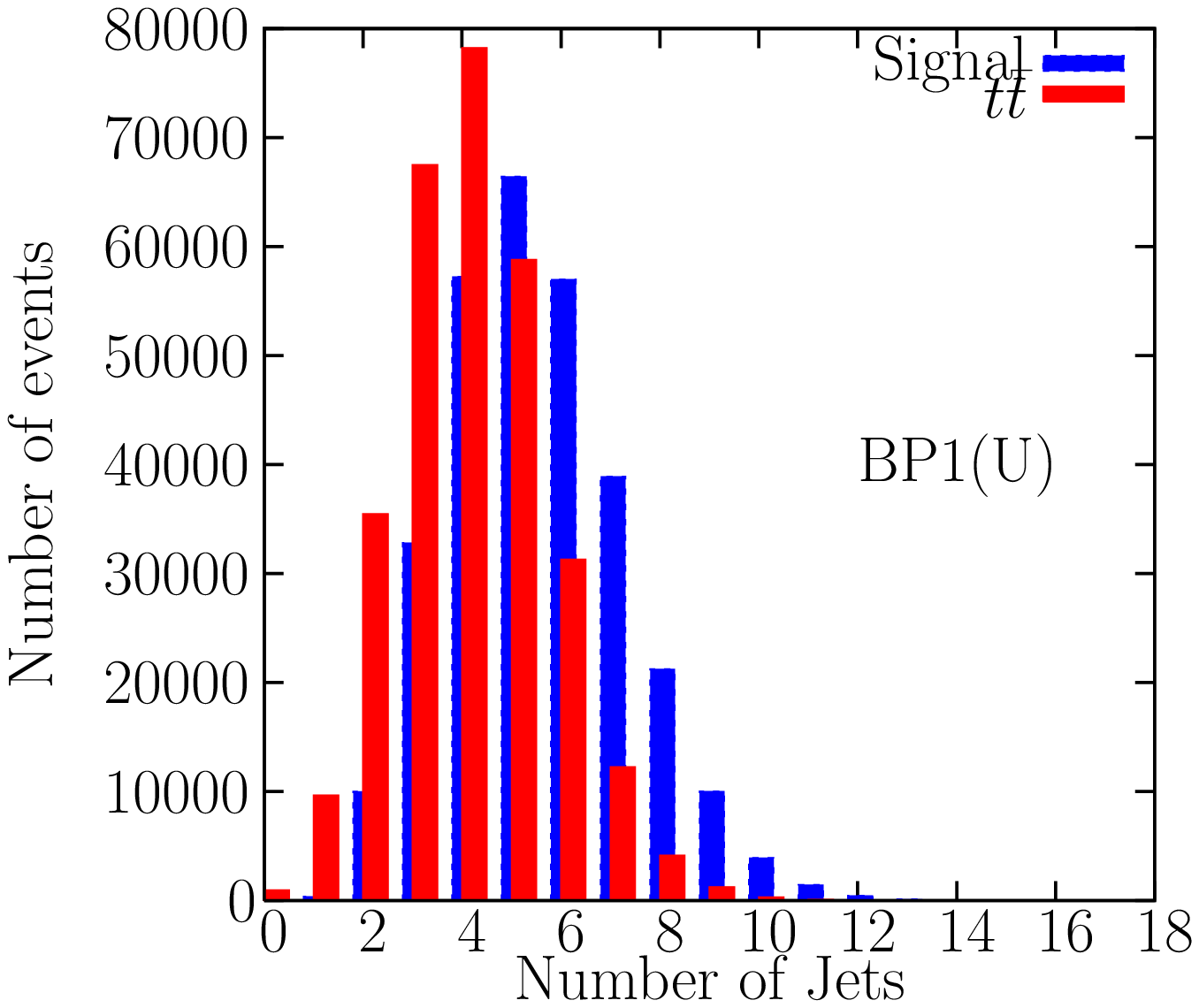,width=10.0 cm,height=12.0cm,angle=-0.0}}
\hskip -12pt 
\hspace*{-1in}{\epsfig{file=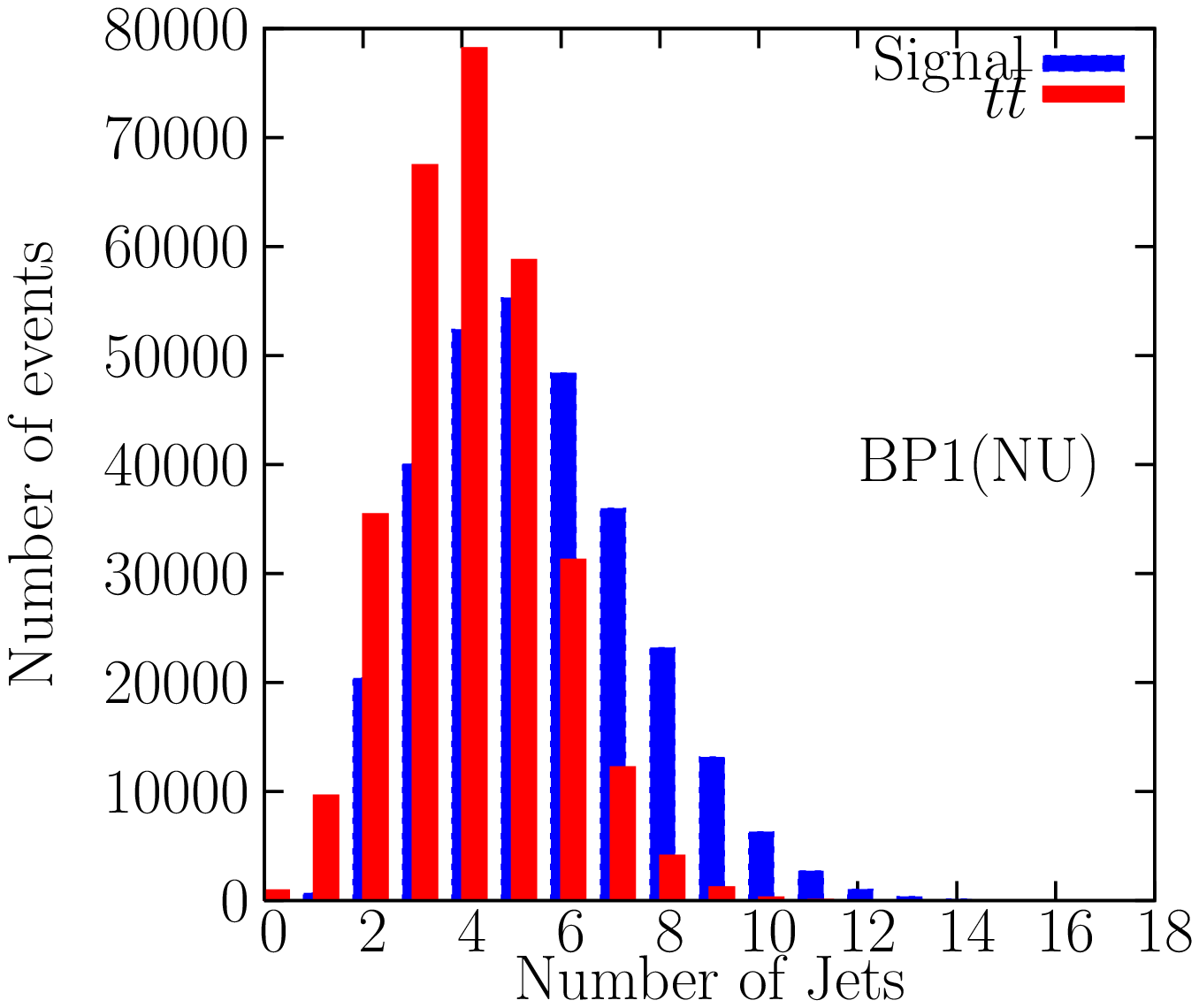,width=10.0cm,height=12.0cm,angle=-0.0}}}
\vskip -20pt
\vspace*{-6.cm}
\hspace*{-1in}{\epsfig{file=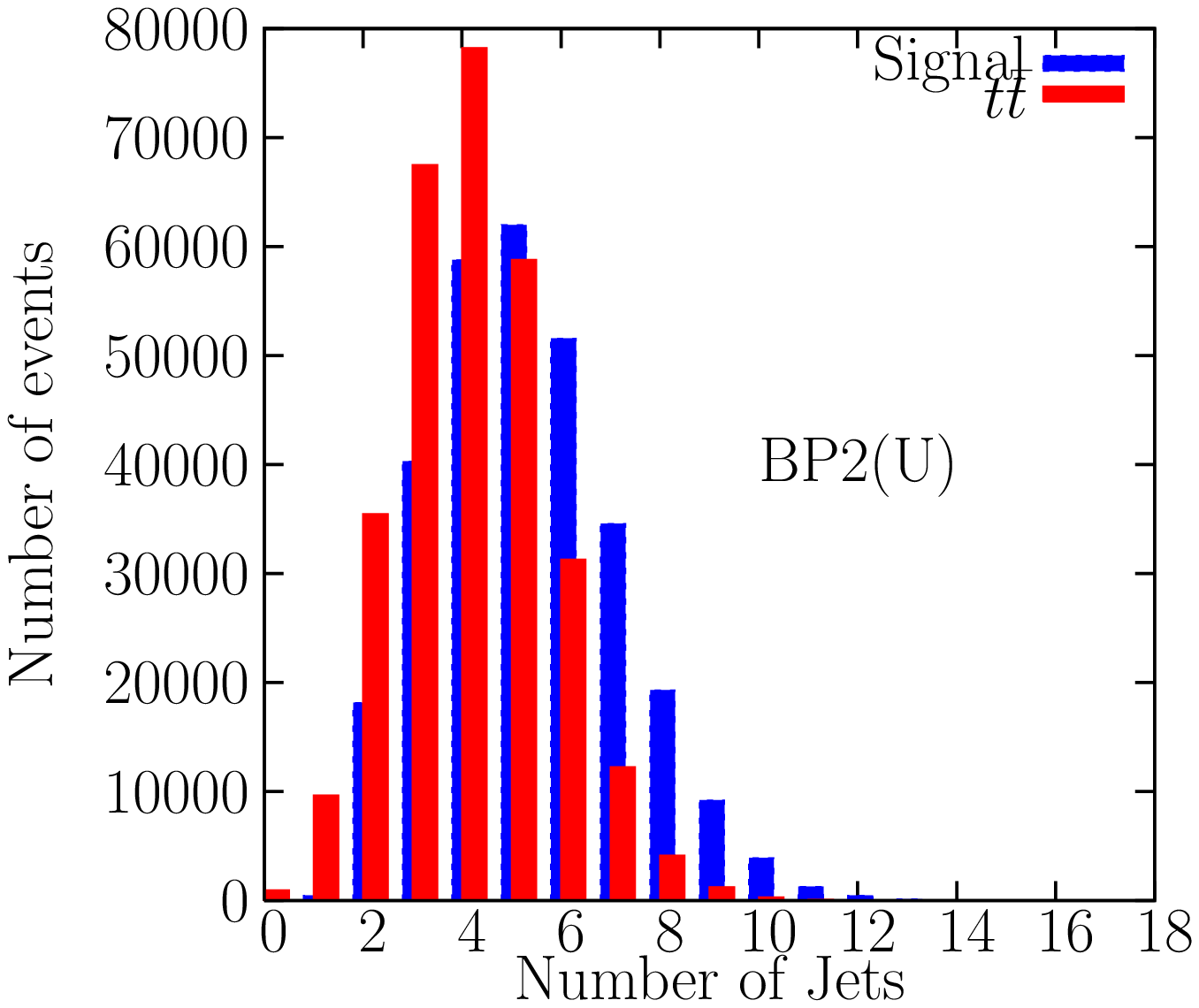,width=10.0 cm,height=12.0cm,angle=-0.0}}
\hskip -12pt 
\hspace*{-1in}{\epsfig{file=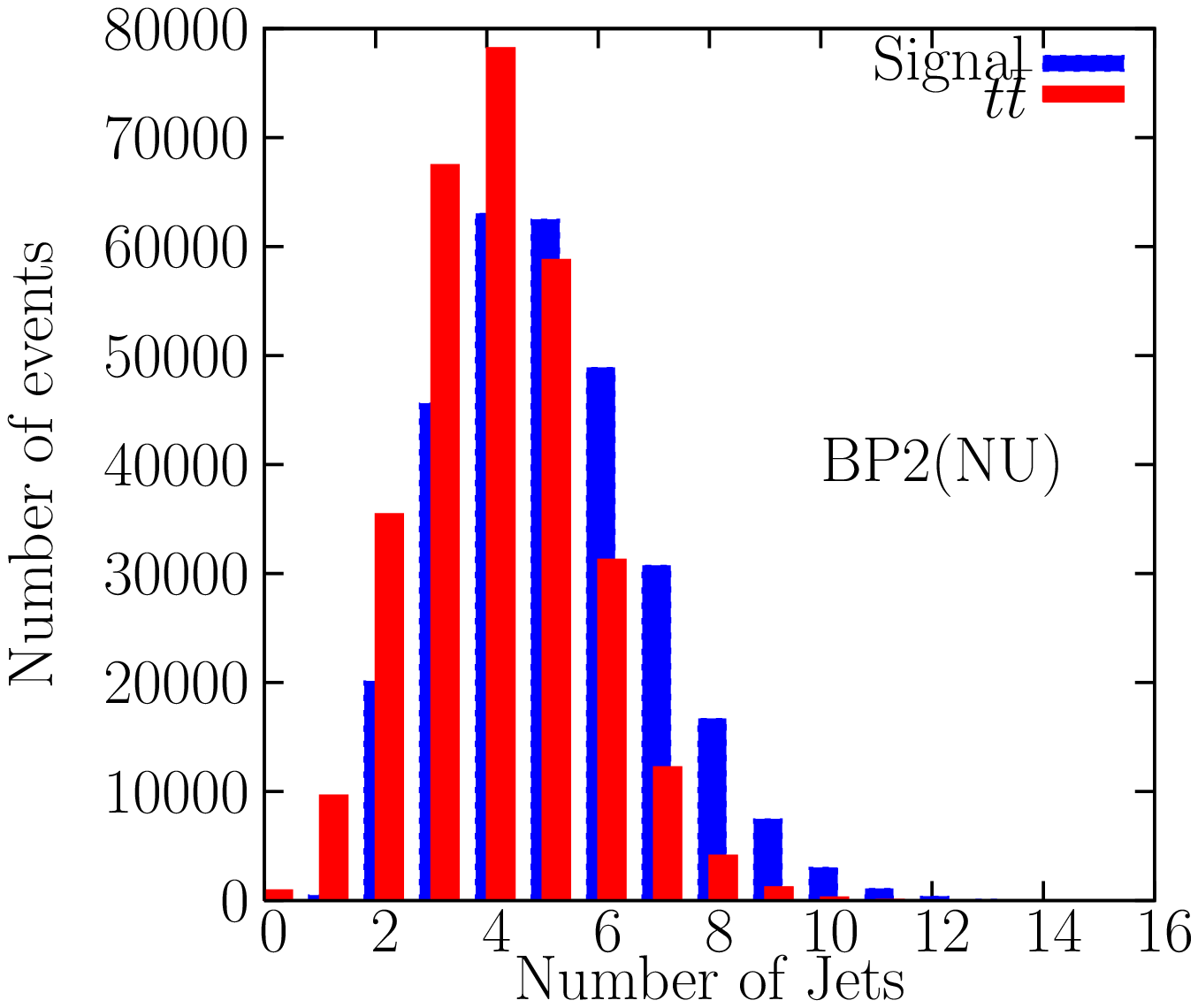,width=10.0cm,height=12.0cm,angle=-0.0}}
\vskip -20pt
\vspace*{-6.cm}
\hspace*{-1in}{\epsfig{file=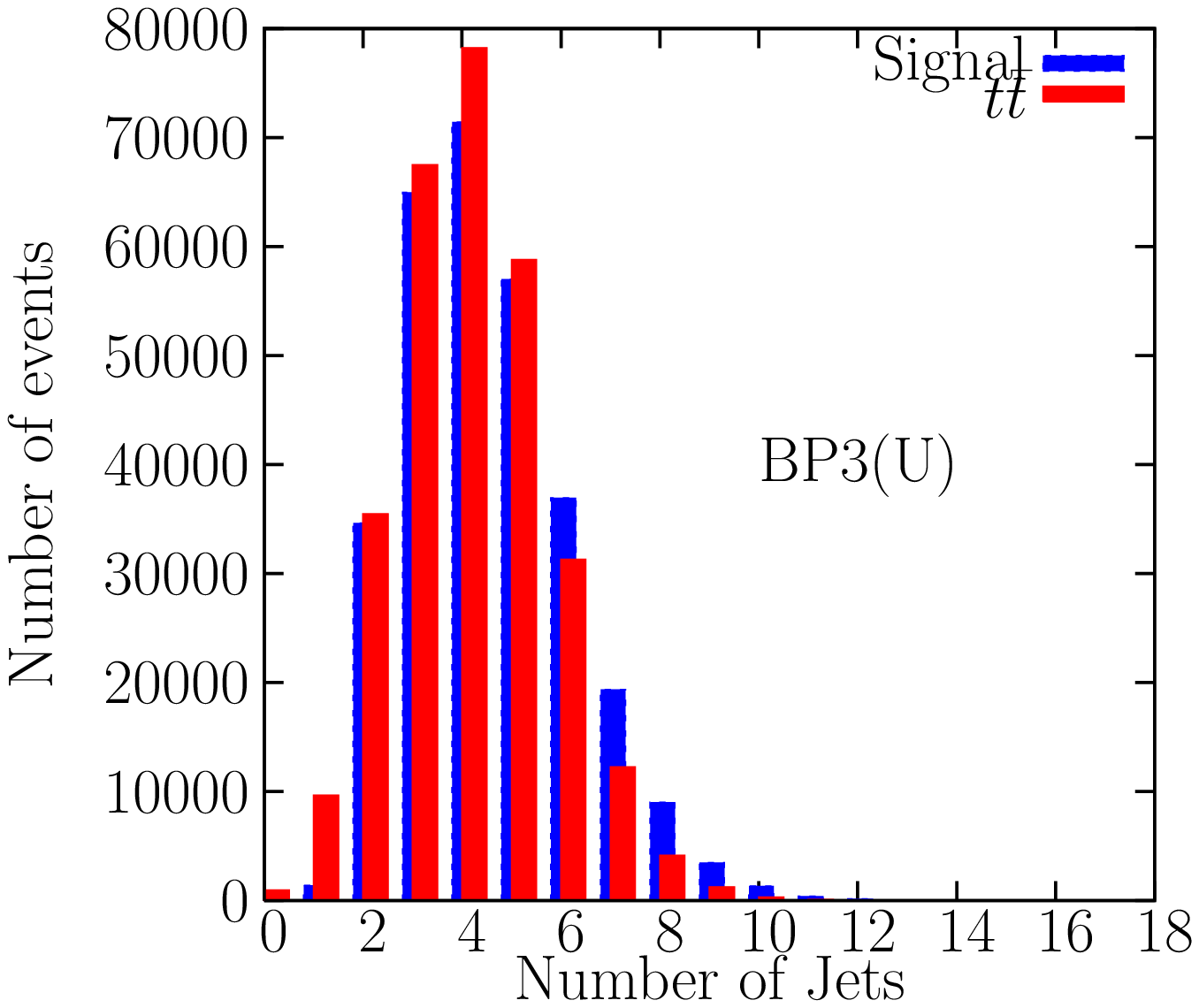,width=10.0 cm,height=12.0cm,angle=-0.0}}
\hskip -12pt 
\hspace*{-1in}{\epsfig{file=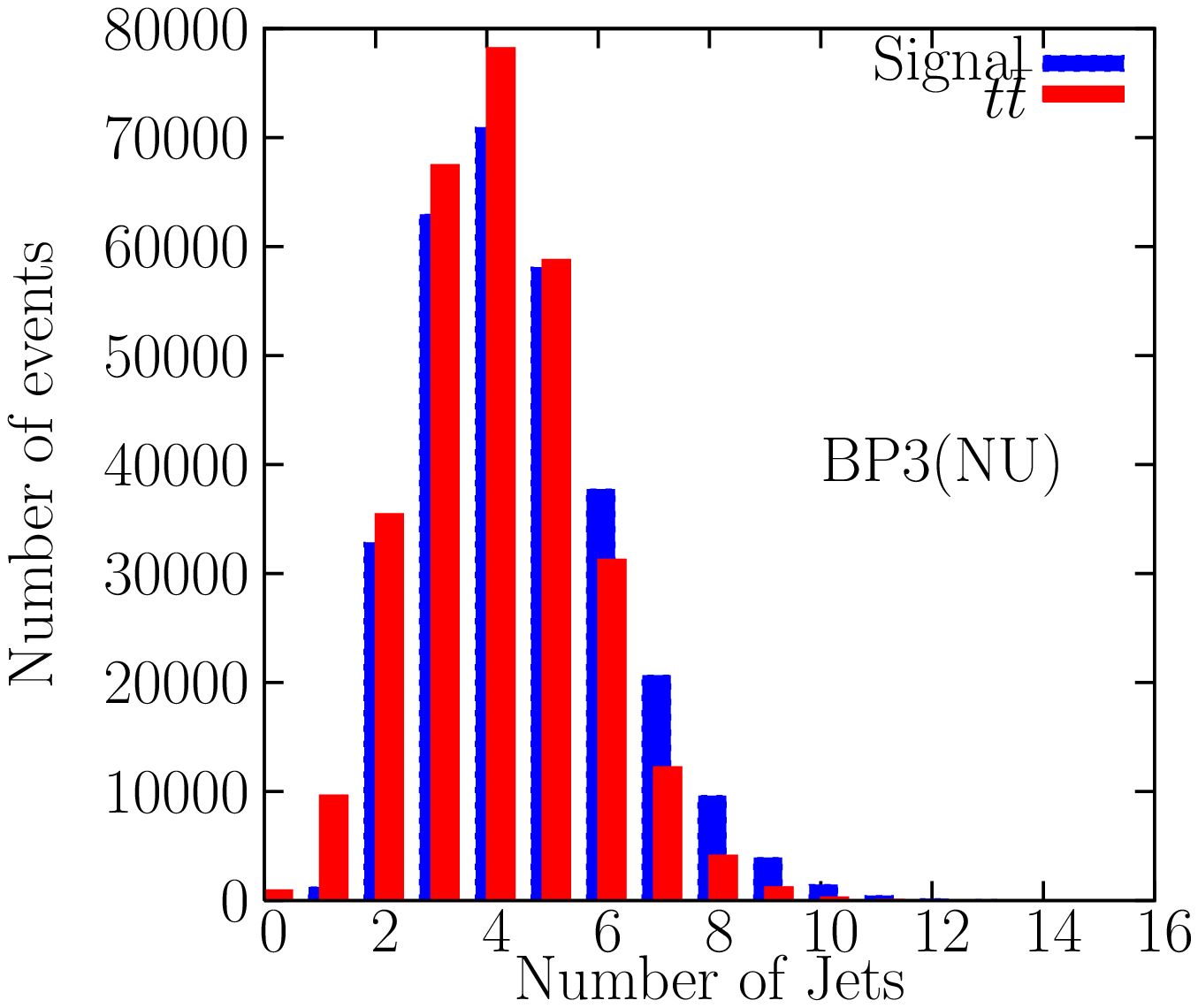,width=10.0cm,height=12.0cm,angle=-0.0}}
\vskip -20pt
\vspace*{-6.cm}
\hspace*{-1in}{\epsfig{file=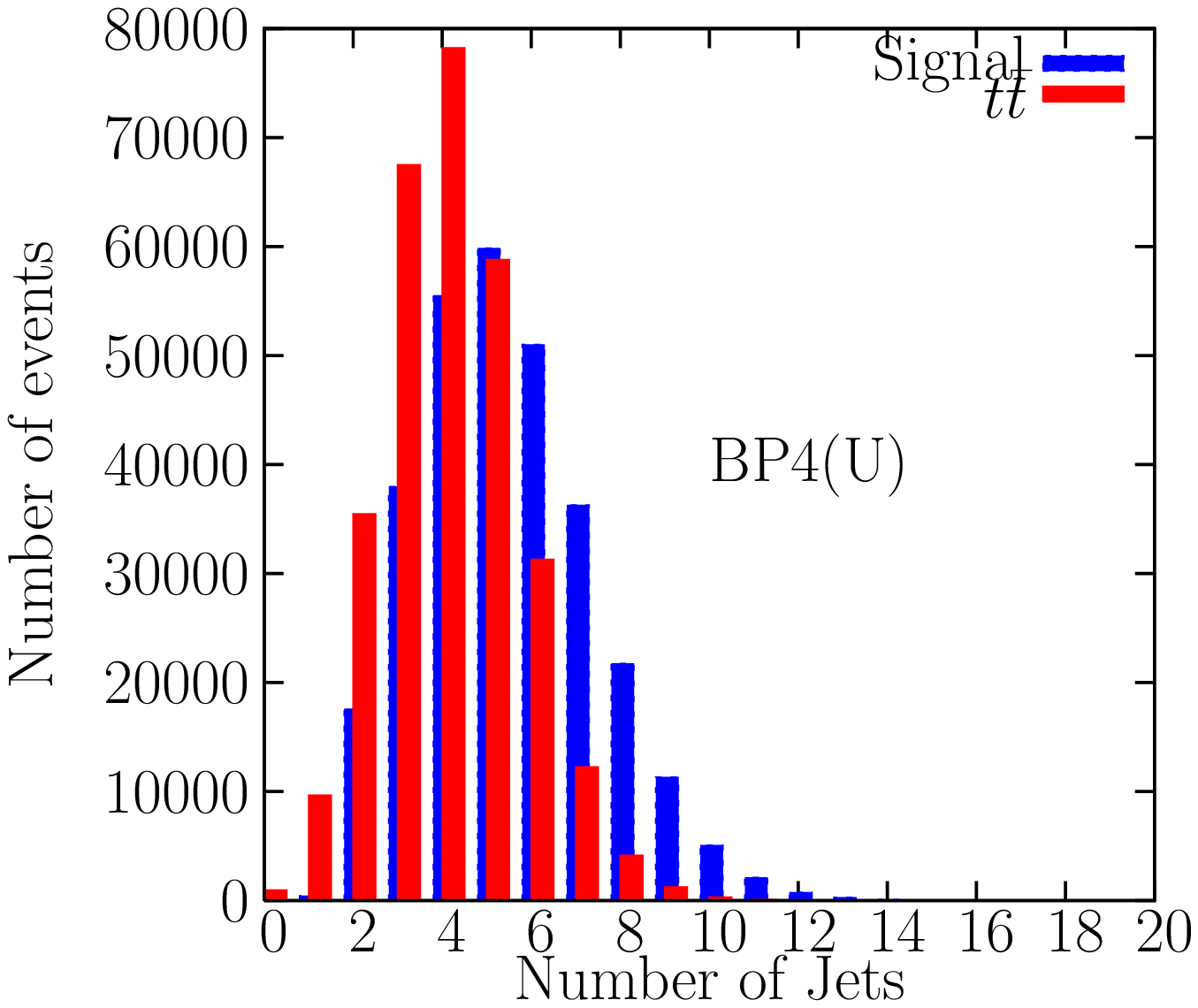,width=10.0 cm,height=12.0cm,angle=-0.0}}
\hskip -12pt 
\hspace*{-1in}{\epsfig{file=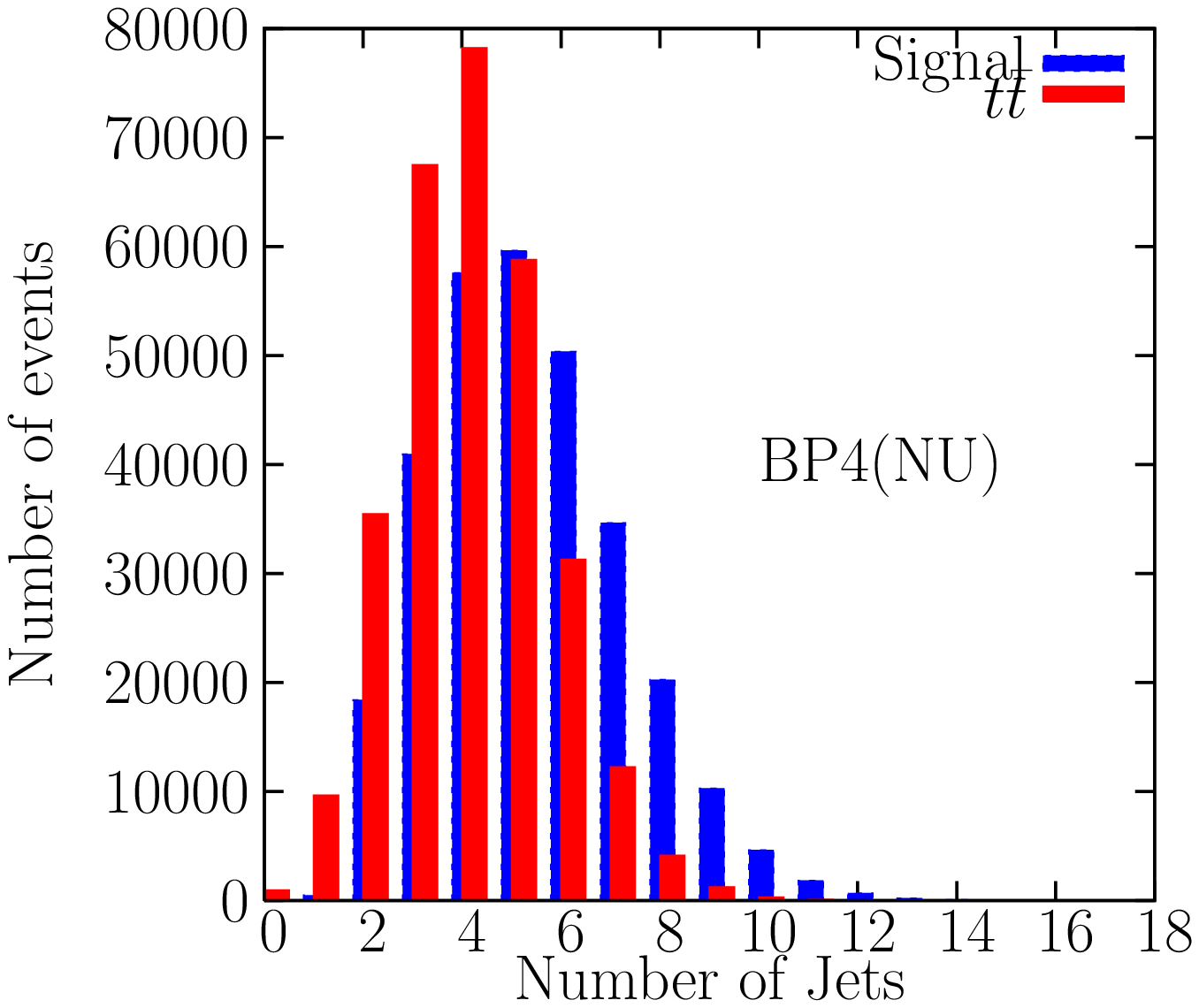,width=10.0cm,height=12.0cm,angle=-0.0}}
\vspace*{-5.5cm}
\caption{Jet multiplicity distribution for universal (left) and non-universal (right) scenarios } 
\end{center}
\label{fig11}
\vspace*{-1.0cm}
\end{figure}

\begin{figure}[hbtp]
\begin{center}
\vspace*{-2.2cm}
{\hspace*{-1in}{\epsfig{file=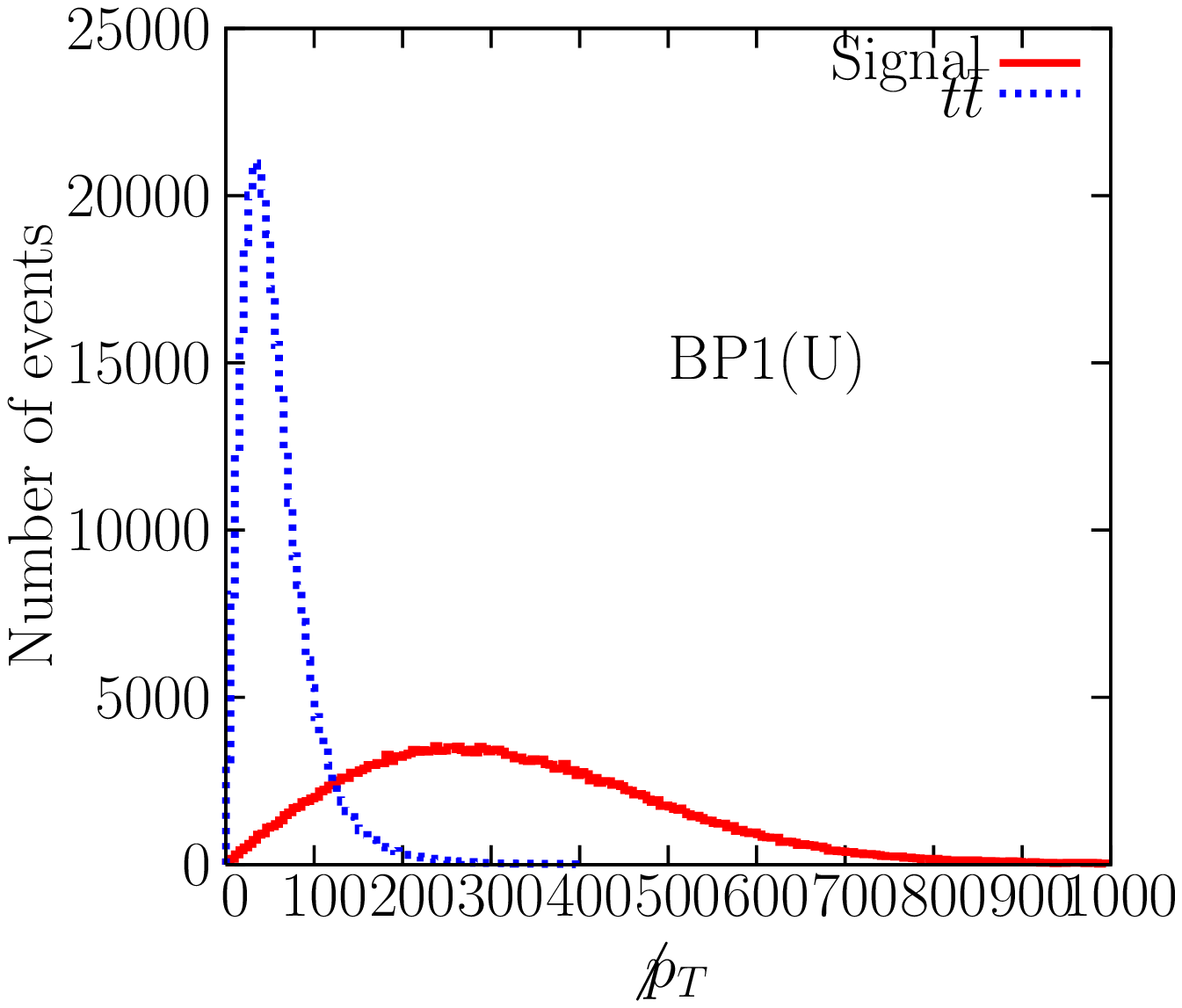,width=10.0 cm,height=12.0cm,angle=-0.0}}
\hskip -12pt 
\hspace*{-1in}{\epsfig{file=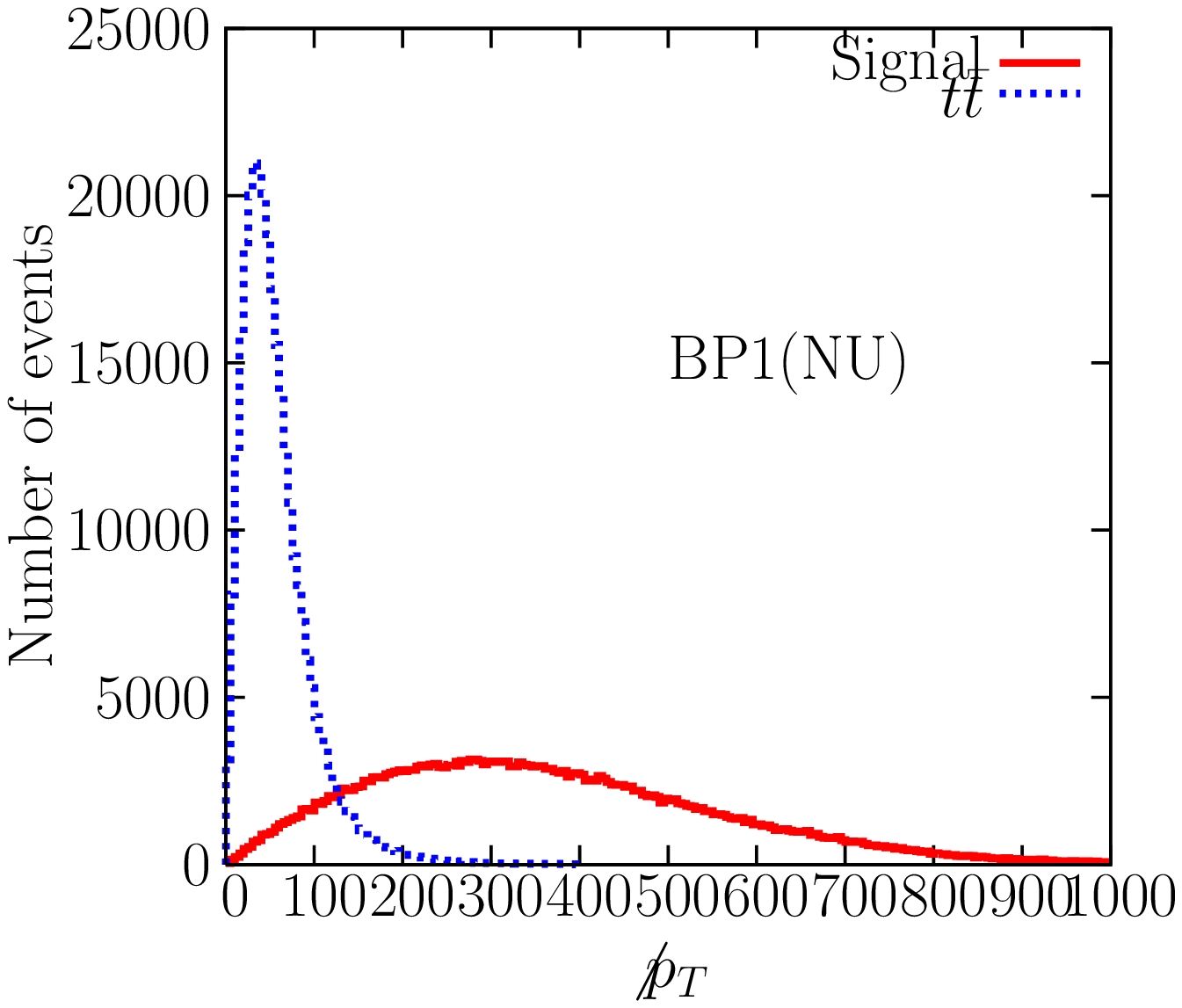,width=10.0cm,height=12.0cm,angle=-0.0}}}
\vskip -20pt
\vspace*{-6.cm}
\hspace*{-1in}{\epsfig{file=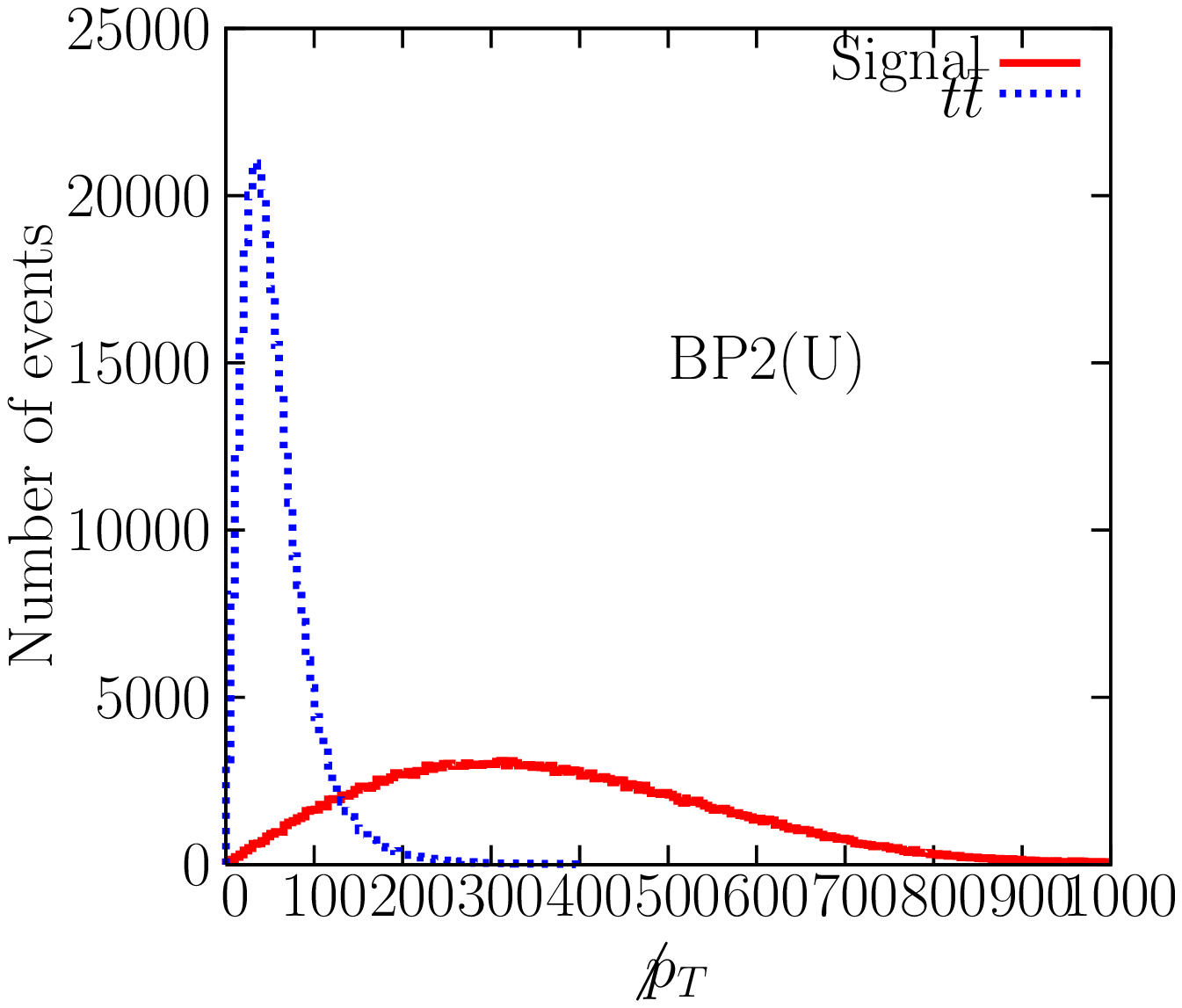,width=10.0 cm,height=12.0cm,angle=-0.0}}
\hskip -12pt 
\hspace*{-1in}{\epsfig{file=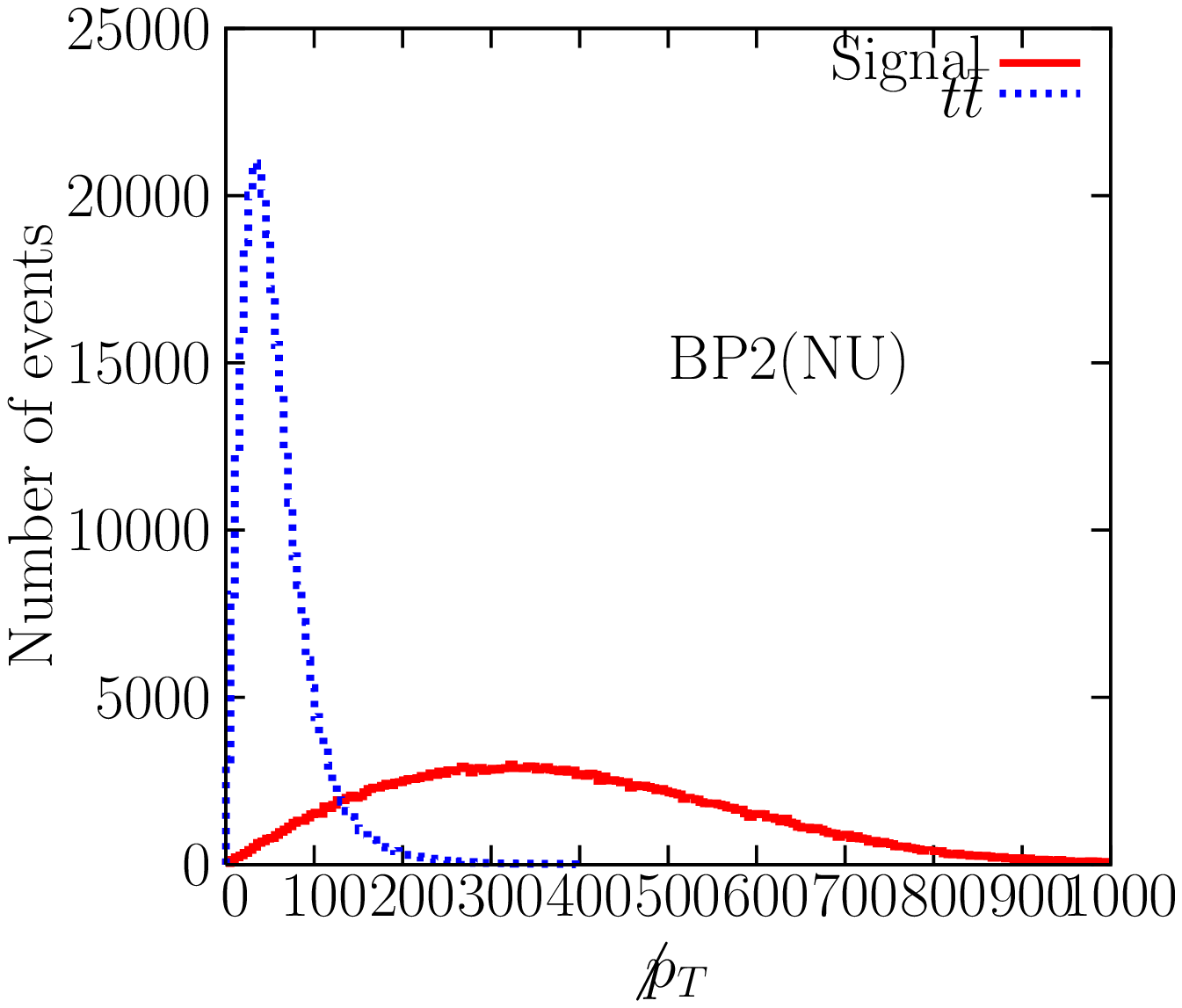,width=10.0cm,height=12.0cm,angle=-0.0}}
\vskip -20pt
\vspace*{-6.cm}
\hspace*{-1in}{\epsfig{file=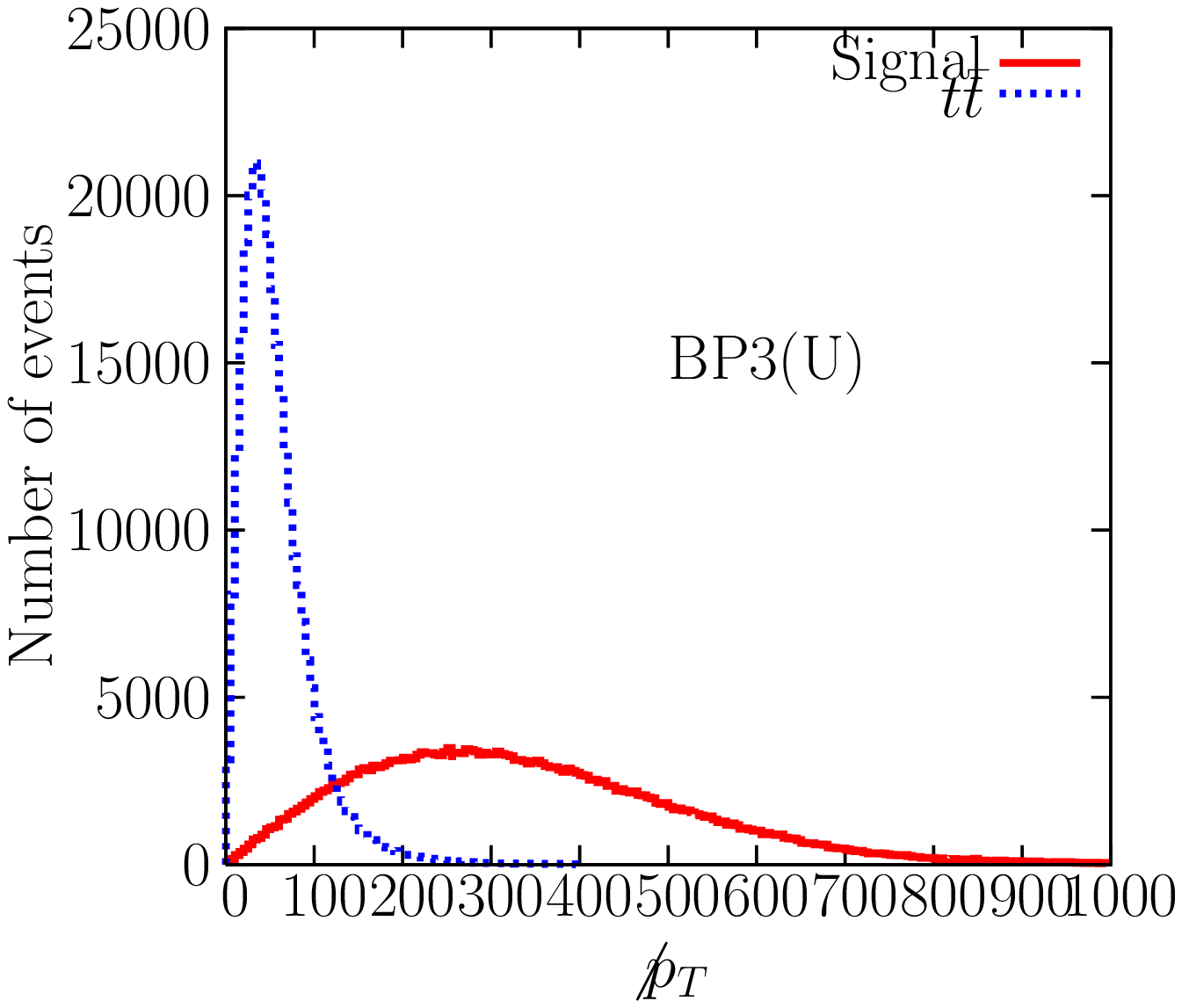,width=10.0 cm,height=12.0cm,angle=-0.0}}
\hskip -12pt 
\hspace*{-1in}{\epsfig{file=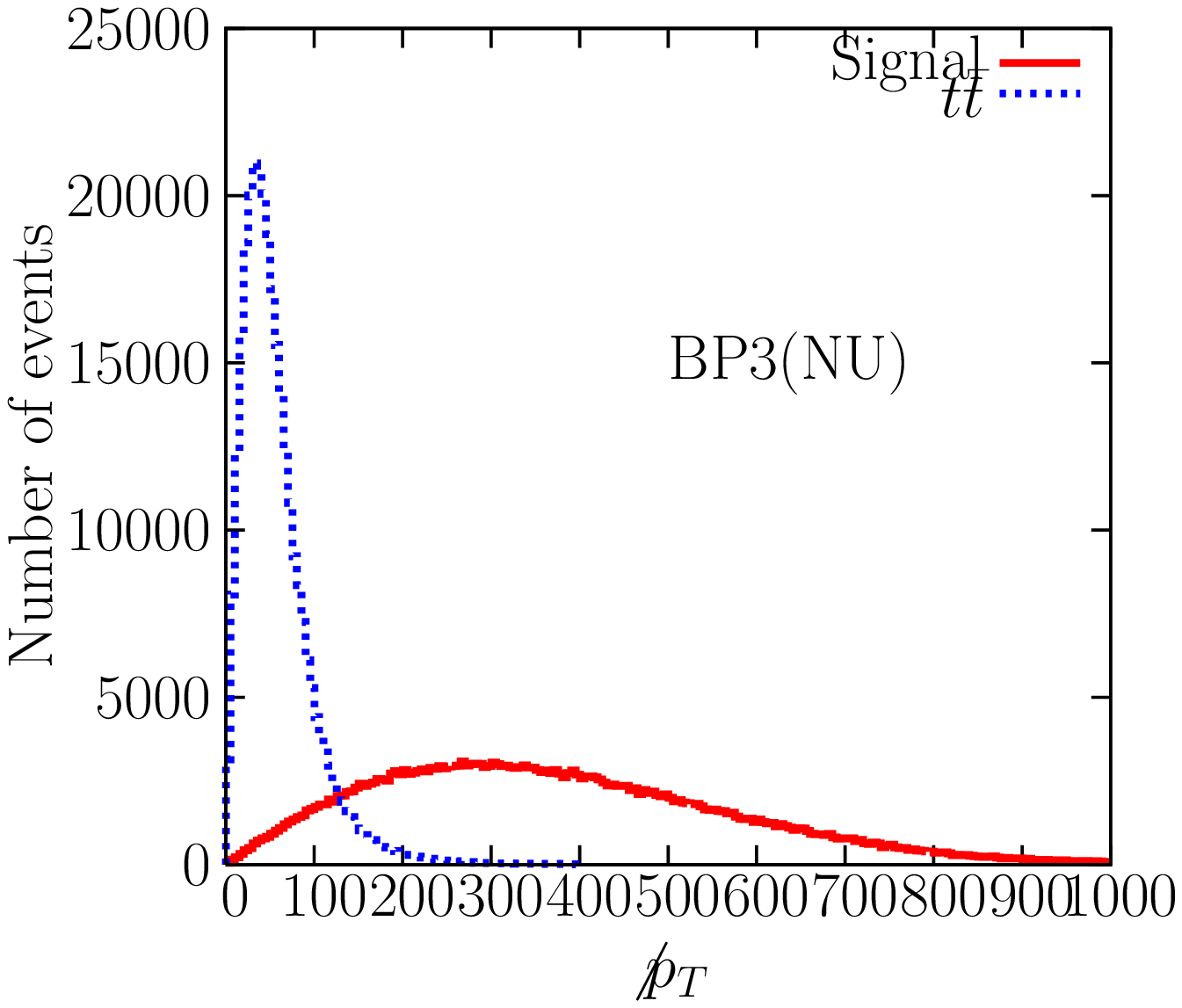,width=10.0cm,height=12.0cm,angle=-0.0}}
\vskip -20pt
\vspace*{-6.cm}
\hspace*{-1in}{\epsfig{file=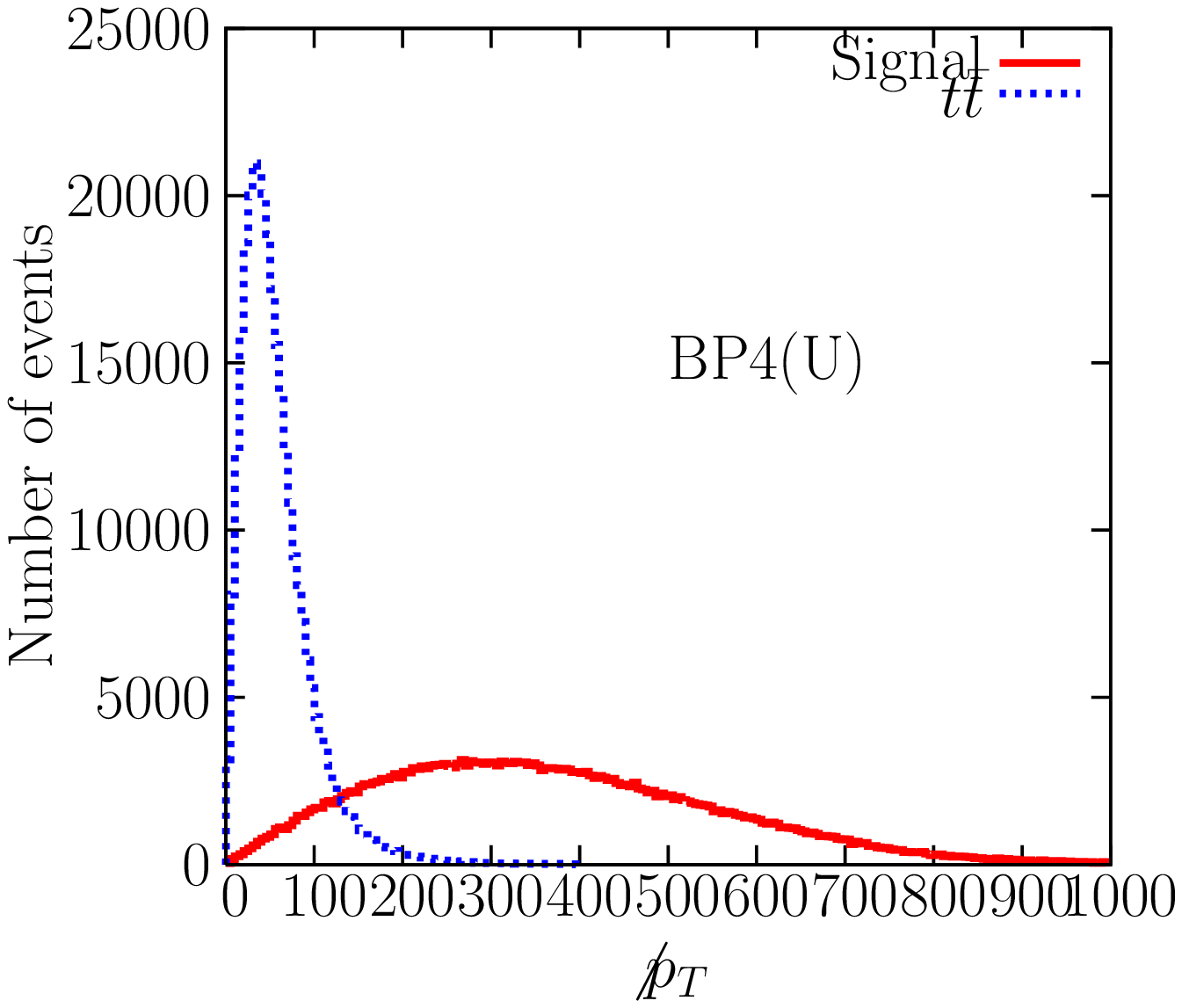,width=10.0 cm,height=12.0cm,angle=-0.0}}
\hskip -12pt 
\hspace*{-1in}{\epsfig{file=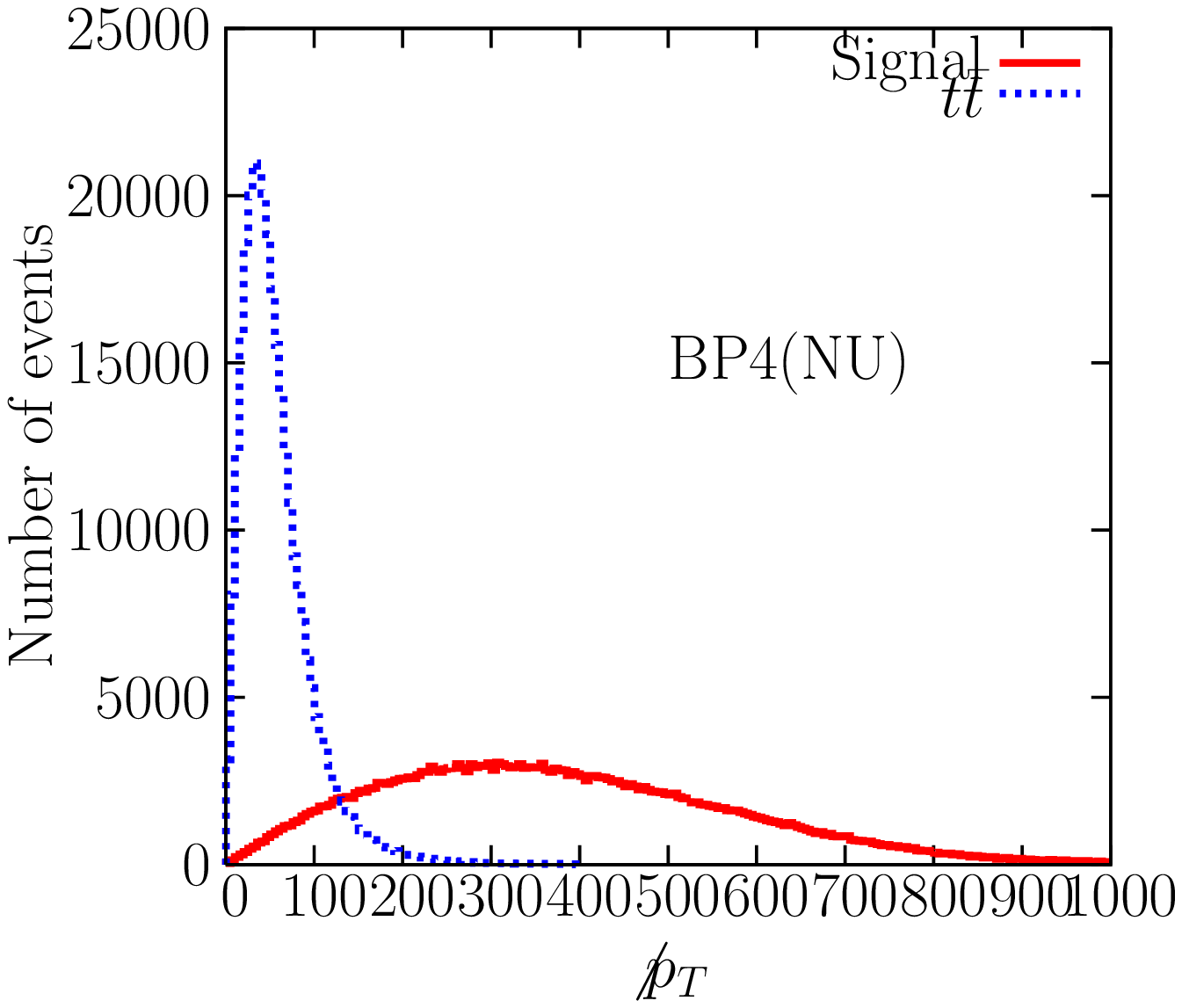,width=10.0cm,height=12.0cm,angle=-0.0}}
\vspace*{-5.5cm}
\caption{$\not{p_T}$ (in GeV) distribution for universal(left) and non-non-universal(right) scenarios } 
\end{center}
\label{fig11}
\vspace*{-1.0cm}
\end{figure}

\begin{figure}[hbtp]
\begin{center}
\vspace*{-2.2cm}
{\hspace*{-1in}{\epsfig{file=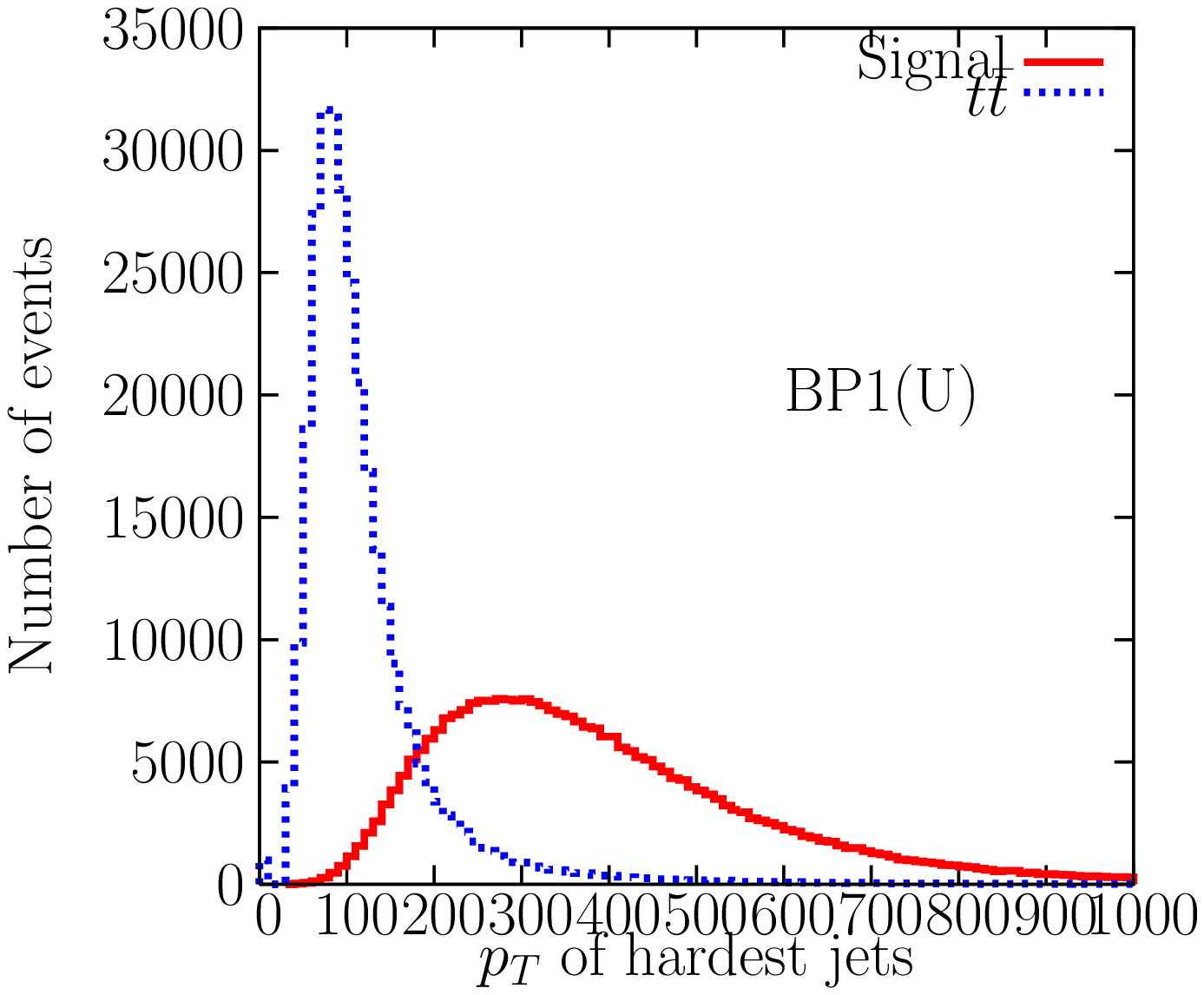,width=10.0 cm,height=12.0cm,angle=-0.0}}
\hskip -12pt 
\hspace*{-1in}{\epsfig{file=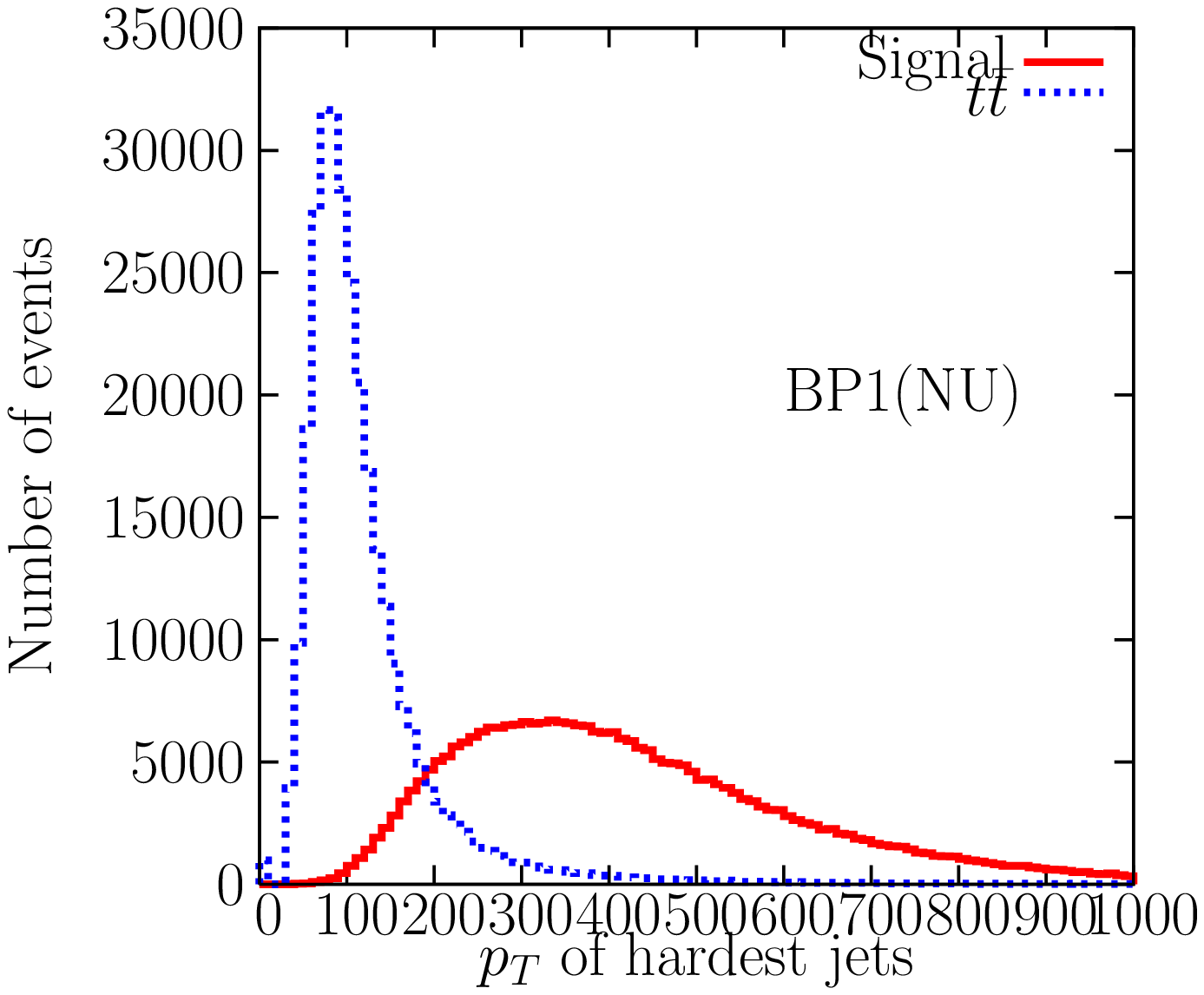,width=10.0cm,height=12.0cm,angle=-0.0}}}
\vskip -20pt
\vspace*{-6.cm}
\hspace*{-1in}{\epsfig{file=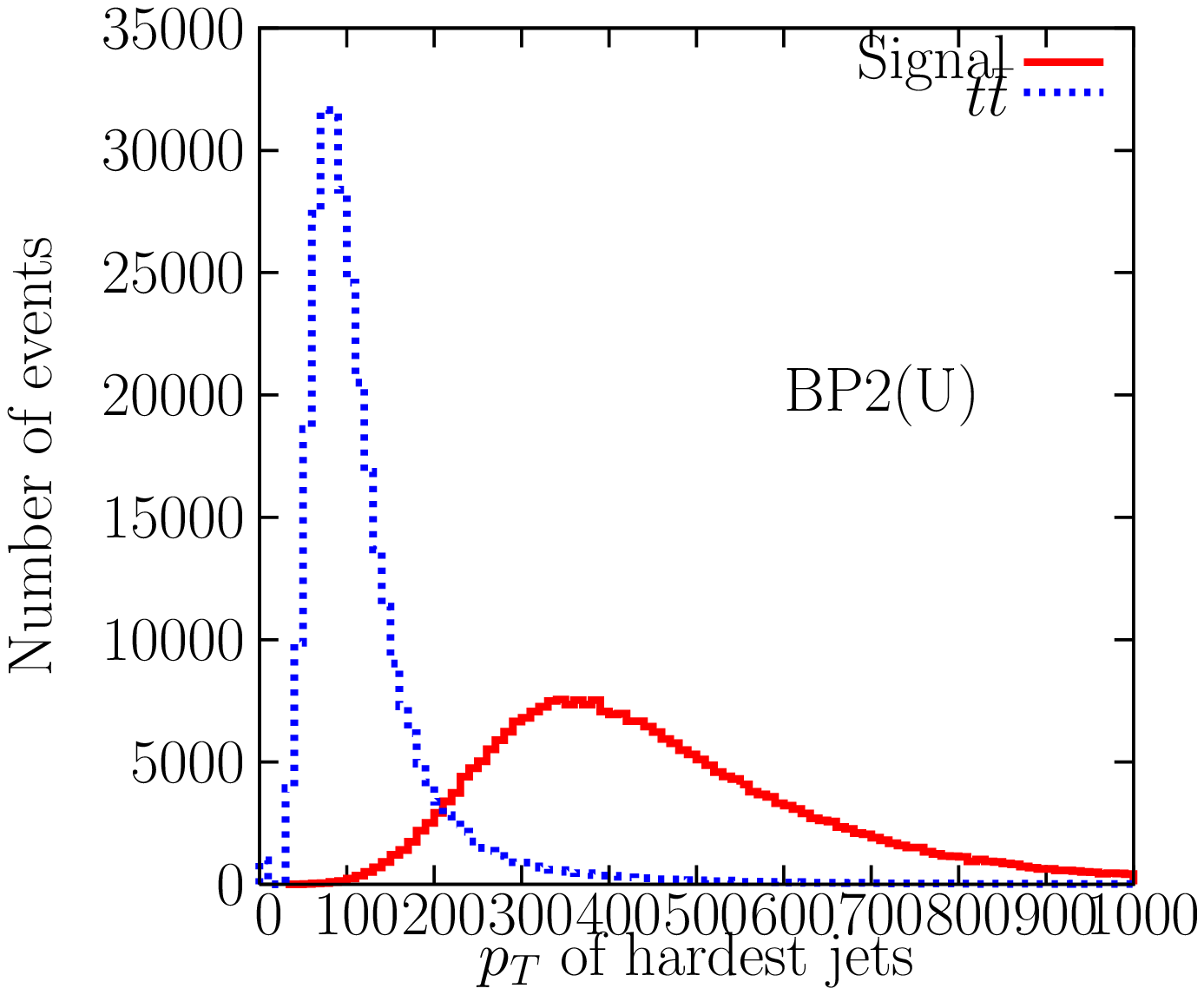,width=10.0 cm,height=12.0cm,angle=-0.0}}
\hskip -12pt 
\hspace*{-1in}{\epsfig{file=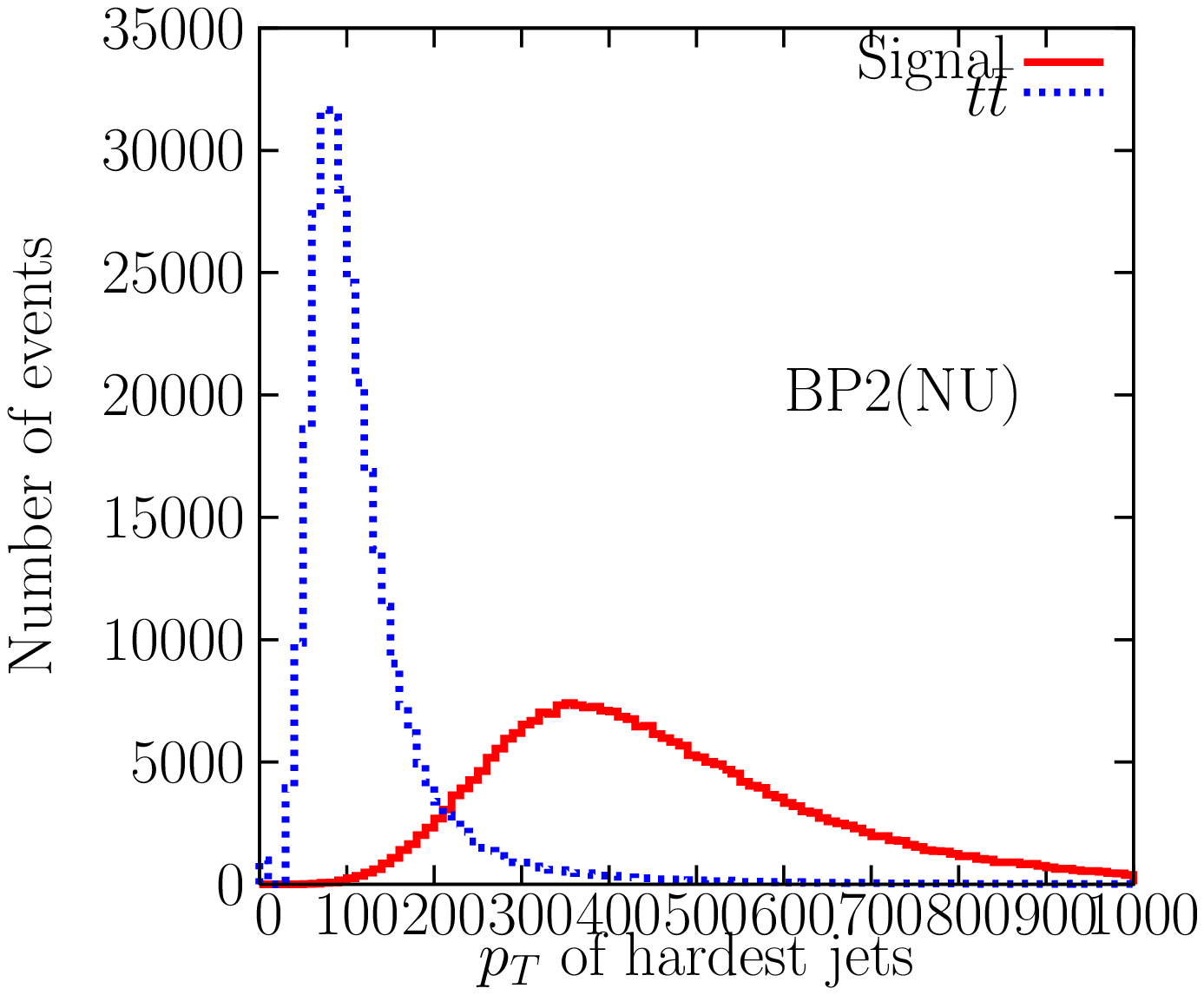,width=10.0cm,height=12.0cm,angle=-0.0}}
\vskip -20pt
\vspace*{-6.cm}
\hspace*{-1in}{\epsfig{file=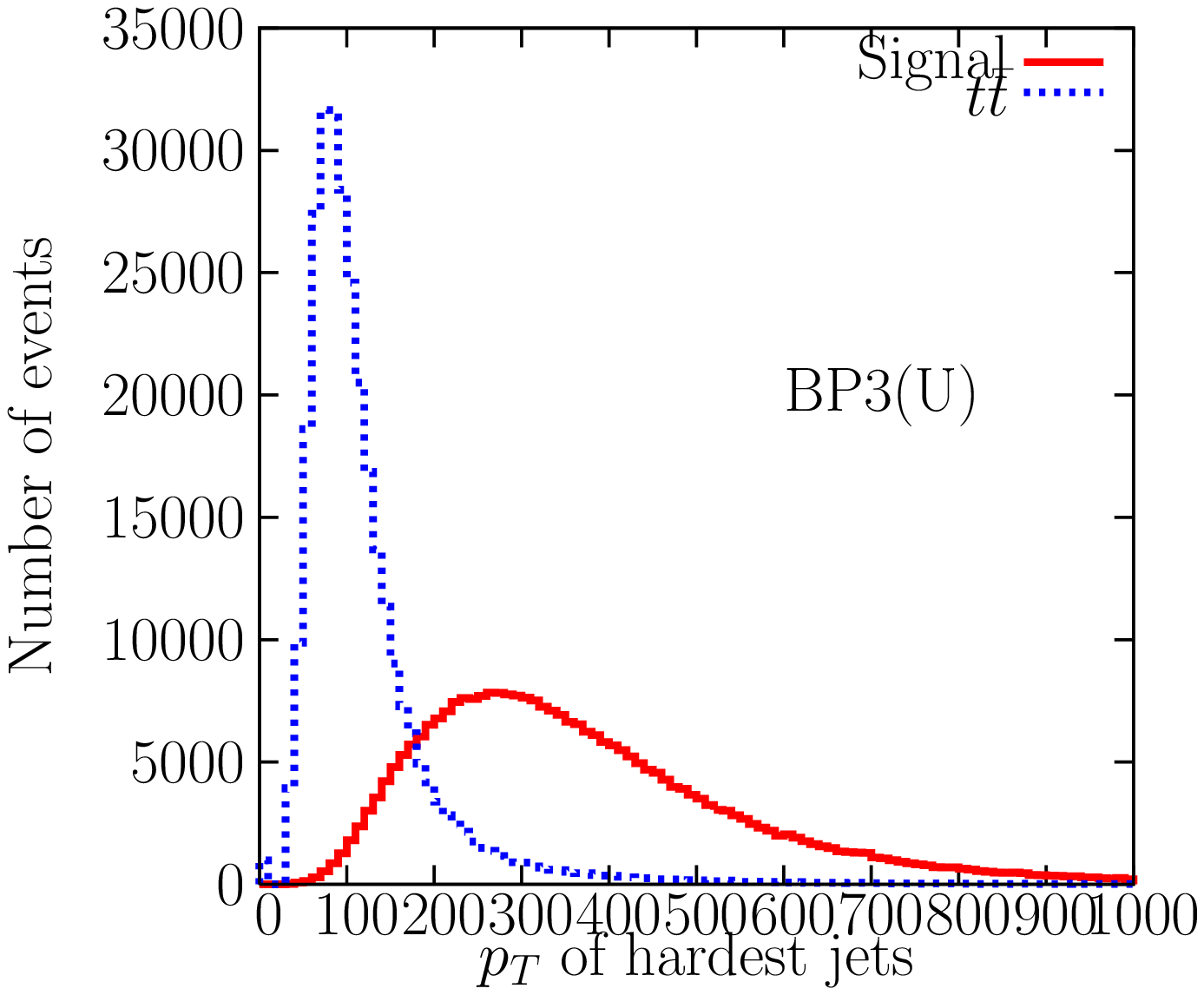,width=10.0 cm,height=12.0cm,angle=-0.0}}
\hskip -12pt 
\hspace*{-1in}{\epsfig{file=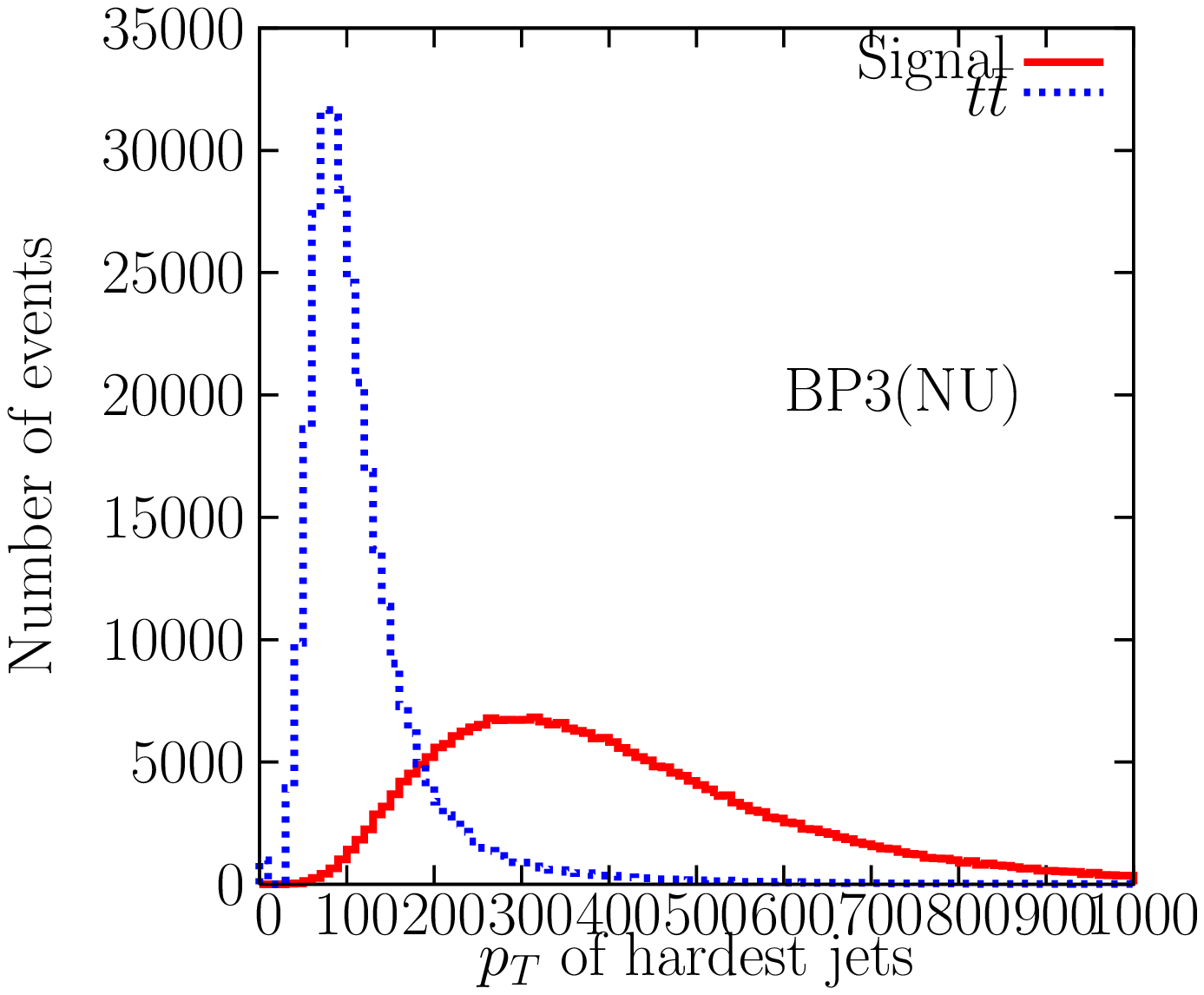,width=10.0cm,height=12.0cm,angle=-0.0}}
\vskip -20pt
\vspace*{-6.cm}
\hspace*{-1in}{\epsfig{file=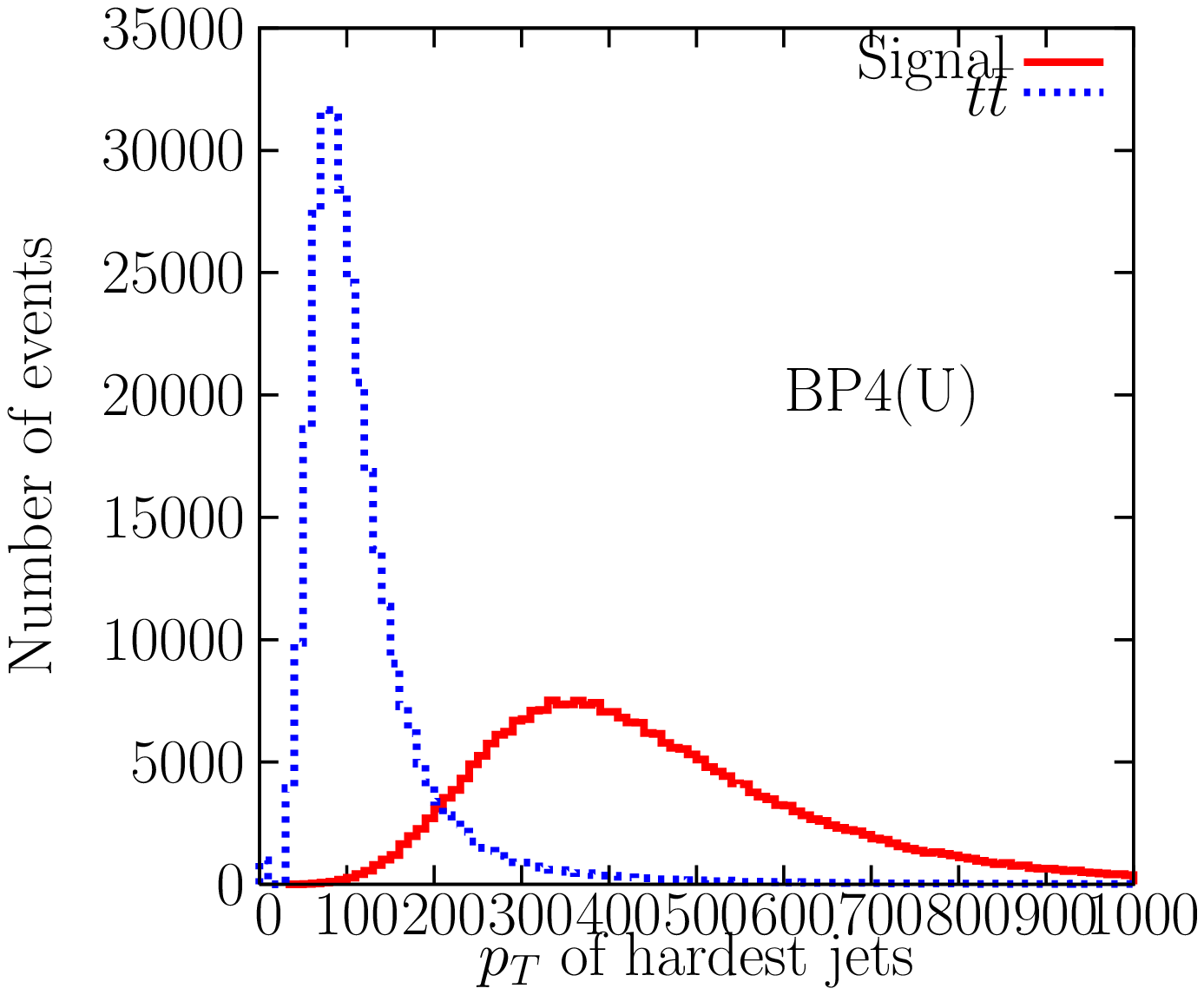,width=10.0 cm,height=12.0cm,angle=-0.0}}
\hskip -12pt 
\hspace*{-1in}{\epsfig{file=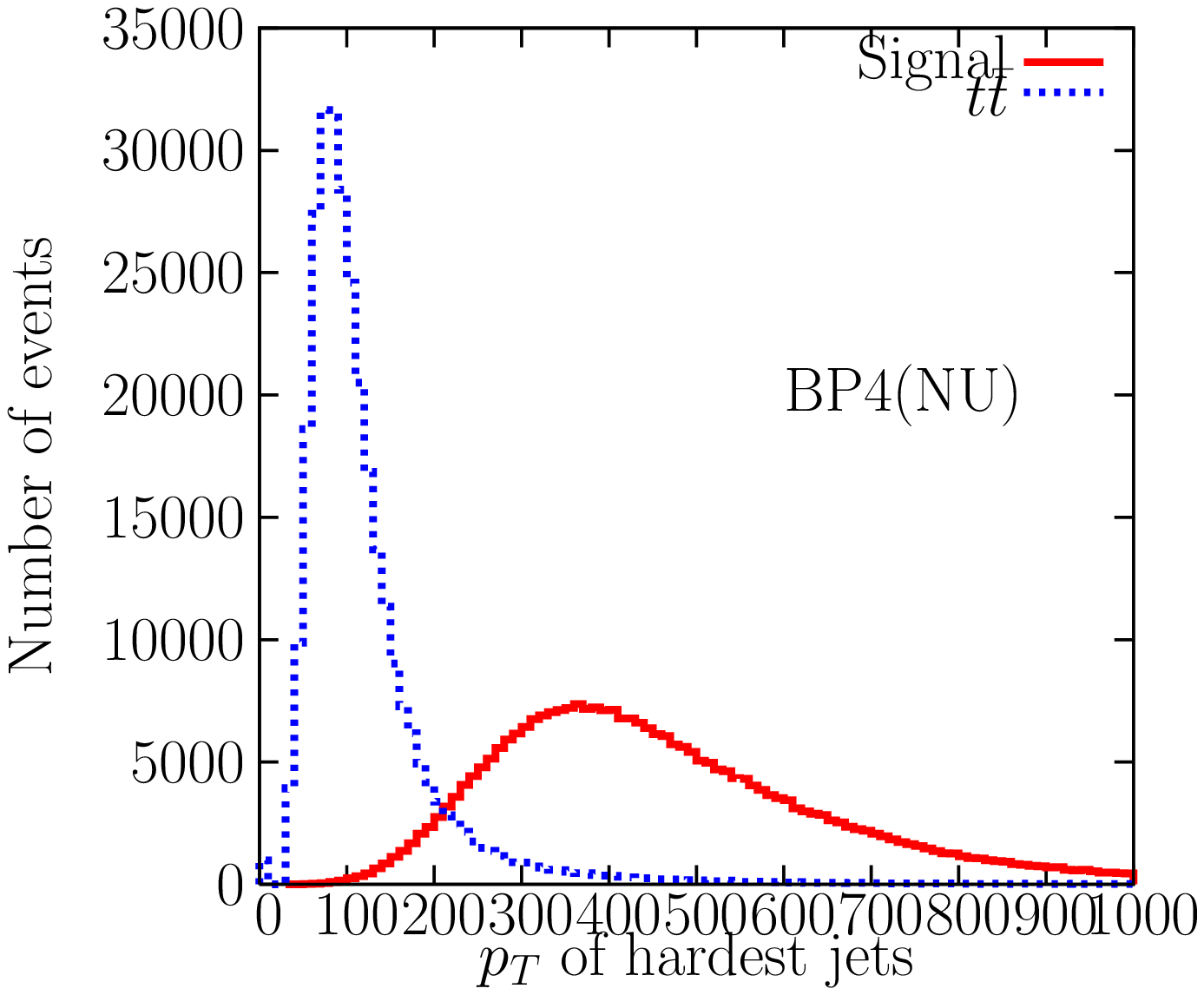,width=10.0cm,height=12.0cm,angle=-0.0}}
\vspace*{-5.5cm}
\caption{Hardest Jet pt (in GeV) distribution for universal (left) and non-non-universal(right) scenarios } 
\end{center}
\label{fig11}
\vspace*{-1.0cm}
\end{figure}
\begin{figure}[hbtp]
\begin{center}
\vspace*{-2.2cm}
{\hspace*{-1in}{\epsfig{file=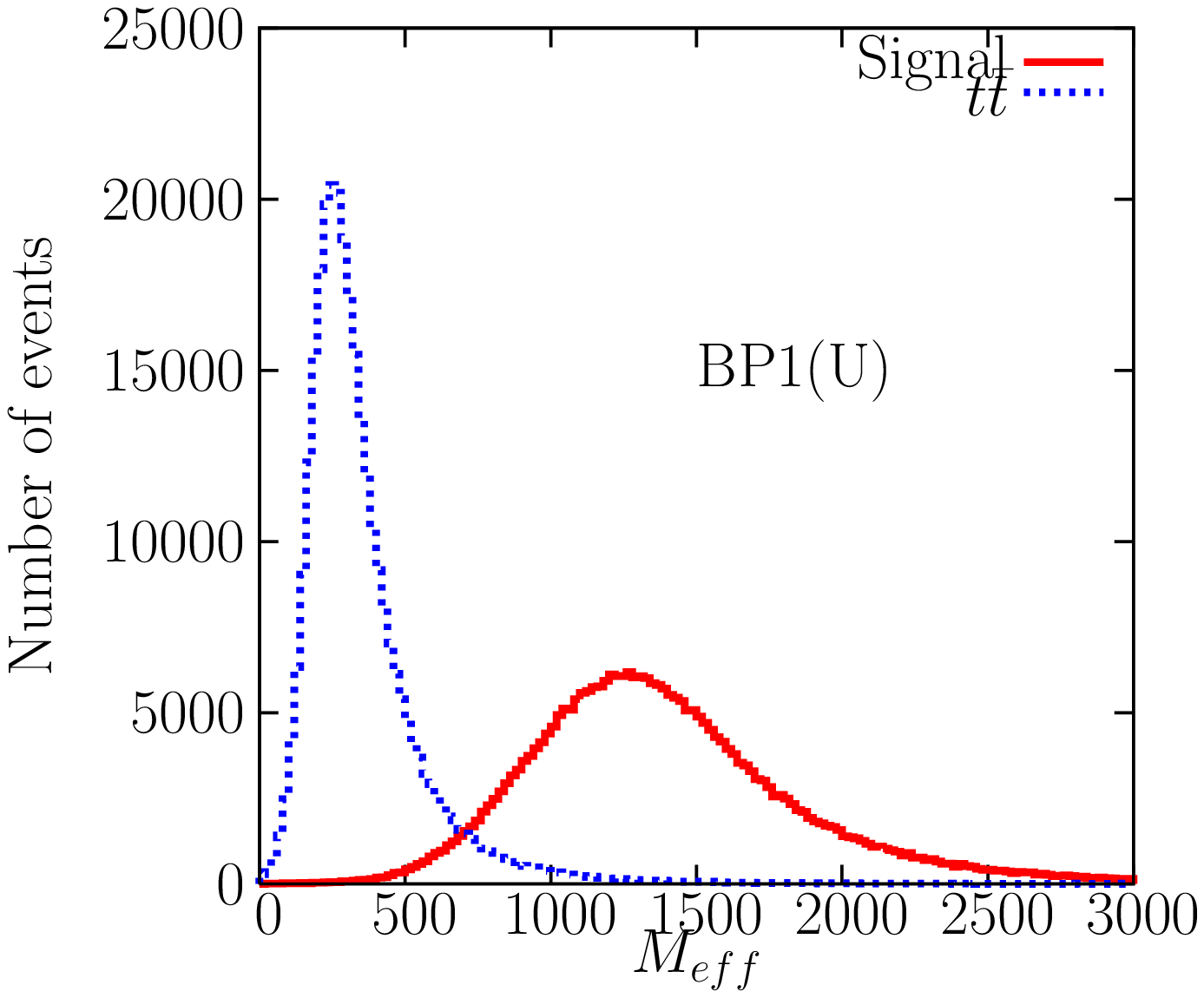,width=10.0 cm,height=12.0cm,angle=-0.0}}
\hskip -12pt 
\hspace*{-1in}{\epsfig{file=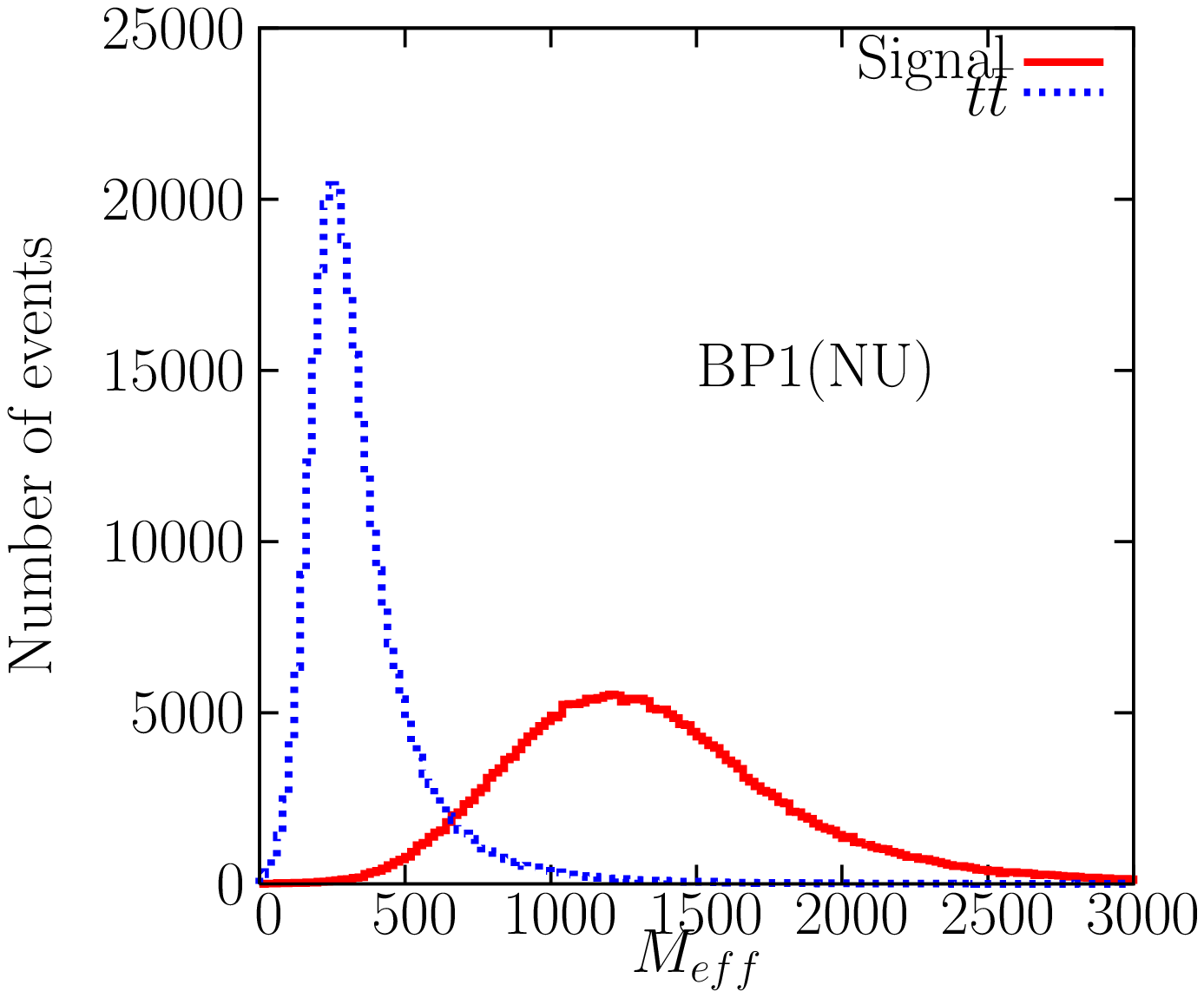,width=10.0cm,height=12.0cm,angle=-0.0}}}
\vskip -20pt
\vspace*{-6.cm}
\hspace*{-1in}{\epsfig{file=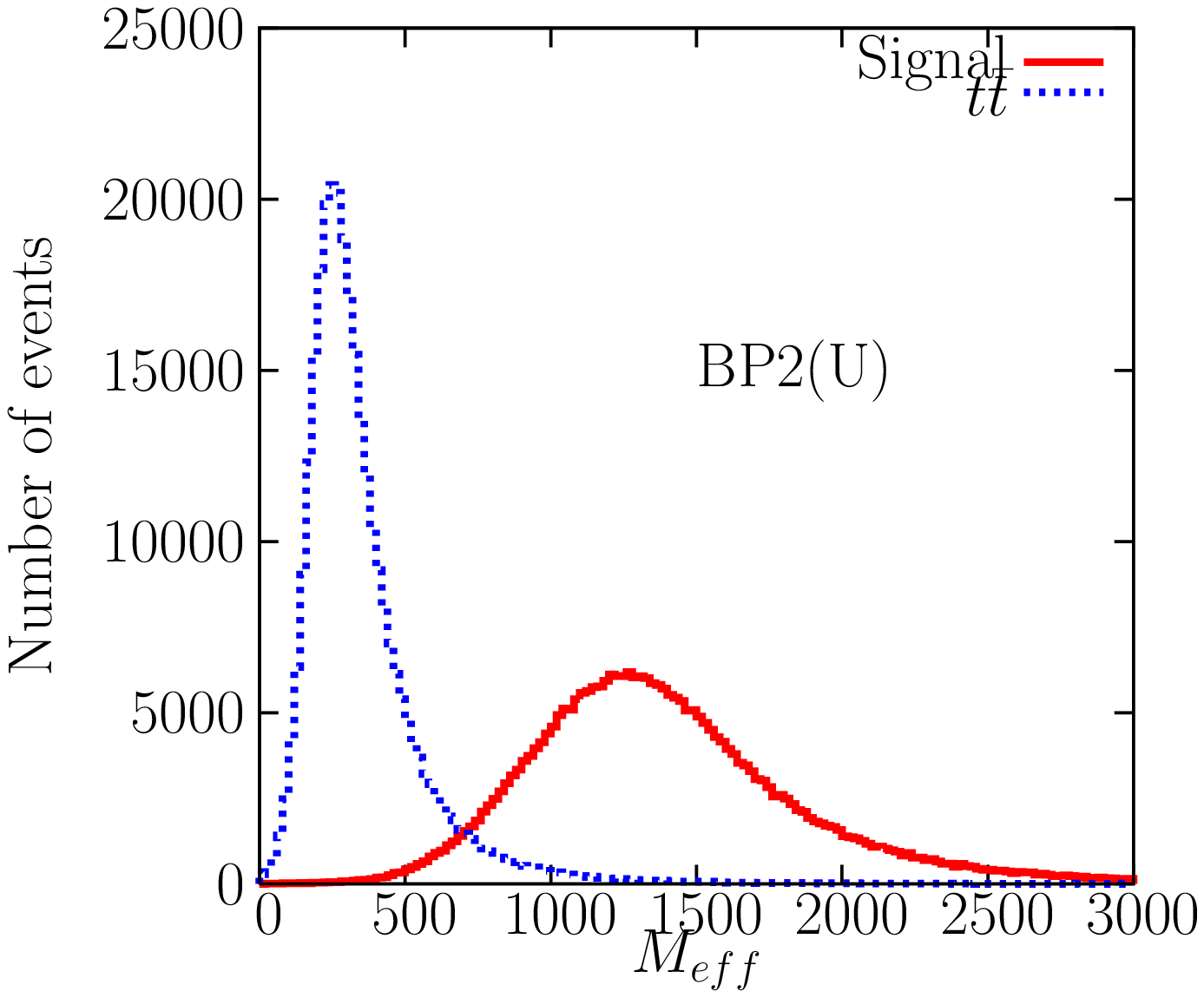,width=10.0 cm,height=12.0cm,angle=-0.0}}
\hskip -12pt 
\hspace*{-1in}{\epsfig{file=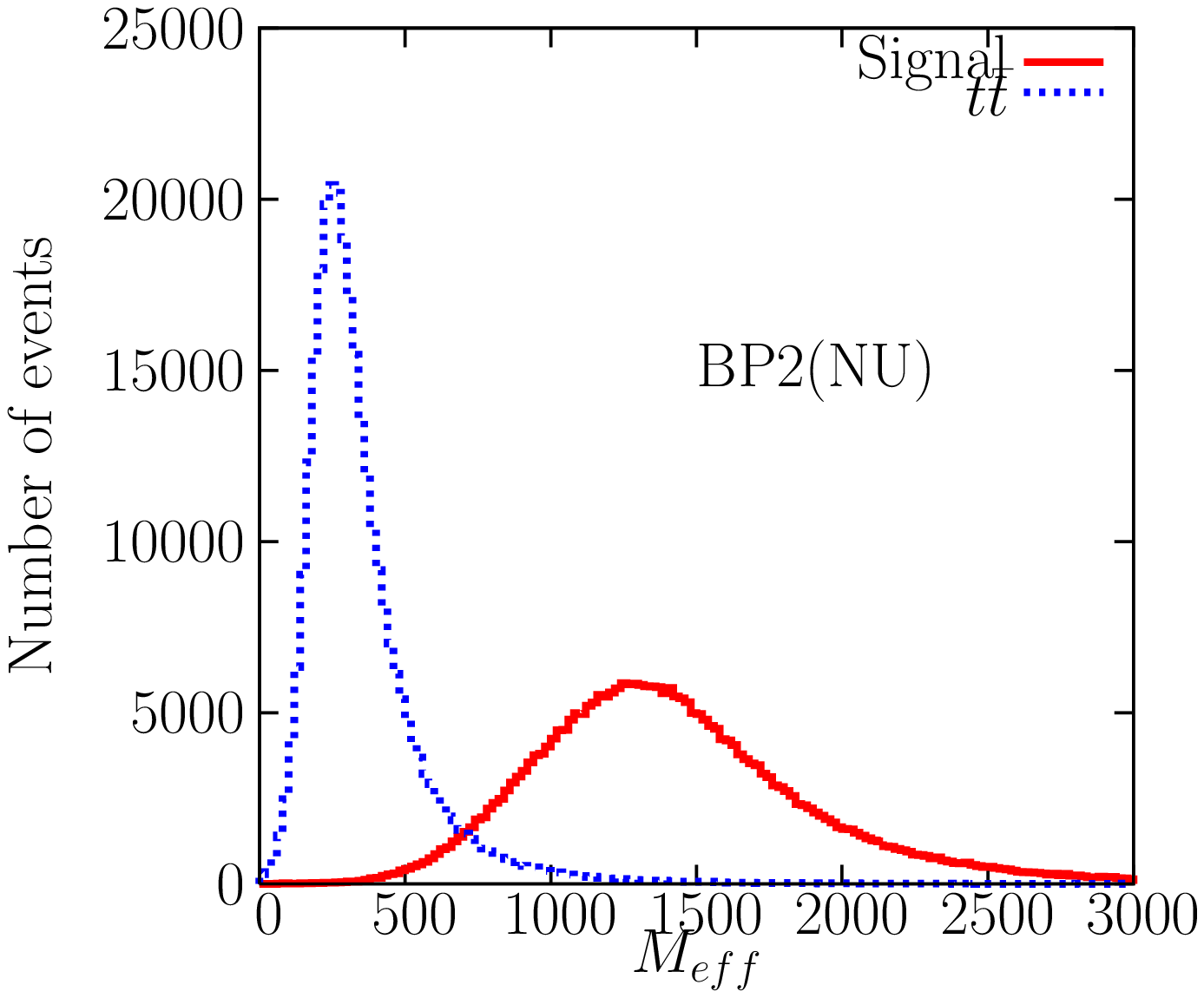,width=10.0cm,height=12.0cm,angle=-0.0}}
\vskip -20pt
\vspace*{-6.cm}
\hspace*{-1in}{\epsfig{file=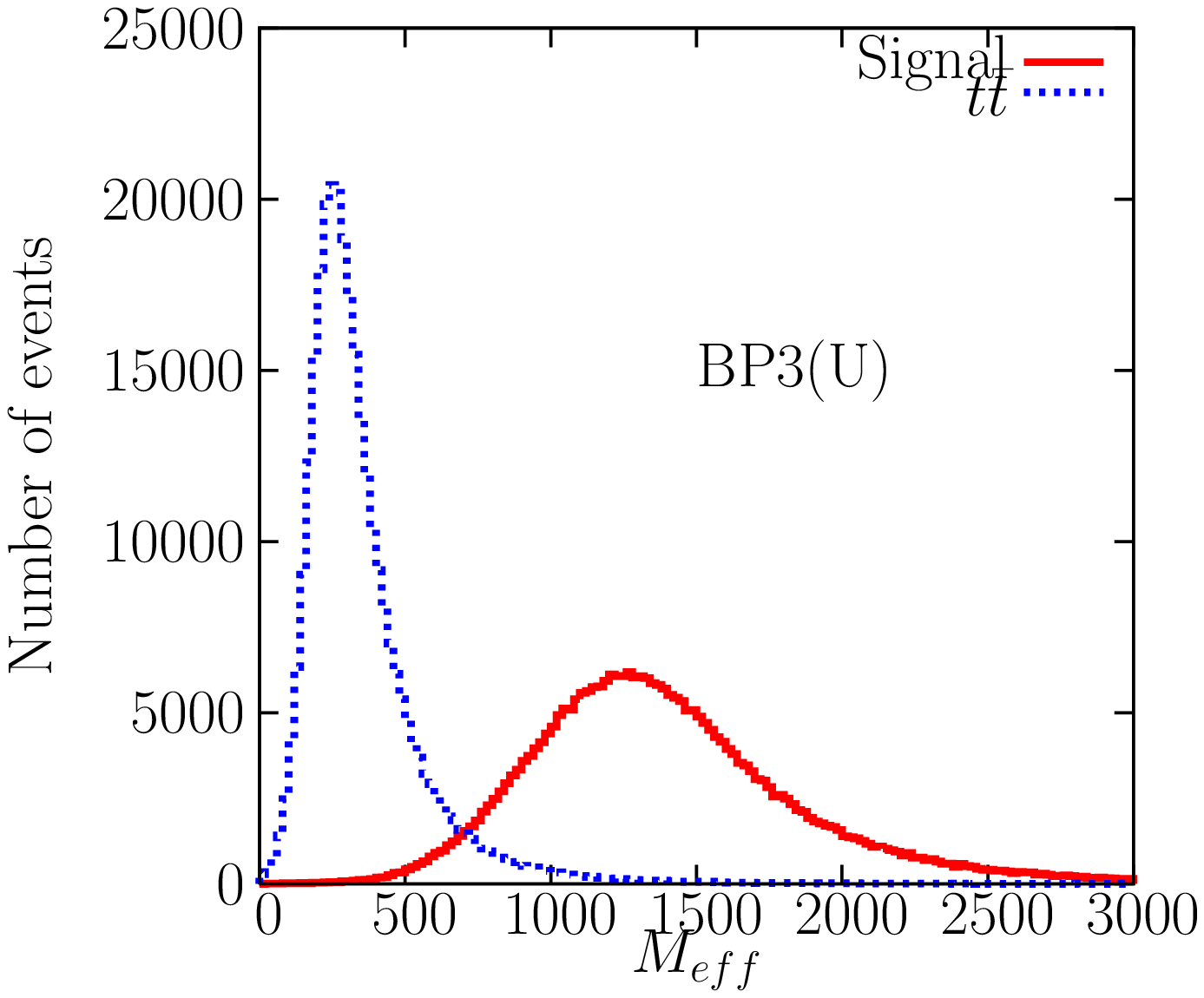,width=10.0 cm,height=12.0cm,angle=-0.0}}
\hskip -12pt 
\hspace*{-1in}{\epsfig{file=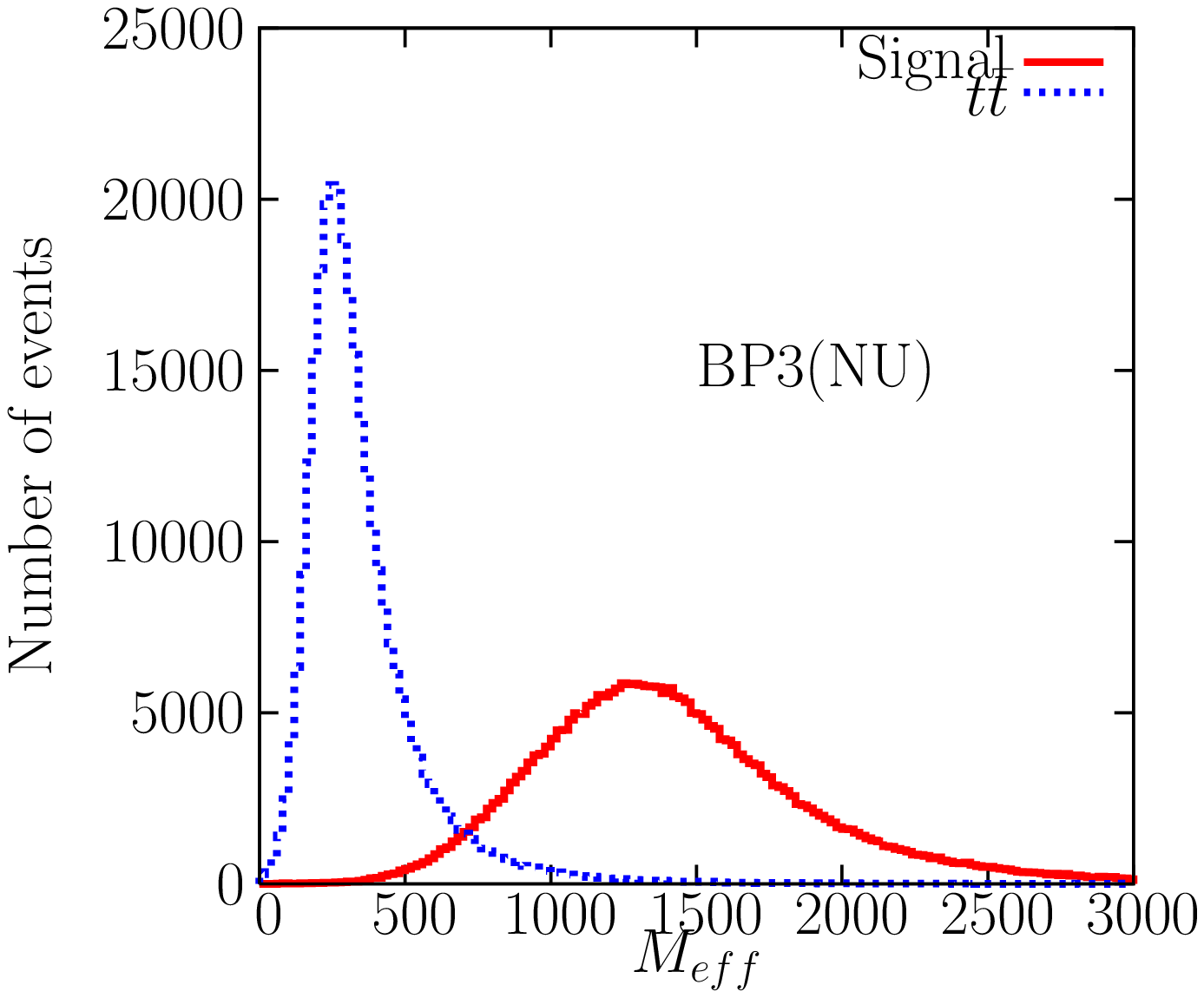,width=10.0cm,height=12.0cm,angle=-0.0}}
\vskip -20pt
\vspace*{-6.cm}
\hspace*{-1in}{\epsfig{file=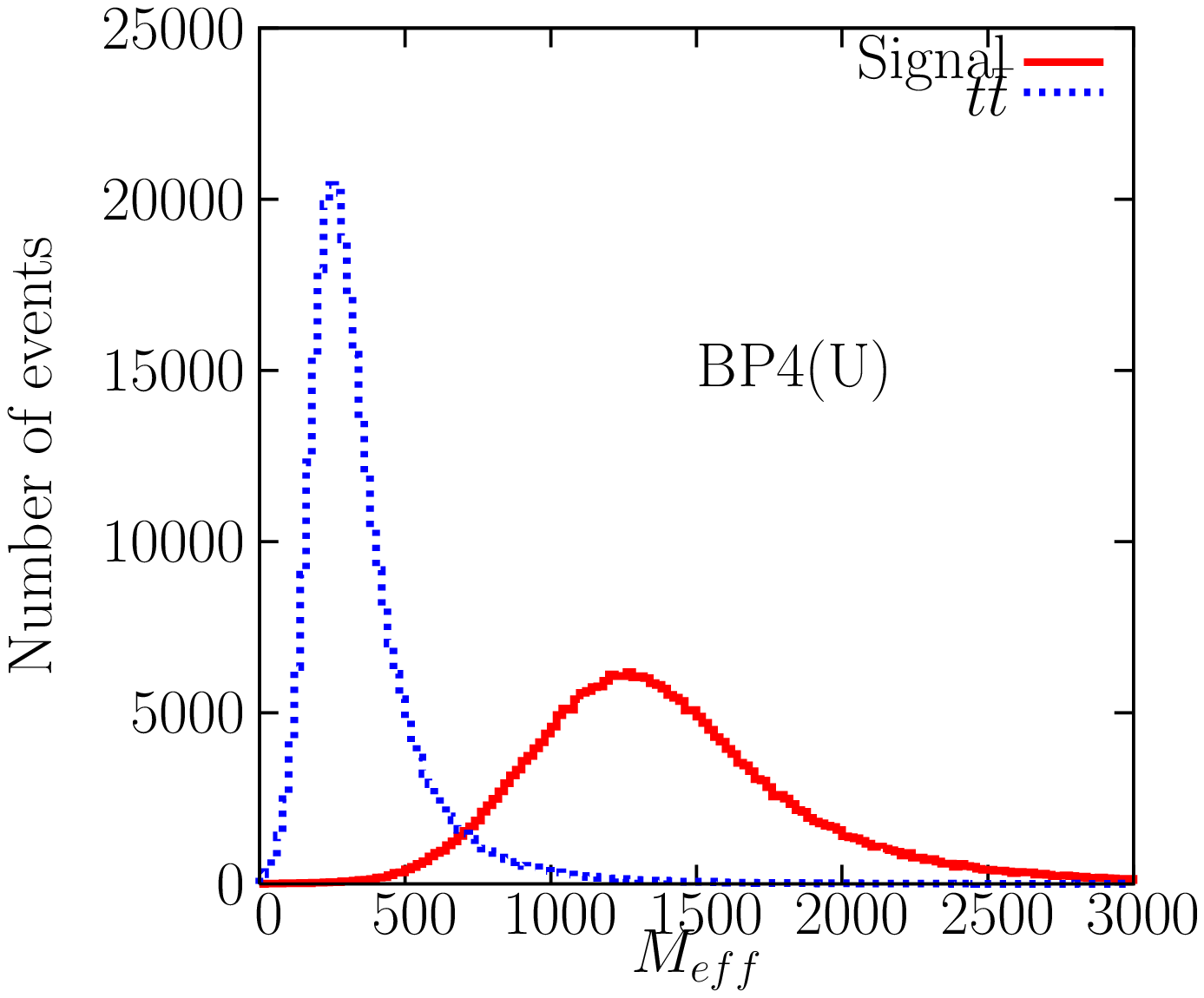,width=10.0 cm,height=12.0cm,angle=-0.0}}
\hskip -12pt 
\hspace*{-1in}{\epsfig{file=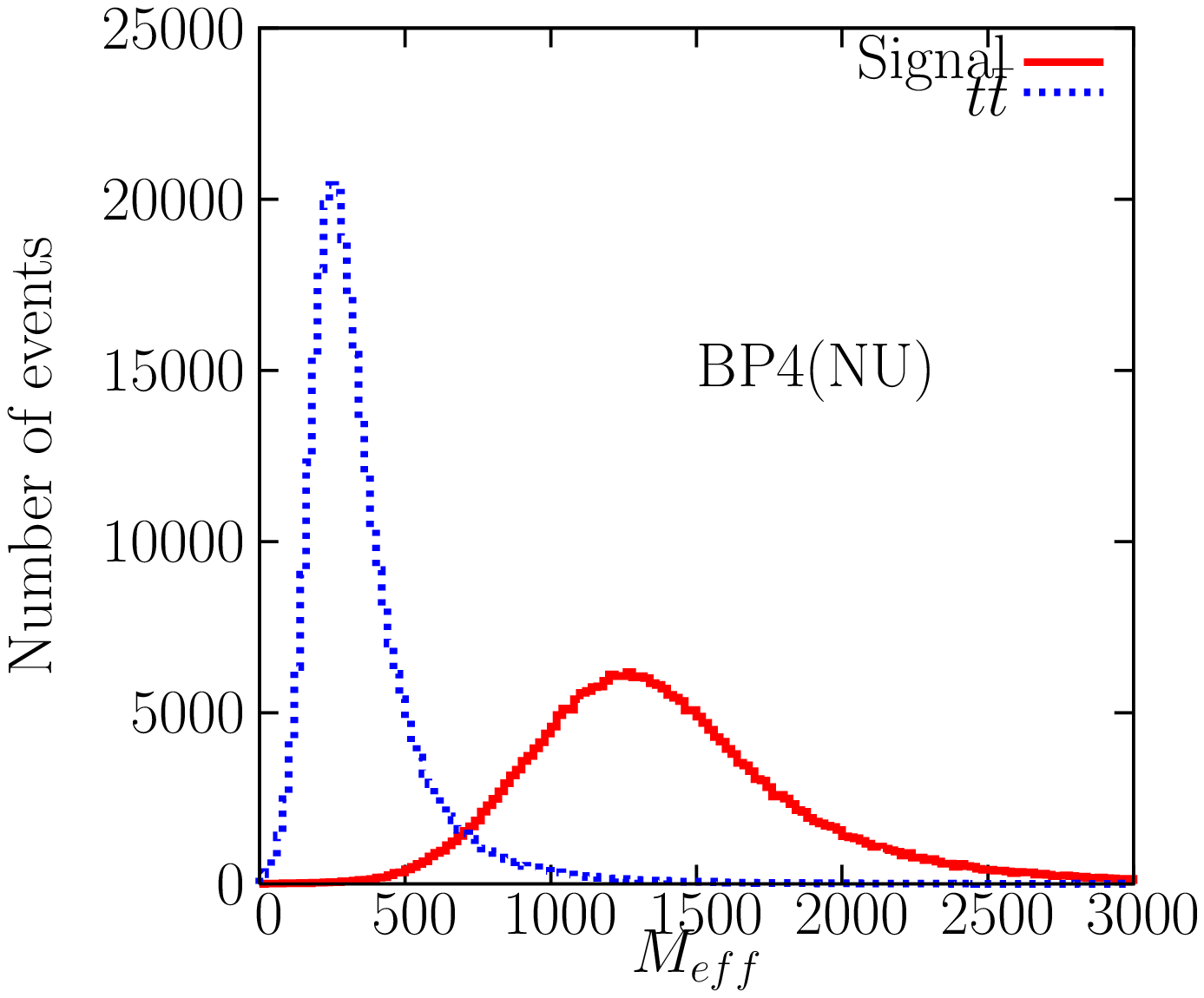,width=10.0cm,height=12.0cm,angle=-0.0}}
\vspace*{-5.5cm}
 \caption{Effective mass (in GeV) distribution for universal(left) and non-non-universal(right) scenarios}  
\end{center}
\label{fig11}
\vspace*{-1.0cm}
\end{figure}

\section{Neutral Higgs bosons under SUSY cascades}
In section 3, while discussing the benchmark scenarios, we discussed our choice of the Higgs spectra. We set the mass of the charged Higgs boson mass (180 GeV and 250 GeV) such that they represent two phenomenologically interesting situations. These in turn determine the masses of the neutral Higgs bosons. As can be seen from Table 3, the mass of the lightest neutral Higgs boson is pretty much fixed at around 109 GeV while the heavier neutral Higgs bosons are somewhat degenerate in each case with $m_A \, (m_H)=162 \, (164)$ GeV and $m_A \, (m_H)=238 \, (239)$ GeV respectively. Thus, the neutral Higgs bosons are to be looked for in two different mass regions in both cases. As we will see in the following discussion, exploring these two mass regions for neutral Higgs bosons simultaneously (for a given charged Higgs mass) would not be an easy task. We would thus attempt to optimize signals for the lightest neutral Higgs boson and the heavier mates separately. 

 In the present analysis we look for neutral Higgs bosons decaying to $b\bar{b}$. So, the generic signal for neutral Higgs boson(s) produced in the SUSY cascades would be 
$n_{jet}\ge 5 + \ptmiss$ including a pair of $b$-jets whose invariant mass is reconstructed to the neutral Higgs mass(es). Thus, tagging of the $b$-jets would be essential.  We use a $b$-tagging efficiency of 50\% \cite{atlas-2,Ball:2007zza} over the entire mass range (for all the Higgs bosons) of our concern, i.e., from 110 GeV to 250 GeV. We also observe that selecting a suitable invariant mass window for a $b$-jet pair to pin down Higgs resonances may not be efficient enough in general. Supplementing such a kinematic criterion further with an optimal window in $p^{b_{jet}}_T$ would aid the search for the neutral Higgses appreciably \cite{Datta:2003iz,Huitu:2008sa}. In the following two subsections we discuss the cases for lightest and the heavier neutral Higgs bosons separately by incorporating the above-mentioned prescriptions for kinematic cuts. 

\subsection{The lightest neutral Higgs boson}
In the case of the lightest neutral Higgs boson with a mass at around 109 GeV, it is observed that a window of $45\,{\mathrm{GeV}}\le{p^{b_{j_1,j_2}}_{T}}\le{70}$ GeV preselects the candidate $b$-jets efficiently which are (Figure 5 substantiates this conclusion) subsequently used to study the invariant mass distribution of the $b$-jet pair that ultimately peaks at the lightest Higgs boson mass. It is also noted that while under favourable situations the preselection can be skipped, the general outcome of imposing the same always improves the signal to background ratio. The pair-wise invariant mass distributions of the tagged $b$-jets are illustrated in Figure 6 and Figure 7 for $m_{H^{\pm}}=180$ GeV and 250 GeV cases, for both the universal and the non-universal scenarios. Note that mass of  the lightest neutral Higgs boson remains almost the same ($m_h\simeq 109$ GeV) for both charged Higgs masses.
\begin{figure}[hbtp]
\begin{center}
{\epsfig{file=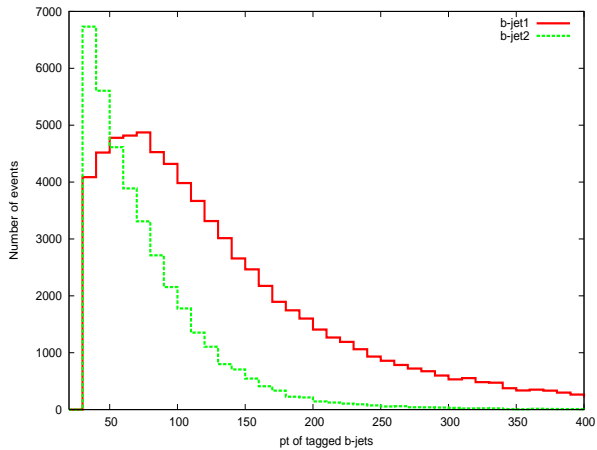,width=6.0cm,height=8.0cm,angle=-90.0}}
\caption{$p_T$ distribution of two tagged $b$-jets in GeV.} 
\end{center}
\label{fig11}
\end{figure}
The corresponding SUSY backgrounds, as defined earlier in section 5.2, are also plotted in each of these graphs. The plots are generated for  an integrated luminosity of $\mathcal{L}=10$ fb$^{-1}$. The red (dark) ones are for the signal final state, i.e., $n_{jet}\ge 5$ (with at least two $b$-jets) + $\not{p_T}$.
as defined earlier in section 5 \& 6  with all the `basic' cuts imposed along
with $45 \, \mathrm{GeV} \, \le{p^{b_{j_1,j_2}}_{T}}\le{70}$ GeV. The green (grey) ones, on the other 
hand, correspond to the SUSY background (i.e., events that pass the same kinematic criteria but not containing the lightest neutral Higgs). 
\begin{figure}[hbtp]
\begin{center}
{\epsfig{file=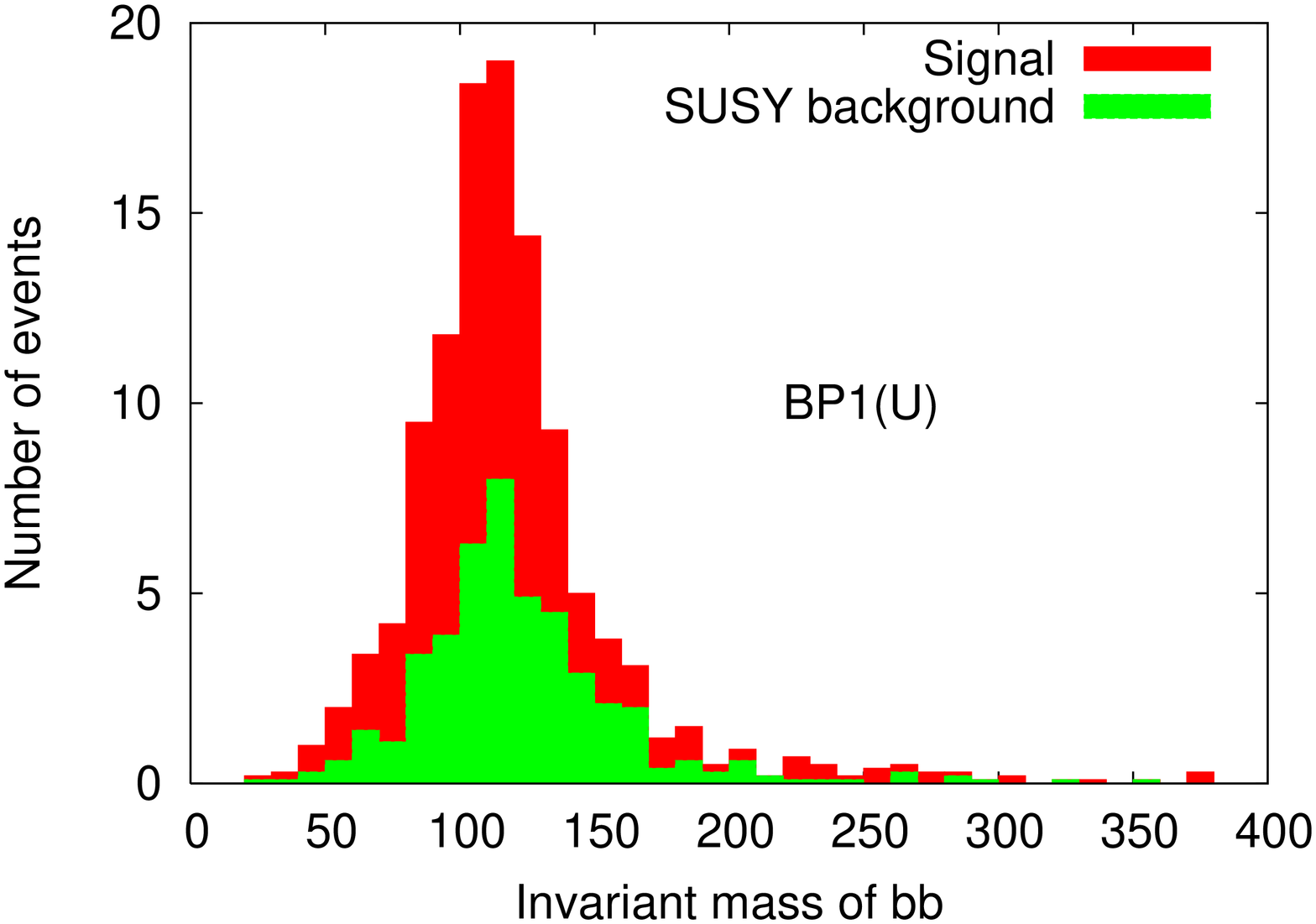,width=5.0 cm,height=7.0cm,angle=-90.0}}
\hskip -12pt 
{\epsfig{file=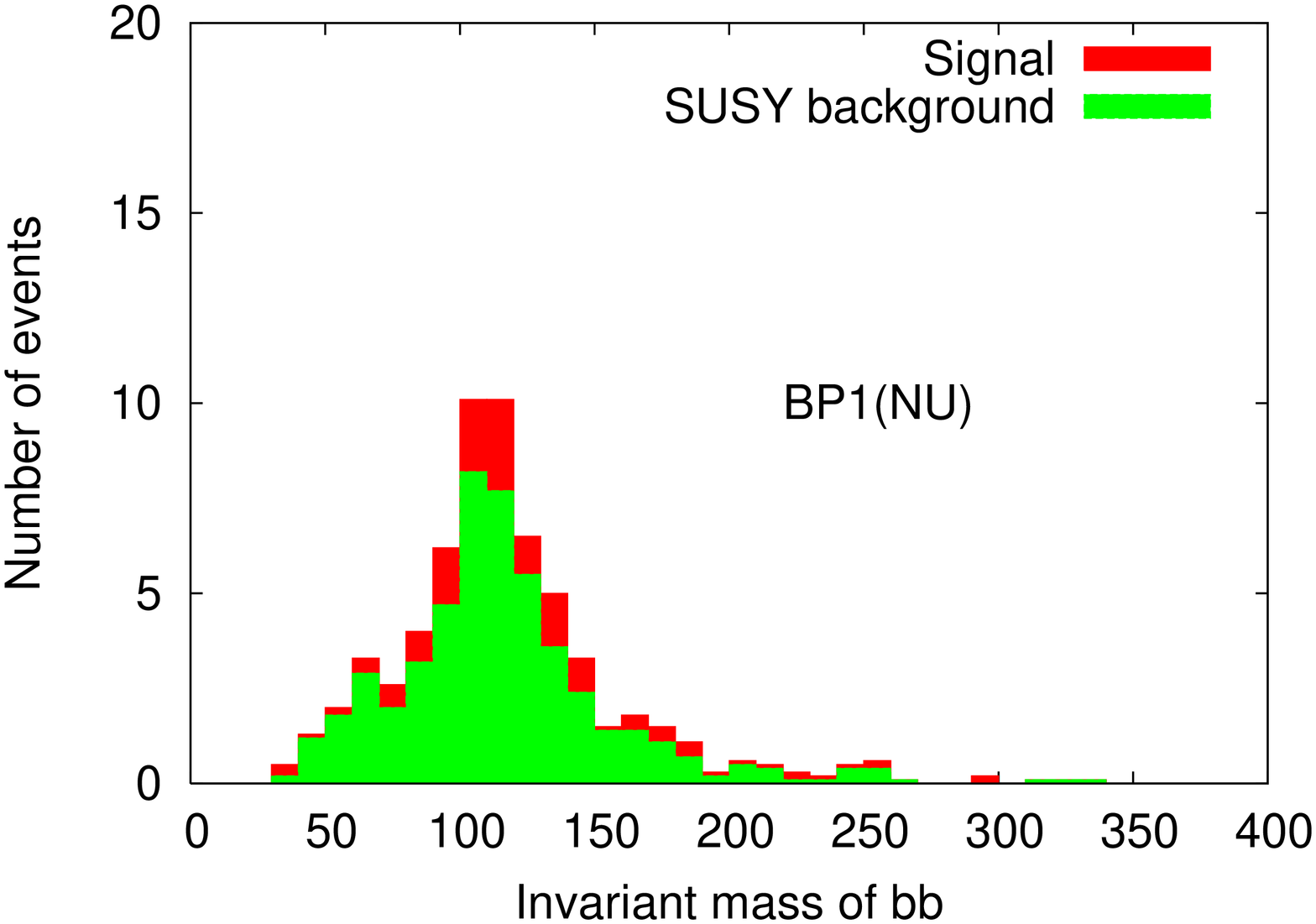,width=5.0cm,height=7.0cm,angle=-90.0}}
{\epsfig{file=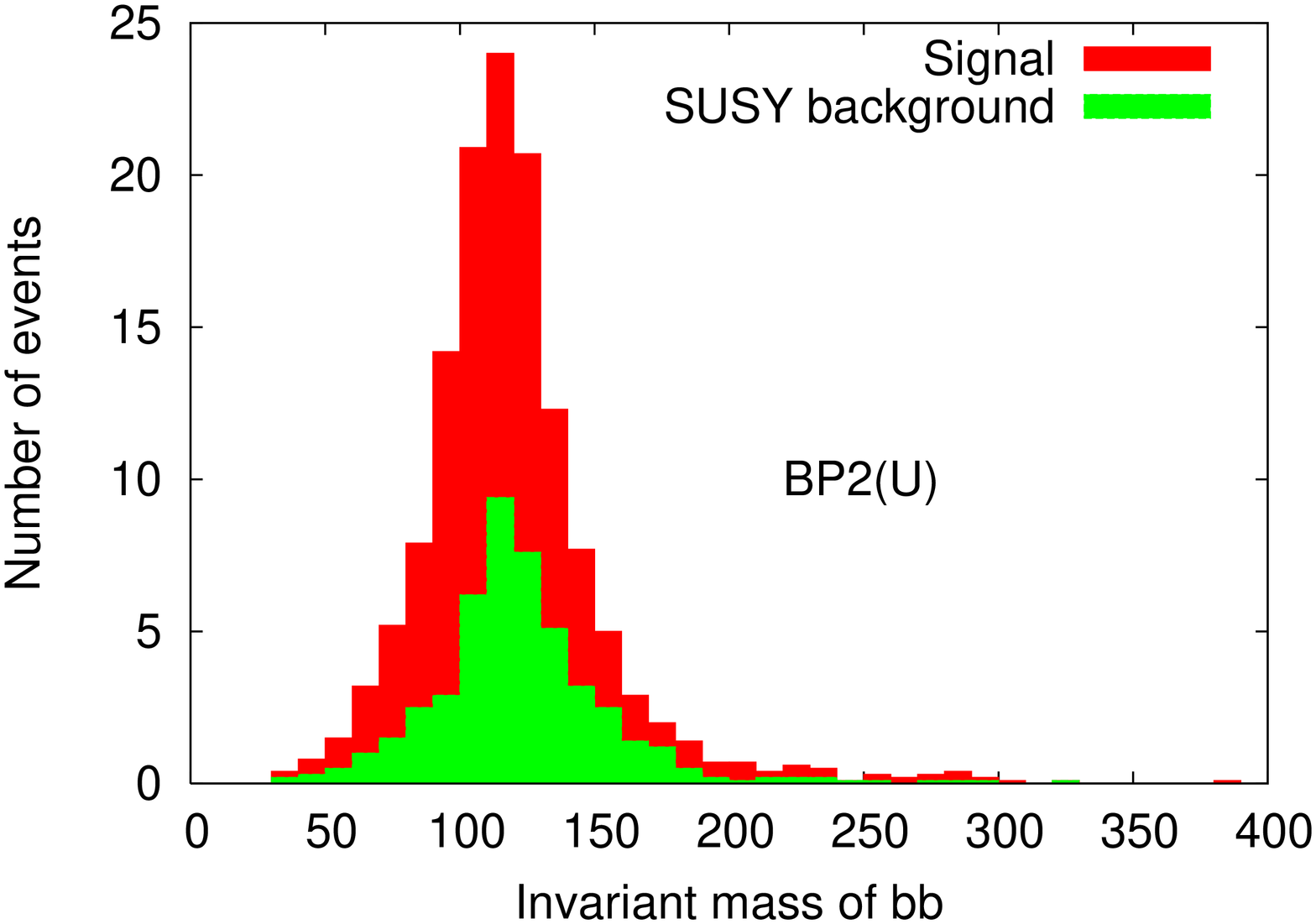,width=5.0 cm,height=7.0cm,angle=-90.0}}
\hskip -12pt 
{\epsfig{file=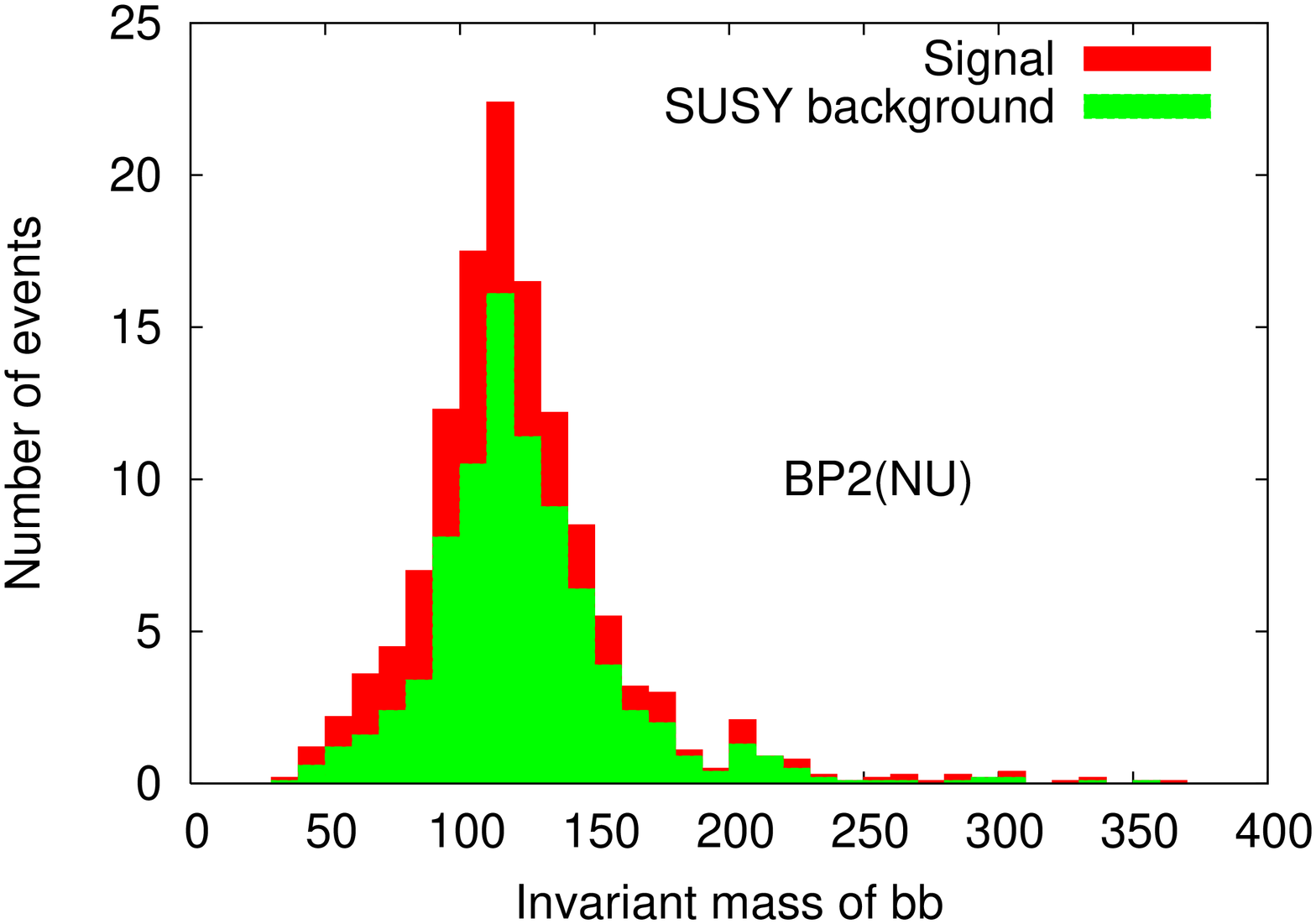,width=5.0cm,height=7.0cm,angle=-90.0}}
{\epsfig{file=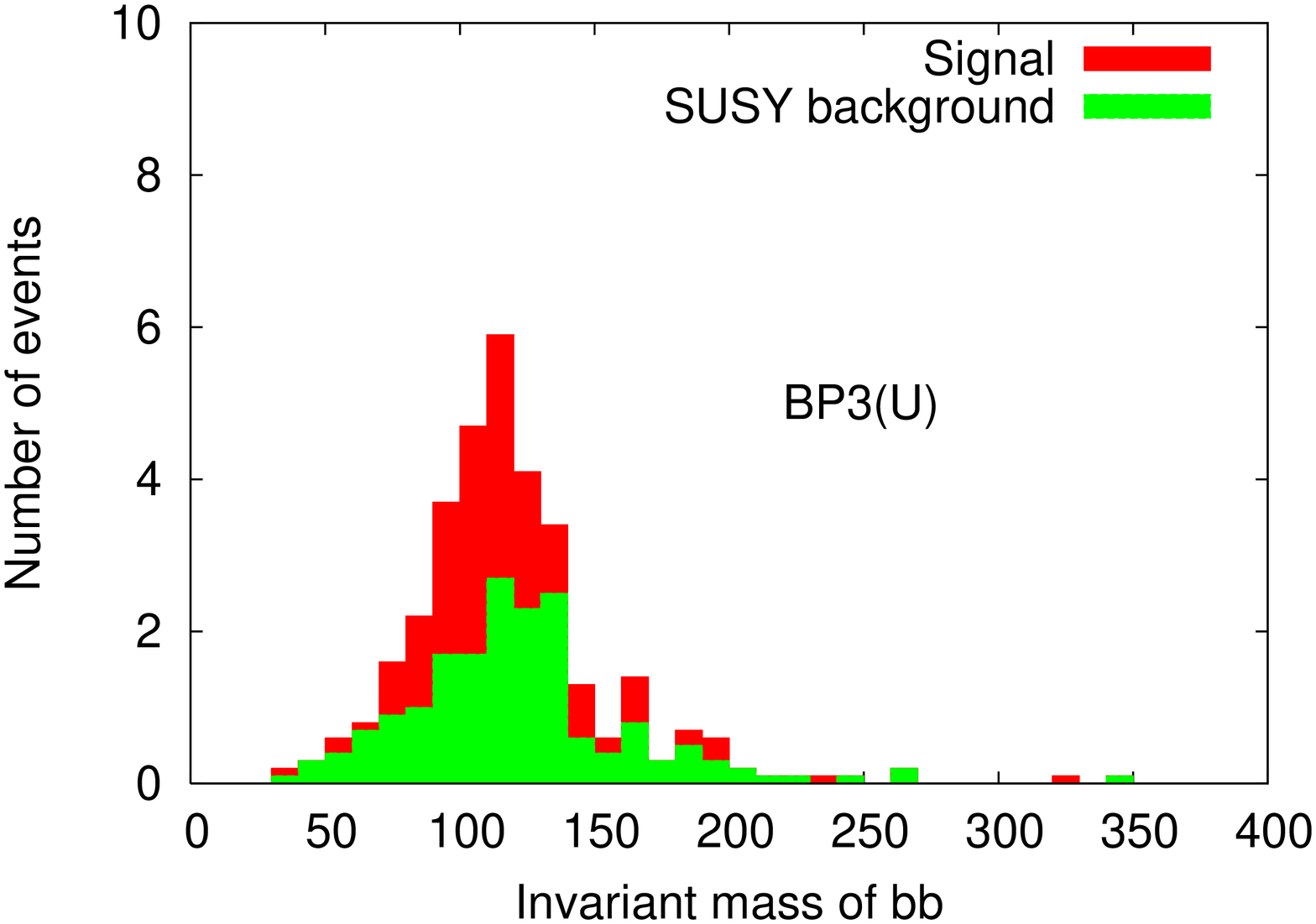,width=5.0 cm,height=7.0cm,angle=-90.0}}
\hskip -12pt 
{\epsfig{file=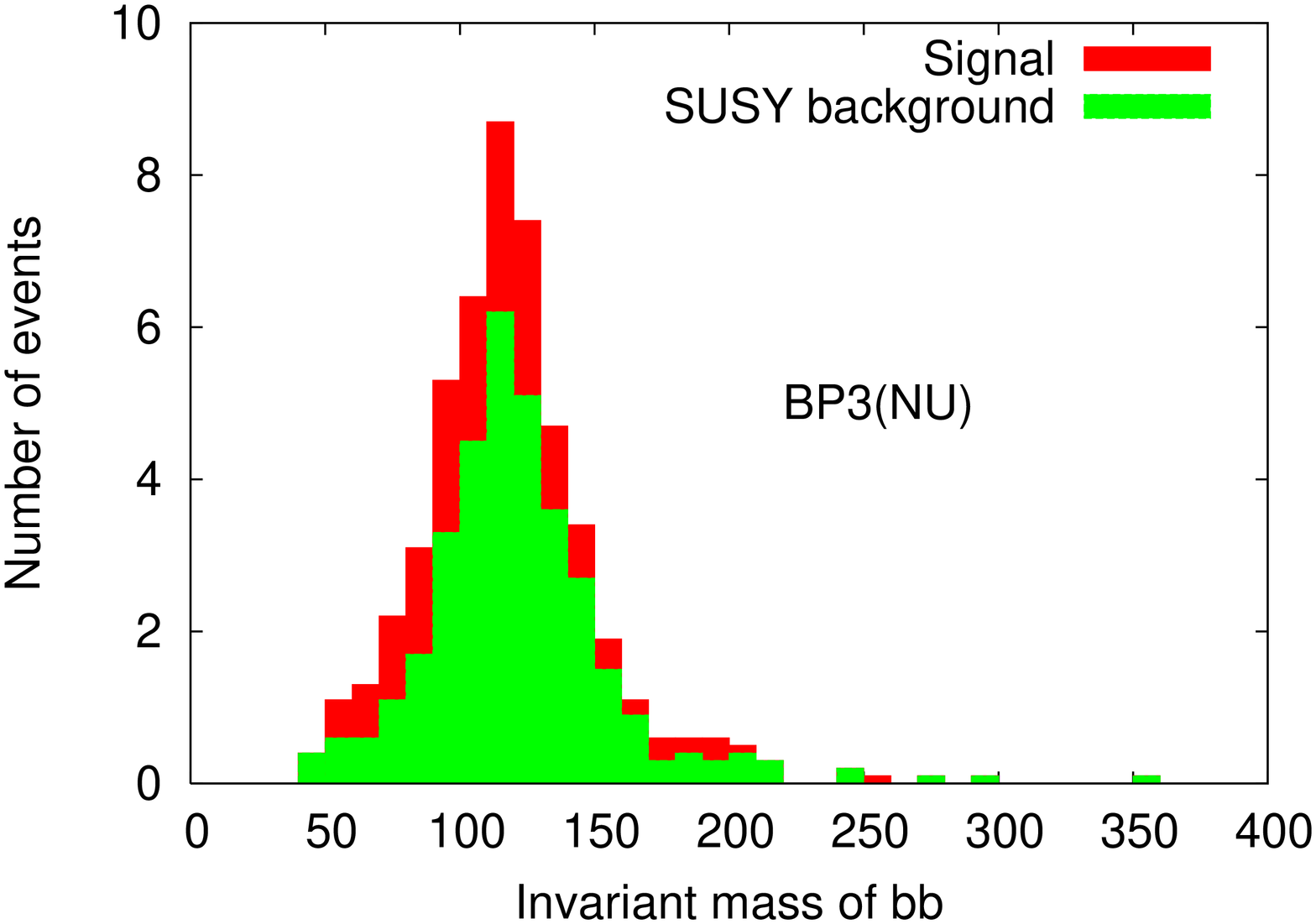,width=5.0cm,height=7.0cm,angle=-90.0}}
{\epsfig{file=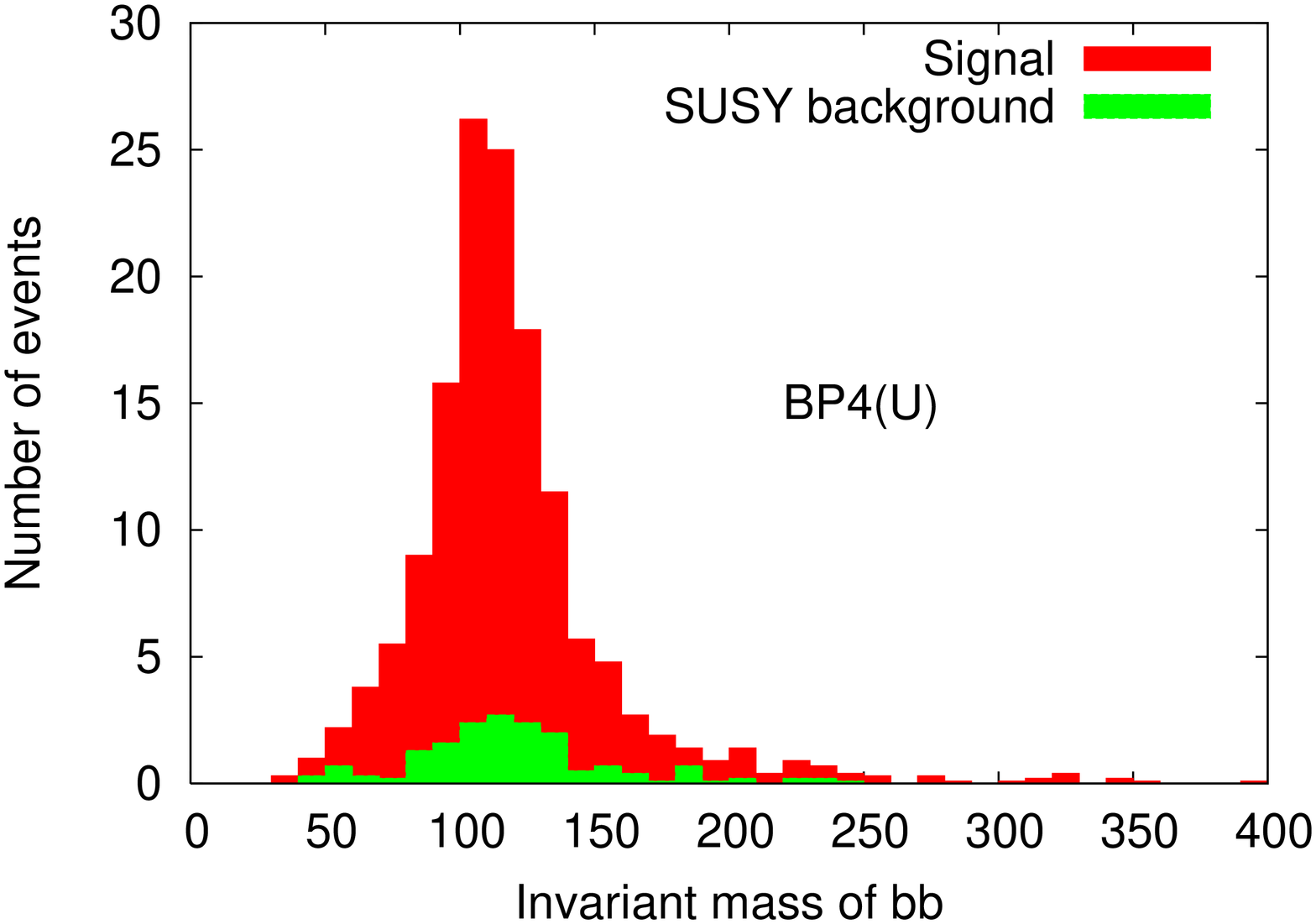,width=5.0 cm,height=7.0cm,angle=-90.0}}
\hskip -12pt 
{\epsfig{file=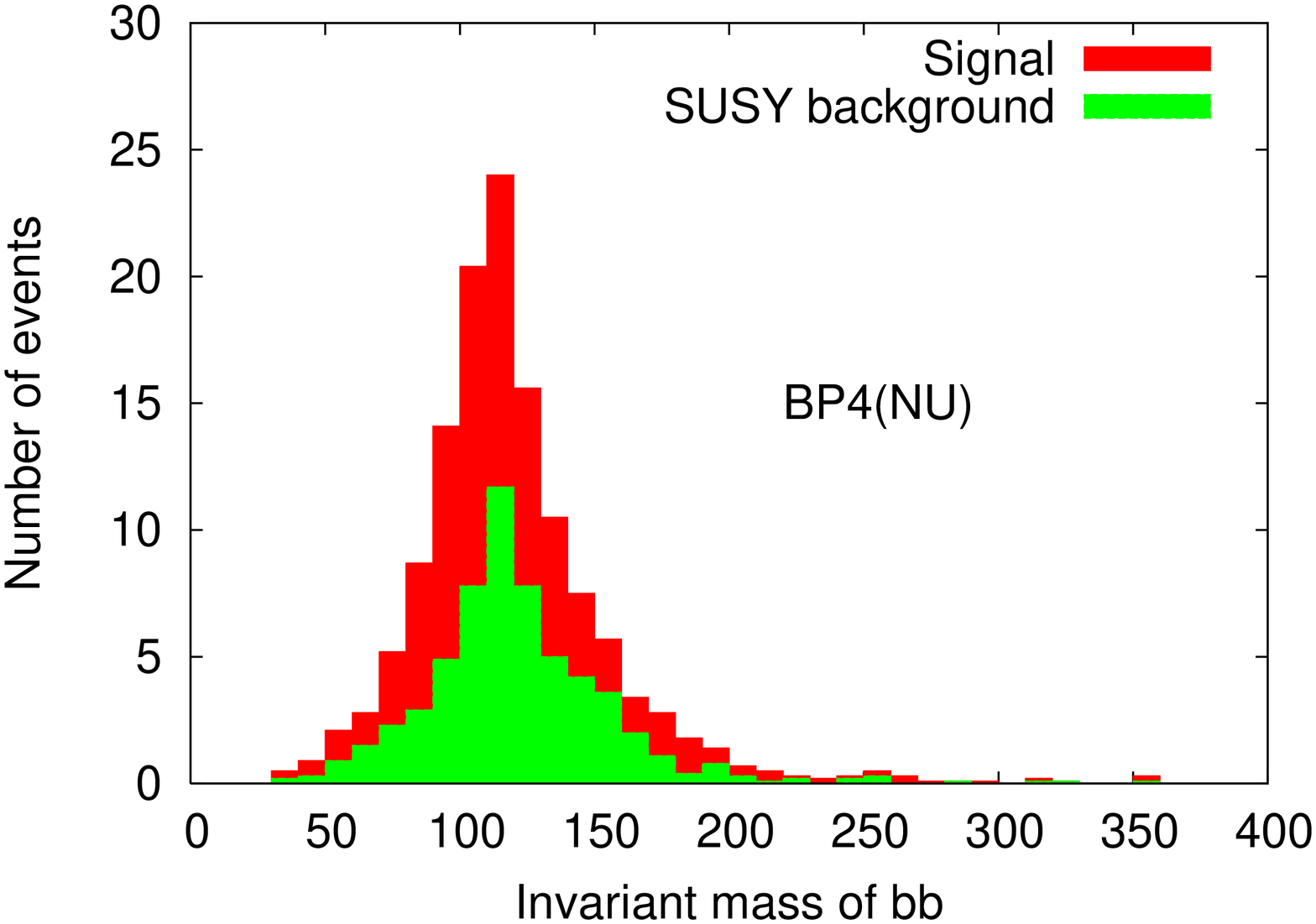,width=5.0cm,height=7.0cm,angle=-90.0}}
\caption{Invarinat mass (in GeV) distribution for universal(left) and non-non-universal(right) for lightest Higgs for $m_{H^{\pm}}=180$ GeV. } 
\end{center}
\label{fig11}
\vspace*{-1.0cm}
\end{figure}
\begin{figure}[hbtp]
\begin{center}
{\epsfig{file=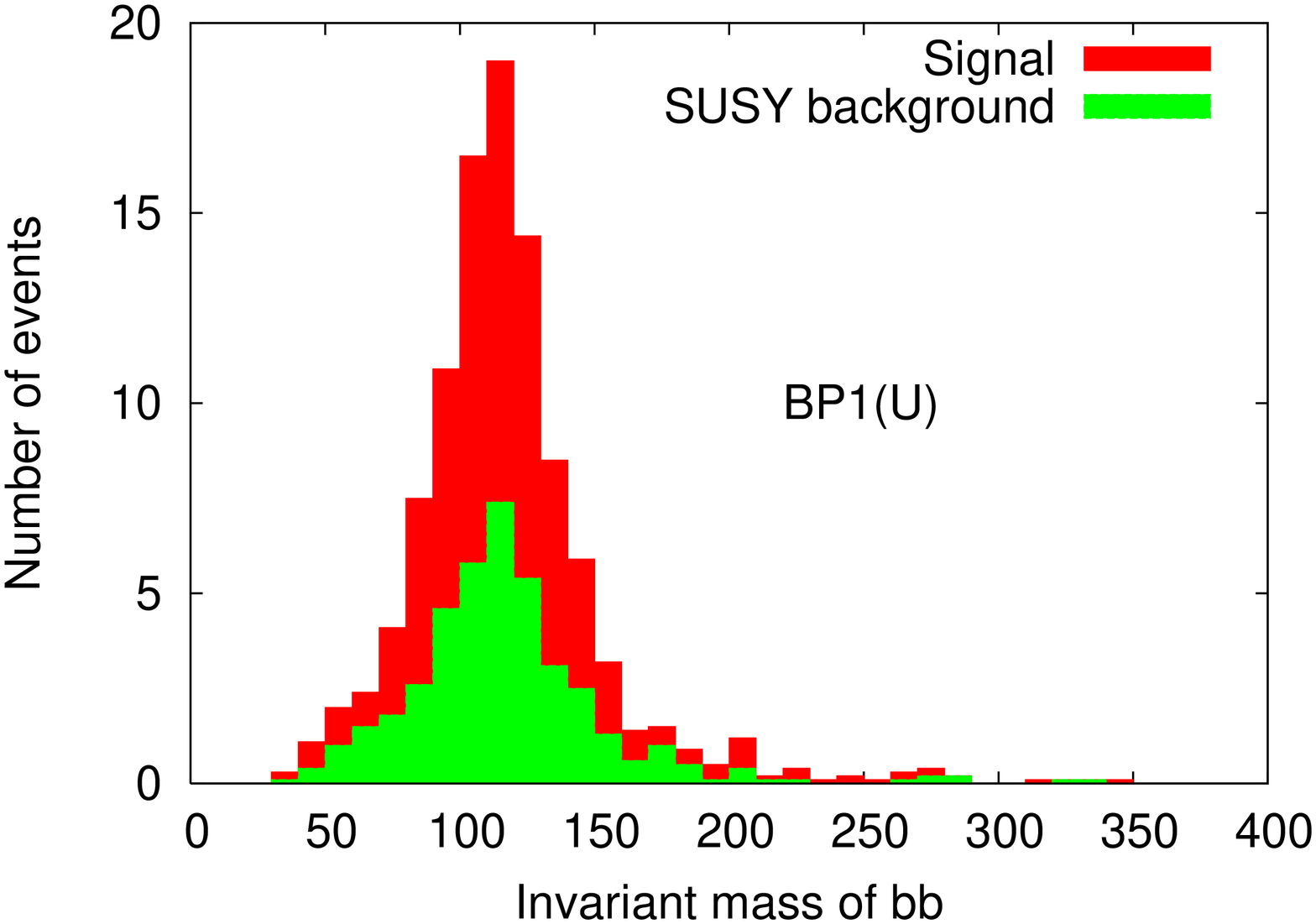,width=5.0 cm,height=7.0cm,angle=-90.0}}
\hskip -12pt 
{\epsfig{file=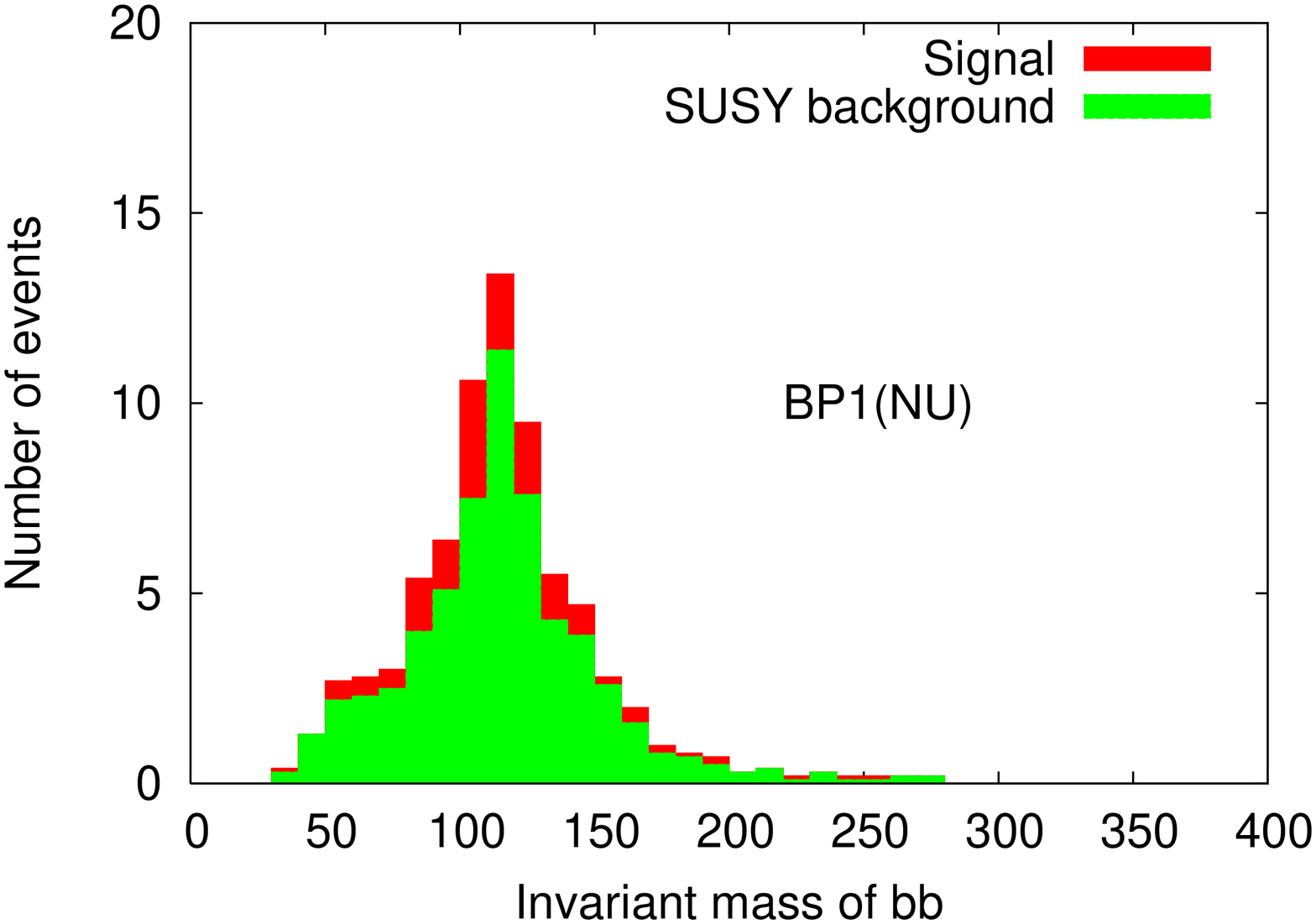,width=5.0cm,height=7.0cm,angle=-90.0}}
{\epsfig{file=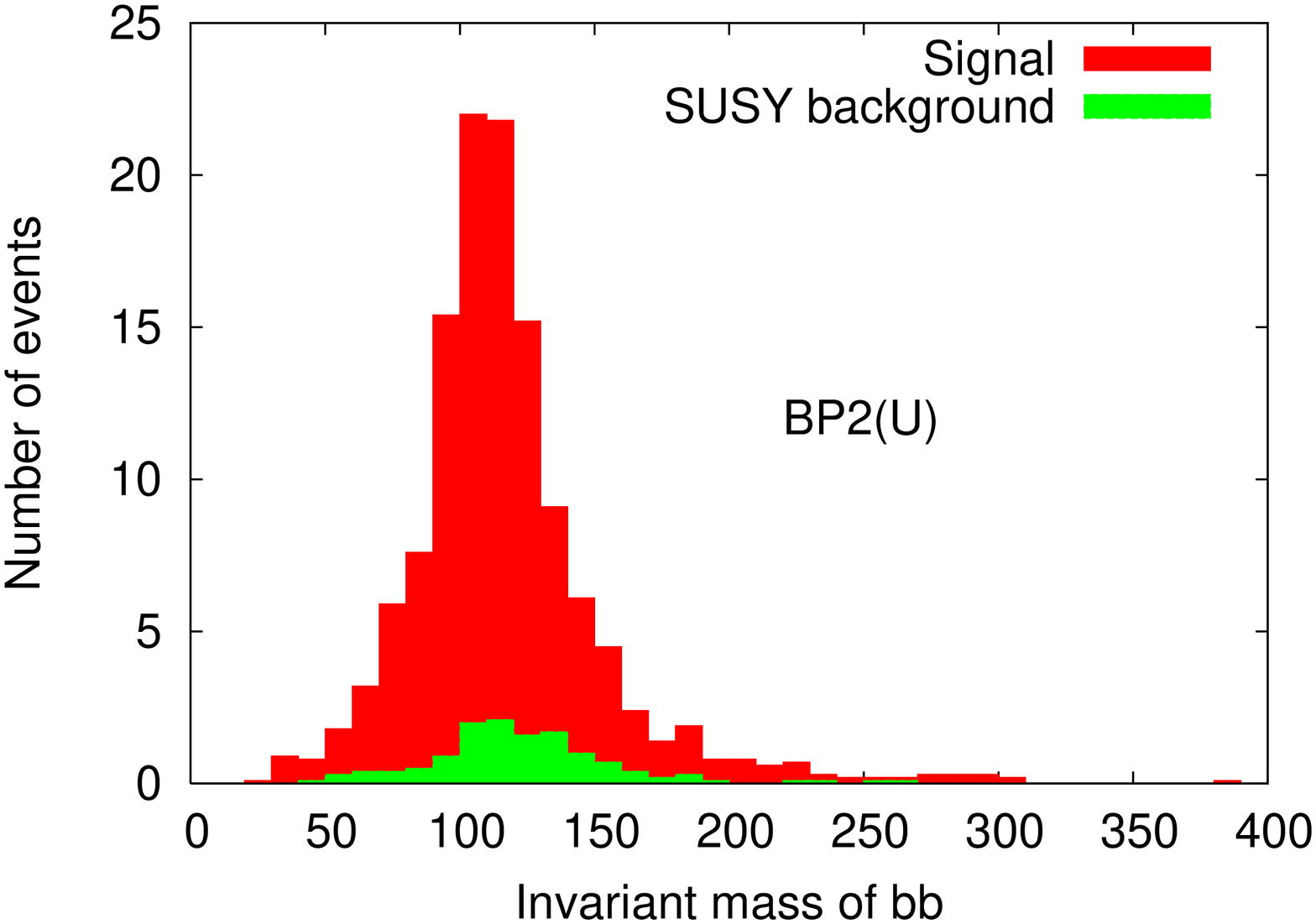,width=5.0 cm,height=7.0cm,angle=-90.0}}
\hskip -12pt 
{\epsfig{file=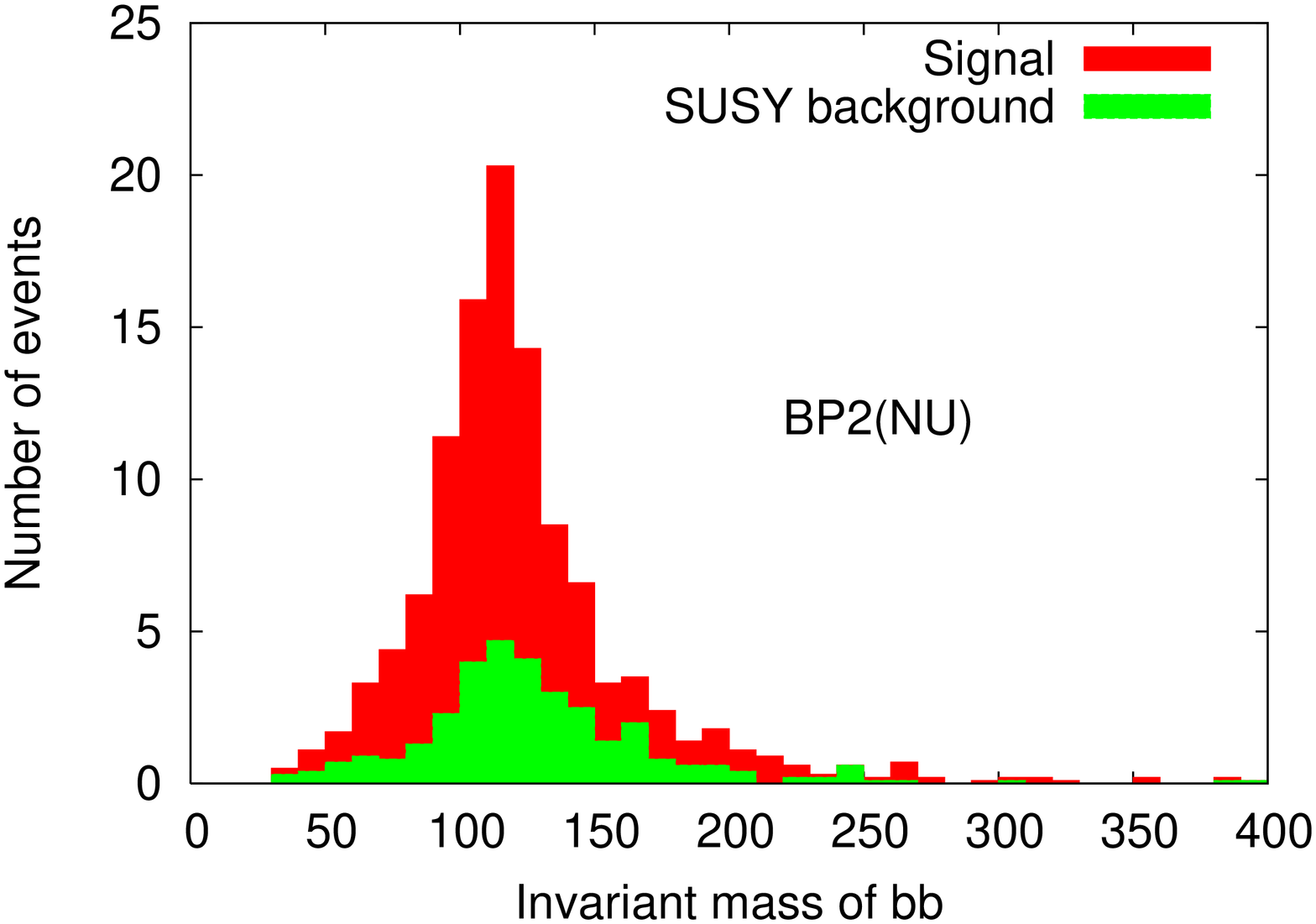,width=5.0cm,height=7.0cm,angle=-90.0}}
{\epsfig{file=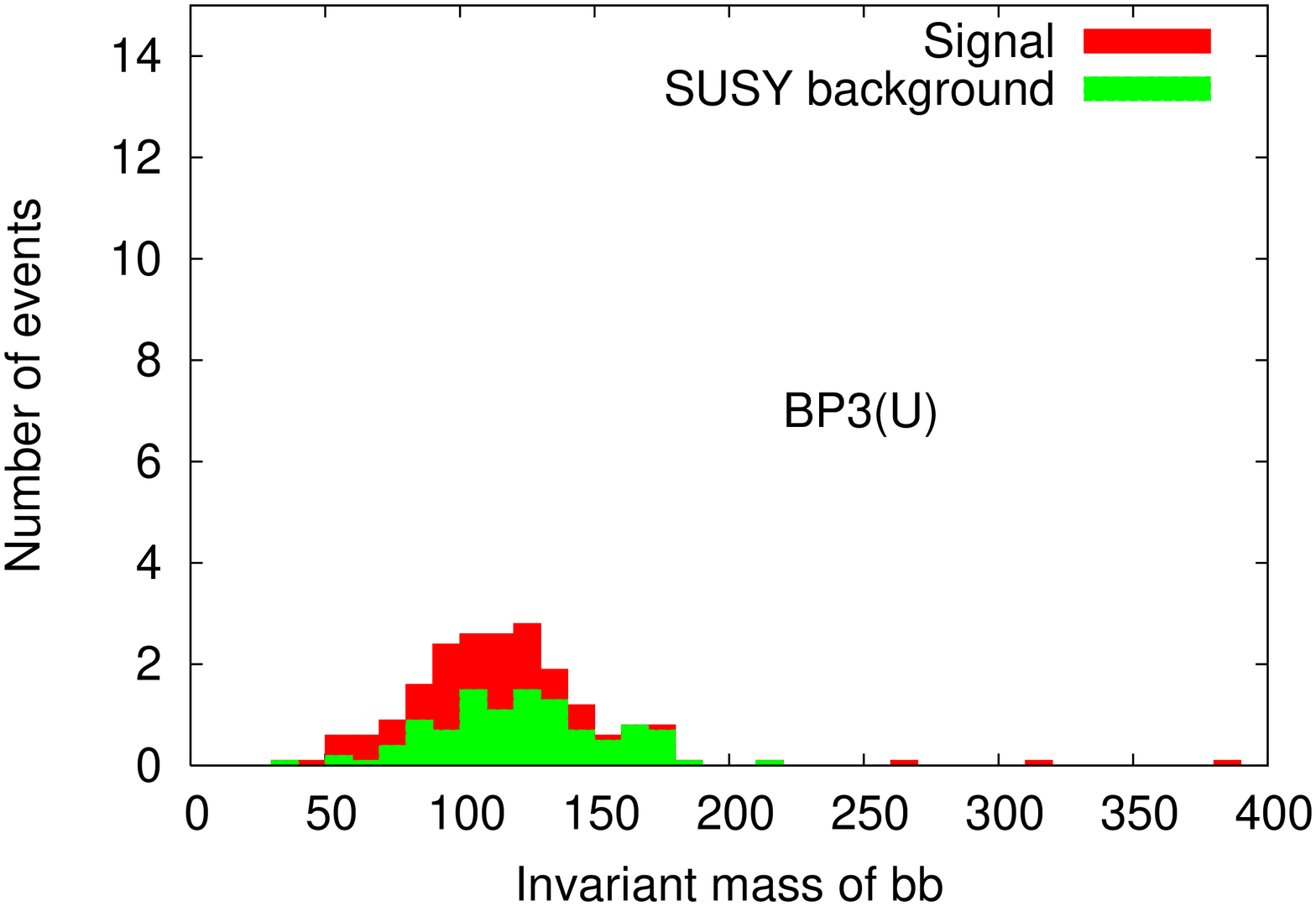,width=5.0 cm,height=7.0cm,angle=-90.0}}
\hskip -12pt 
{\epsfig{file=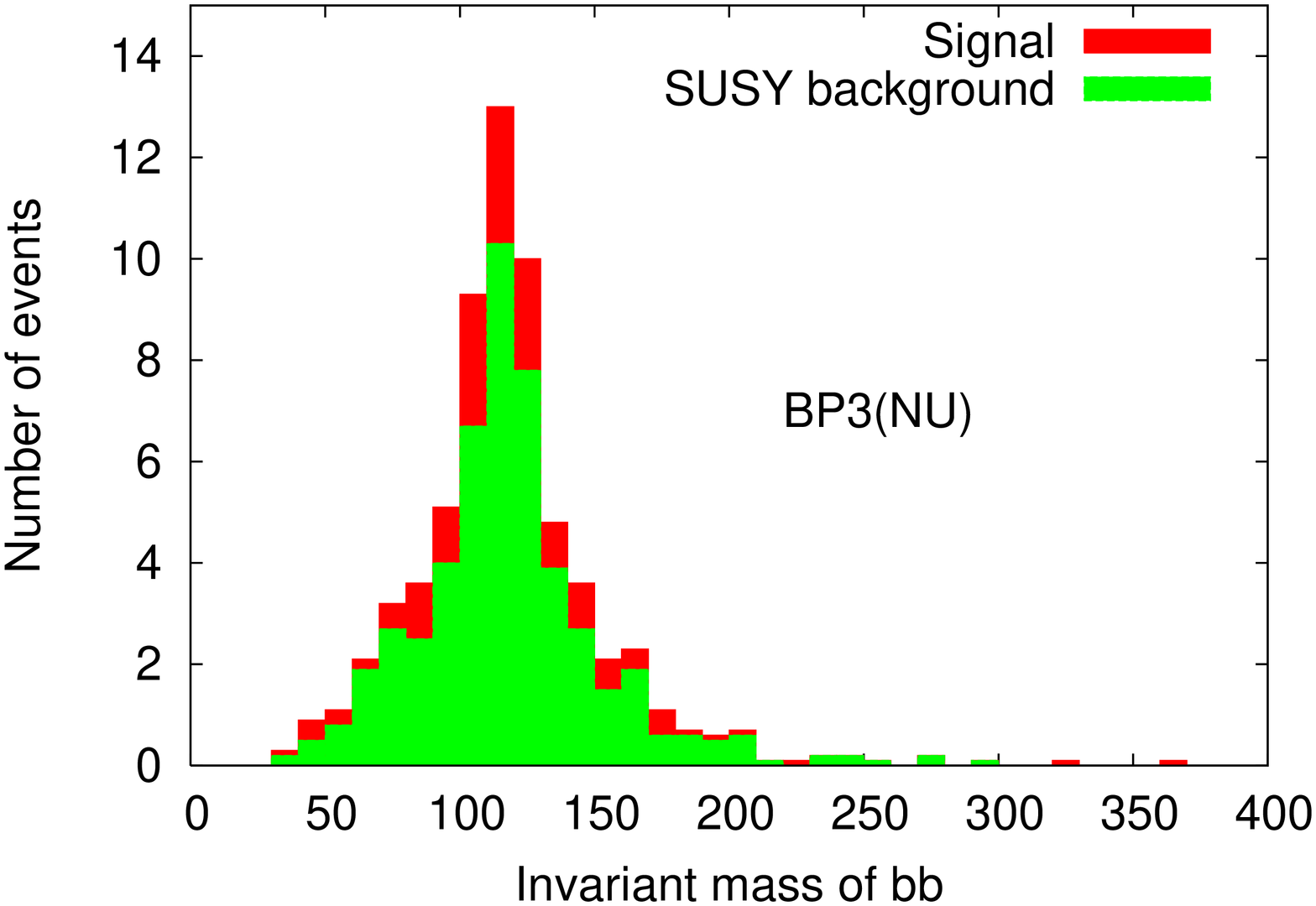,width=5.0cm,height=7.0cm,angle=-90.0}}
{\epsfig{file=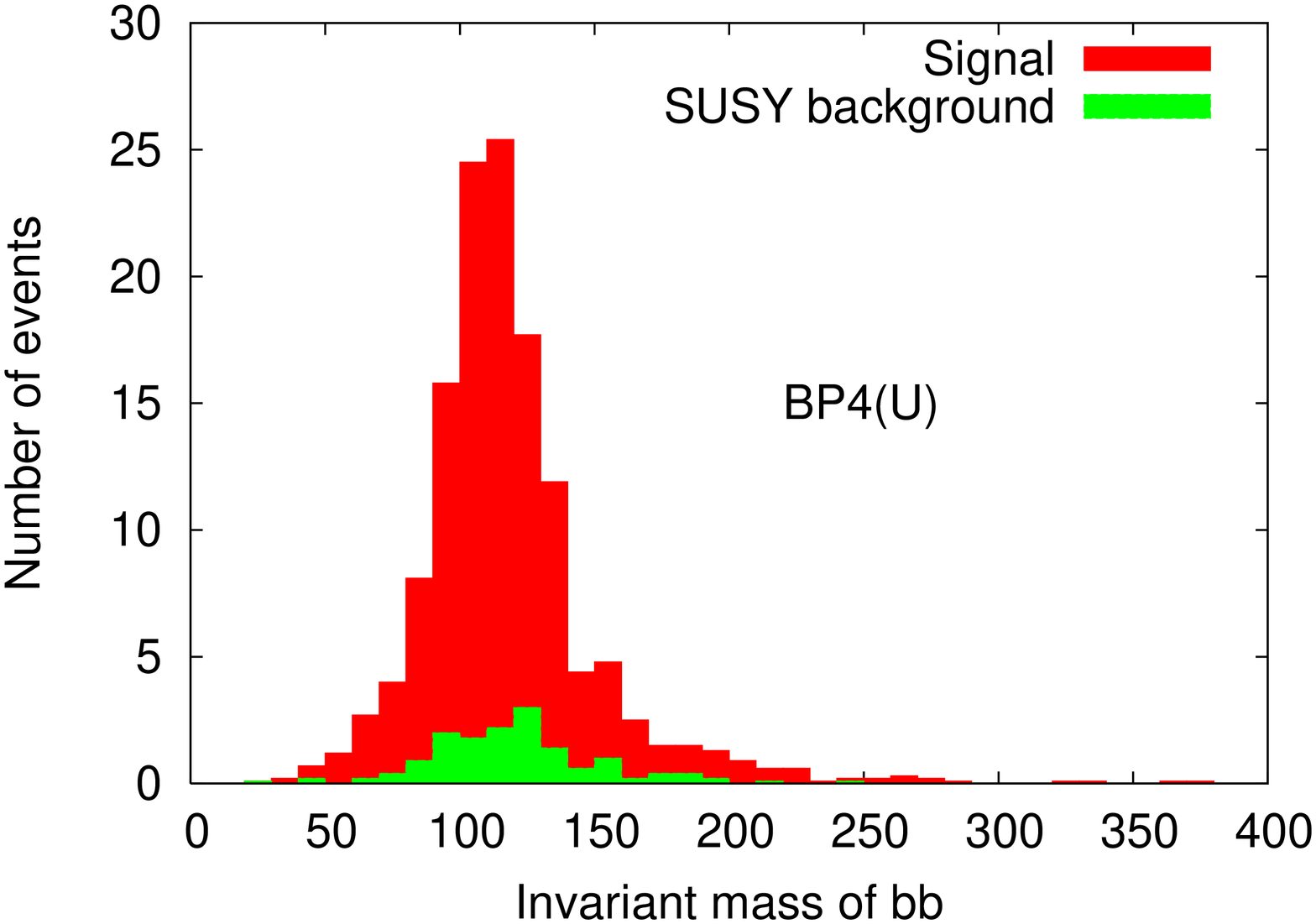,width=5.0 cm,height=7.0cm,angle=-90.0}}
\hskip -12pt 
{\epsfig{file=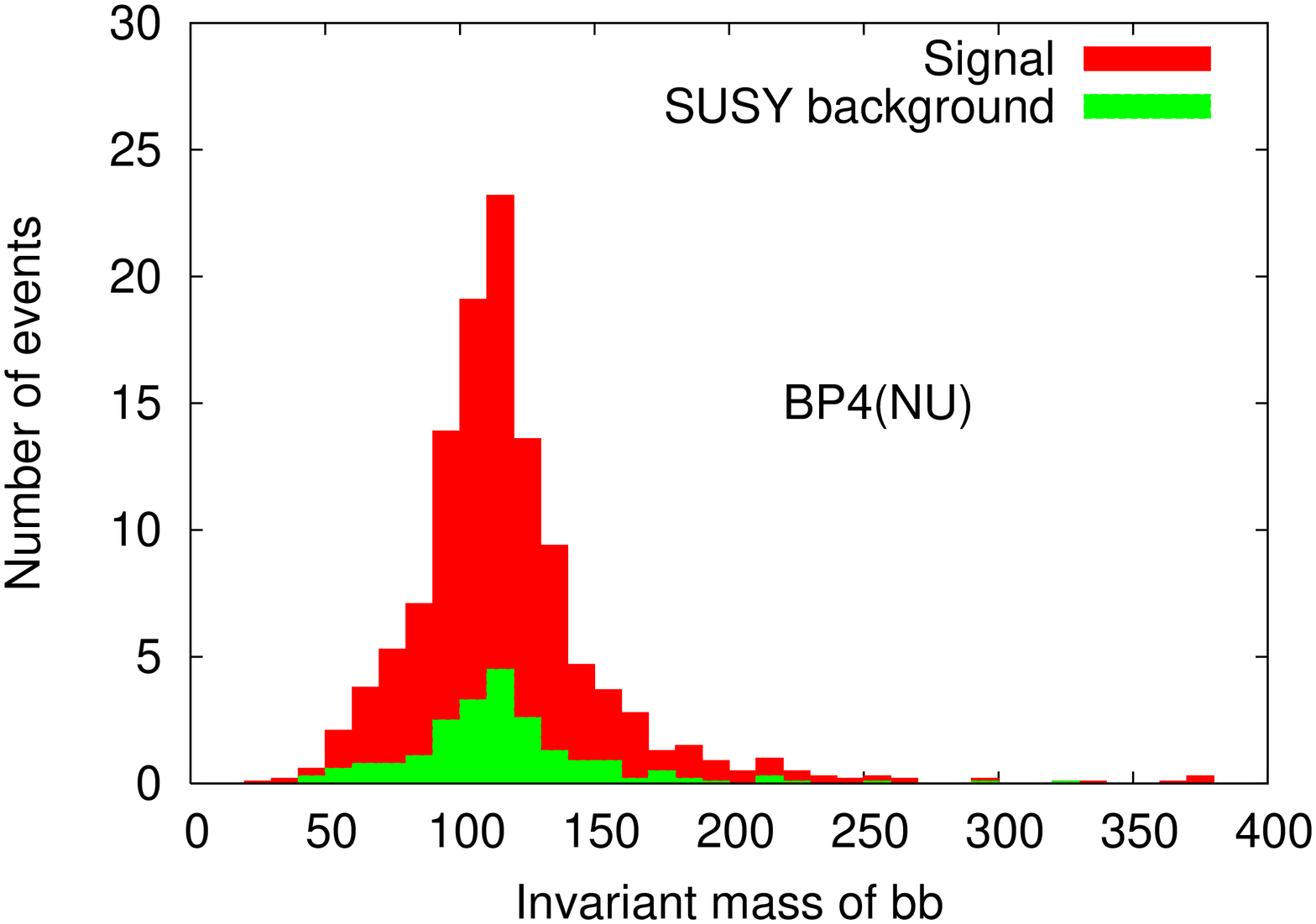,width=5.0cm,height=7.0cm,angle=-90.0}}
\caption{Invariant mass (in GeV) distribution for universal(left) and non-non-universal(right) for lightest Higgs for $m_{H^{\pm}}=250$ GeV. } 
\end{center}
\label{fig11}
\vspace*{-1.0cm}
\end{figure}
The generic observation from the set of plots in Figure 6 is that we can choose a suitable window in the invariant mass of the $b$-jet pairs to optimize the signal. The tails of the distributions are likely to get a significant contribution from the `wrong' combinations of $b$-jets, i.e., the ones which do not come from the lightest Higgs boson. Having said that we also note that such imposing such a window may not help in some situations where not only the signal to background ratios are small but also the absolute rates are rather poor. In any case, thus, the observation in this respect is that the universal scenarios would, in general, respond more favourably under such an invariant mass cut. 

In the present study, we choose an invariant mass window of $95 \, {\mathrm {GeV}}\le m_{b_{j_1},b_{j_2}}\le 140$ GeV. Note that the window is asymmetric about the lightest Higgs mass which is taken to be 109 GeV. Thus, the choice is basically guided by the distributions themselves. The rates for the signal, the SUSY background and $t\bar{t}$ events are presented in Table 8 before and after (within parenthesis) imposing this invariant mass window corresponding to 10 fb$^{-1}$ of data. In Table 8 the term `signal' is used in a little different, but in a more direct sense, when compared to that in the text. The term `signal' in Table 8 corresponds to the actual excess over the combined background. The significance of the signal is defined as 
\[ S = \frac{\mathrm{number \, of \,excess \,
events \, over \, the \, combined \, background}}{\sqrt{\mathrm{combined \, background}}}. \]
\noindent
 On incorporation of the indicated invariant mass window for $b$-jet pairs, we see from Table 8 that there are several situation where we get signal significance of $5\sigma$ or more. As pointed out earlier in this section the situation with the universal scenarios is a little more optimistic with, for example, signal significance being in the range $5- 12 \sigma$ are obtainable for BP1, BP2 and BP4. On the other hand the non-universal scenarios stand less of a chance with only BP4 reaching an optimal signal significance of $5\sigma$. This is more or less expected from the production rates presented in Table 5 in section 4.  This, in general, corresponds to the region of parameter space in $M_2-\mu$ plane where the rates for the lightest neutral Higgs boson is less than that for the charged Higgs. Exception to this may arise since the SUSY background could also be affected simultaneously. However, the dearth of events in non-universal scenario by itself is not a problem, since, this region of parameter space can be probed in a complementary way via the charged Higgs boson  which is produced dominantly.  

It is also noticed that the signal significance gets affected only marginally by further incorporation of the invariant mass window for the $b$-jet pairs. This is not unexpected in view of the preceeding range of $p_T$ of individual $b$-jets used in our analysis. Nonetheless, peak-hunting in different invariant mass windows is  a more direct technique to find resonances. By choosing a narrower window in invariant mass we can, in principle, improve the quality of the signal. However, with already low event-counts as presented in Table 8 this turns out to be a futile exercise. However, with increased volume of data this is worth trying.  

For $m_{H^{\pm}}=250$ GeV (Table 9), the signal significance in some cases would get enhanced compared to the previous case. The reason is that, while the mass of the lightest neutral Higgs boson remains the same, the heavier Higgs bosons now become more massive. This results in a suppression of the SUSY background. This could be seen very clearly from BP2 (for both universal and non-universal cases) and BP4 (non-universal case), where the significances have gone upto $\sim 10\sigma$, $\sim 5.8\sigma$ and $ \sim 7.5 \sigma$ respectively.

\begin{table}[hbtp]
\begin{center}
\begin{tabular}{||c|c|c|c|c|c|c|c|c||} \hline\hline
&\multicolumn{4}{|c|}{ Universal}&\multicolumn{4}{|c|}{Non-universal}\\\hline
Process & BP1 &BP2&BP3&BP4 & BP1 &BP2&BP3&BP4  \\
\hline
\hline
Signal&69(42)&88(56)&16(10)&125(79)&14(8)&43(25)&16(10)&73(44)\\
\hline\hline
SUSY&45(26)&48(30)&18(11)&17(11)&51(27)&85(52)&35(21)&59(35)\\
Background&&&&&&&&\\
\hline\hline
$t\tbar$&\multicolumn{8}{|c|}{91}\\
Background&\multicolumn{8}{|c|}{(50)}\\
\hline\hline
\end{tabular}
\label{tab1}
\caption{Expected number of events with an integrated luminosity of 10 fb$^{-1}$ for the case of the lightest neutral Higgs boson with $m_{H^{\pm}}=180$ GeV. Numbers within the parenthesis are obtained with an invariant mass cut  of $95 \, \mathrm{GeV} \,\le m_{b_{j_1},b_{j_2}}\le 140$ GeV.}
\end{center}
\end{table}

\begin{table}[hbtp]
\begin{center}
\begin{tabular}{||c|c|c|c|c|c|c|c|c||} \hline\hline
&\multicolumn{4}{|c|}{ Universal}&\multicolumn{4}{|c|}{Non-universal}\\\hline
Process & BP1 &BP2&BP3&BP4 & BP1 &BP2&BP3&BP4  \\
\hline
\hline
Signal&63(40)&113(77)&9(5)&117(74)&14(9)&80(48)&15(9)&96(60)\\
\hline\hline
SUSY&41(24)&13(8)&11(6)&15(9)&60(34)&32(18)&51(31)&21(13)\\
Background&&&&&&&&\\
\hline\hline

$t\tbar$&\multicolumn{8}{|c|}{91}\\
Background&\multicolumn{8}{|c|}{(50)}\\
\hline\hline
\end{tabular}
\label{tab1}
\caption{Expected number of events with an integrated luminosity of 10 fb$^{-1}$ for the case of lightest neutral Higgs boson with $m_{H^{\pm}}= 250$ GeV. Numbers within the parenthesis are obtained with an invariant mass cut of $95 \, \mathrm{GeV} \,\le m_{b_{j_1},b_{j_2}}\le 140$ GeV.}
\end{center}
\end{table}

\subsection{Heavy neutral Higgs bosons}

In this section we will discuss about the signal of heavier neutral Higgs bosons, i.e., the CP-odd Higgs ($A$) and the heavier CP-even  neutral scalar Higgs ($H$). Expectedly it would be very difficult to distinguish the CP-odd Higgs boson from the CP-even one since they are too closely space in their masses as discussed earlier. Still, one would expect to see a combinedbroadened peak of $A$ and $H$ at some higher mass value. For $m_H^{\pm}=180$ GeV, the masses of these heavy neutral scalars are approximately 165 GeV for the choice of our SUSY 
parameters. Here, we would expect to find an invariant mass peak in that vicinity by using a suitable $p_{T}$ window on the $b$-jets as done for the lightest neutral Higgs boson. In this case  we use an optimal window on the tagged $b$-jets of  $70 \, \mathrm{GeV} \, \le{p^{b_{j_1,j_2}}_{T}}\le{90}$ GeV to construct the $b\bar{b}$ invariant mass distribution as shown in Figure 8. 

In Table 10, we present  the final numbers  for the signal, the SUSY background and for the $t\bar{t}$ background at an integrated luminosity of 10 fb$^{-1}$. Again, as before,  the numbers inside the parenthesis are with a $b\bar{b}$ invariant mass cut of $140 \, \mathrm{GeV} \, \le m_{b_{j_1},b_{j_2}}\le 190$ GeV. It is observed that for none of the cases, the signal significance is above $5\sigma$.
The BP2 non-universal case is having the largest significance of  $\sim 3.7\sigma$, whereas the corresponding value for the universal one is $\sim 1.7\sigma$. So, unlike the case for the lightest neutral Higgs boson, it is obvious that 10 fb$^{-1}$ of data is not sufficient to probe the heavy neutral Higgs bosons. One thus has to wait for at least 30 fb$^{-1}$ of data for this purpose. Table 10 shows that while BP1, BP2 and BP4 are changing behaviours in terms of signal efficiency while going from the universal to the non-universal scenarios, BP3  remains a low significance for both the cases.  

 For $m_{H^{\pm}}=250$ GeV (Table 11), the masses of the heavy Higgs neutral bosons are around $\sim 240$ GeV. To probe this mass peak we choose a $p_T$ window for the tagged $b$-jet pairs, of $100 \, {\mathrm{GeV}}\le{p^{b_{j_1,j_2}}_{T}}\le{150}$ GeV. Thus, we reconstruct 
the invariant mass distribution for the heavier neutral Higgs bosons at an integrated luminosity of $\mathcal{L}\sim 10$ fb$^{-1}$. This is shown in Figure 9. We also impose an invariant mass cut of $200 \, \mathrm{GeV} \, \le m_{b_{j_1},b_{j_2}}\le 300$ GeV as before. From Table 11, it is clear that in the universal cases the signal strenth decreases because of the phase space suppression as discussed earlier. On the other hand in the non-universal cases, because of the LSP is around 100 GeV, in some regions of the parameters the heavy Higgs bosons production channels are still open kinematically. This can be well understood from Table 11 which chows that the production channels for the heavier Higgs bosons open up for only BP2 and BP3 non-universal cases, while BP1 case is open for both the cases universal and non-universal. For BP4 the parameter space is not sufficient to help for the any of heavy Higgs channel to open up. 

 The distinguishing thing from that of the lightest Higgs case and $m_{H^{\pm}}=180$ GeV heavy Higgses case, is that here $t\bar{t}$
 background drops down after we put the invariant mass cut, which is clear from the Table 11. Again, 10 fb$^{-1}$ of data is not sufficient
 as lightest neutral Higgs case as the highest significance is $\sim 3.2 \sigma$ for no-universal case of BP2. So need to go 
for $\mathcal{L}\sim 30-50$ fb$^{-1}$ to get a $5\sigma$ significance. 
\begin{figure}[hbtp]
\begin{center}
{\epsfig{file=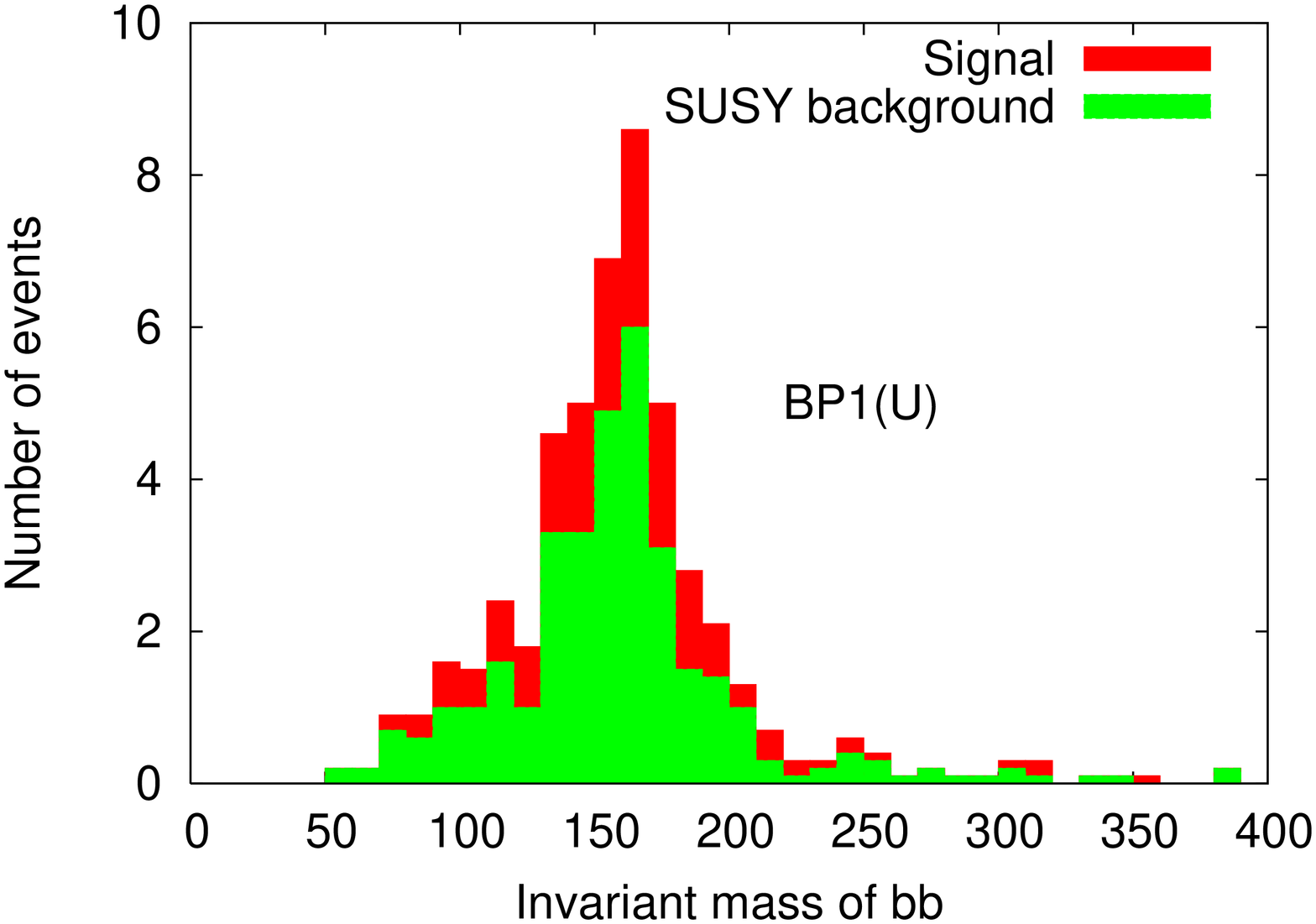,width=5.0 cm,height=7.0cm,angle=-90.0}}
\hskip -12pt 
{\epsfig{file=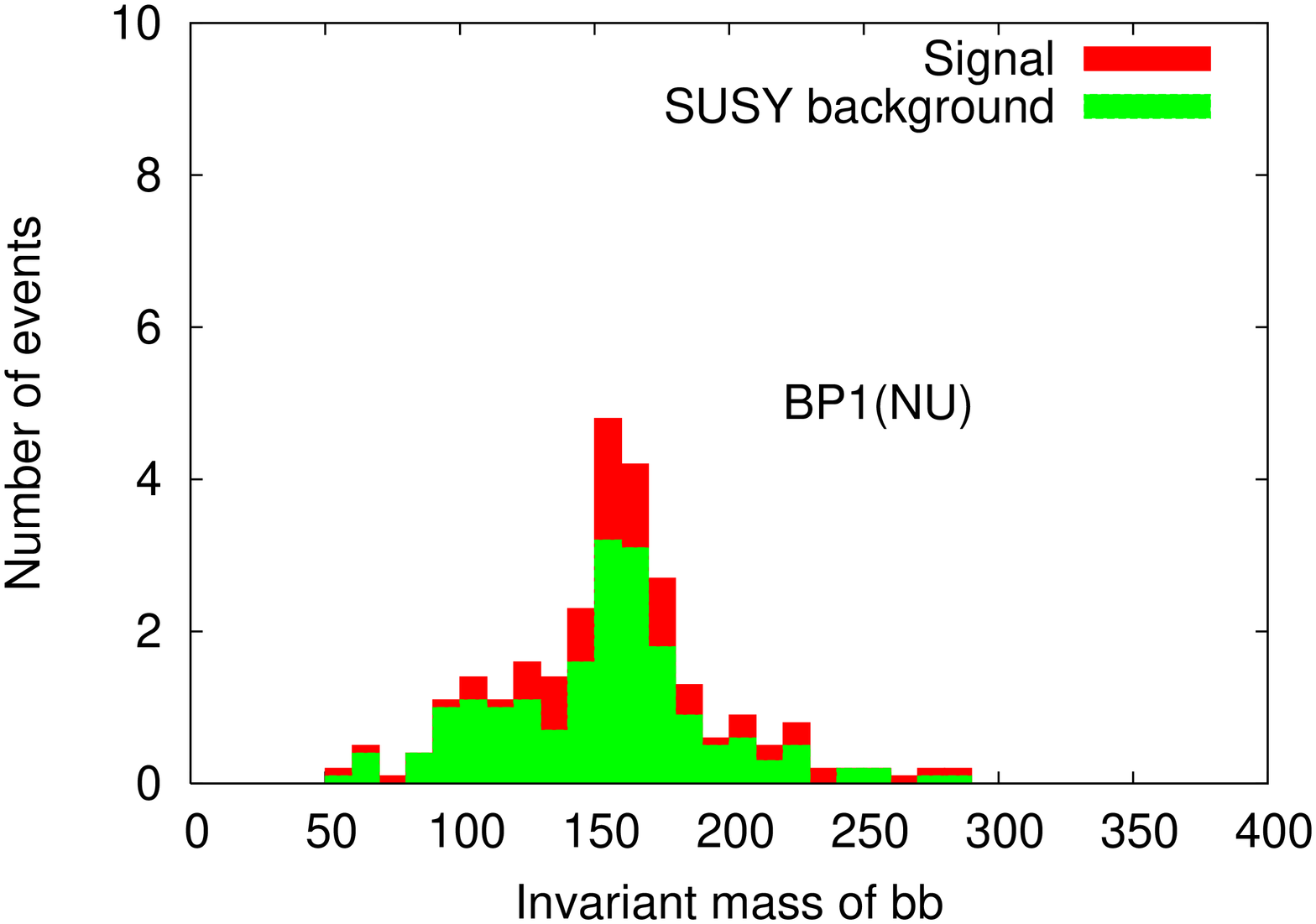,width=5.0cm,height=7.0cm,angle=-90.0}}
{\epsfig{file=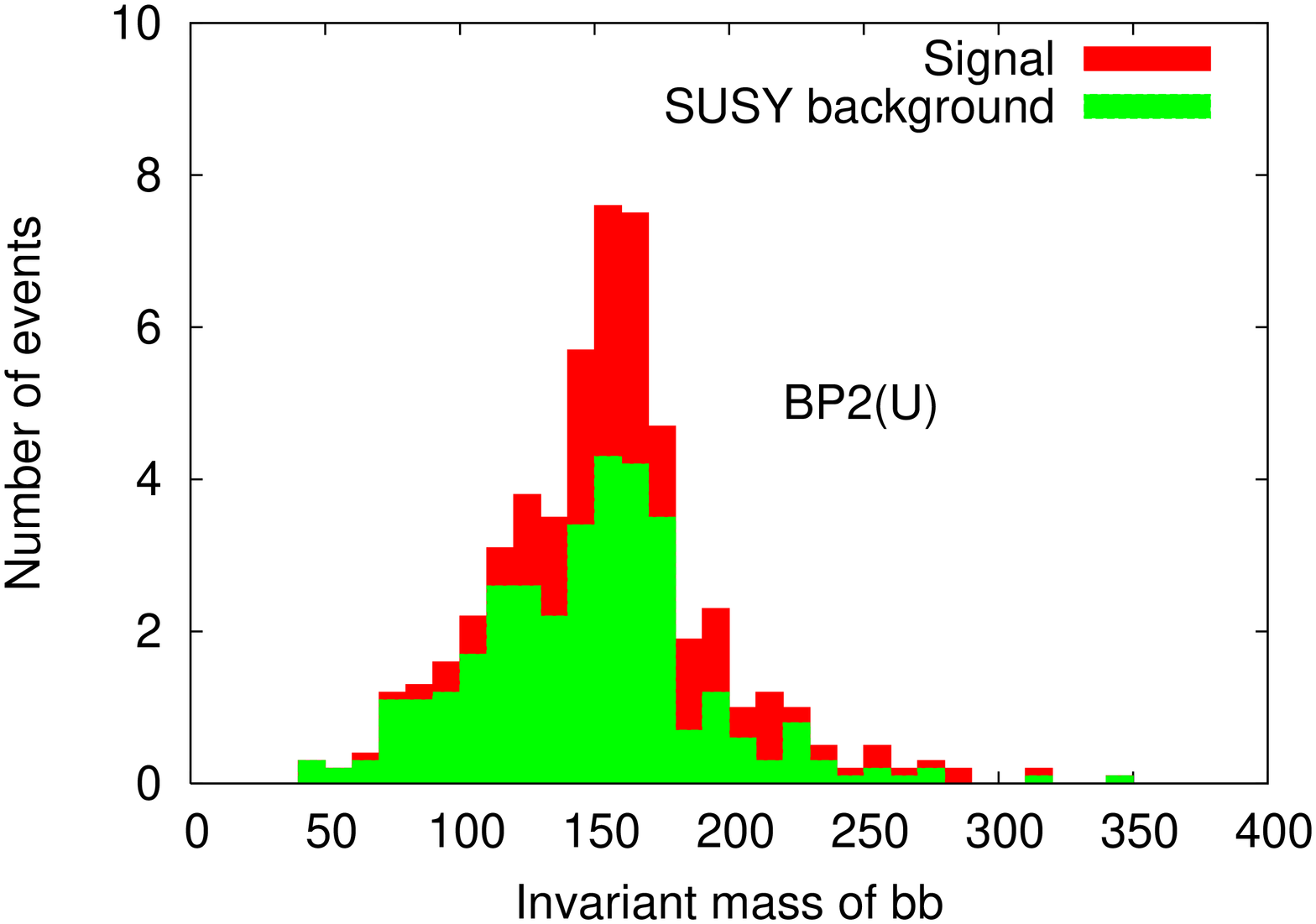,width=5.0 cm,height=7.0cm,angle=-90.0}}
\hskip -12pt 
{\epsfig{file=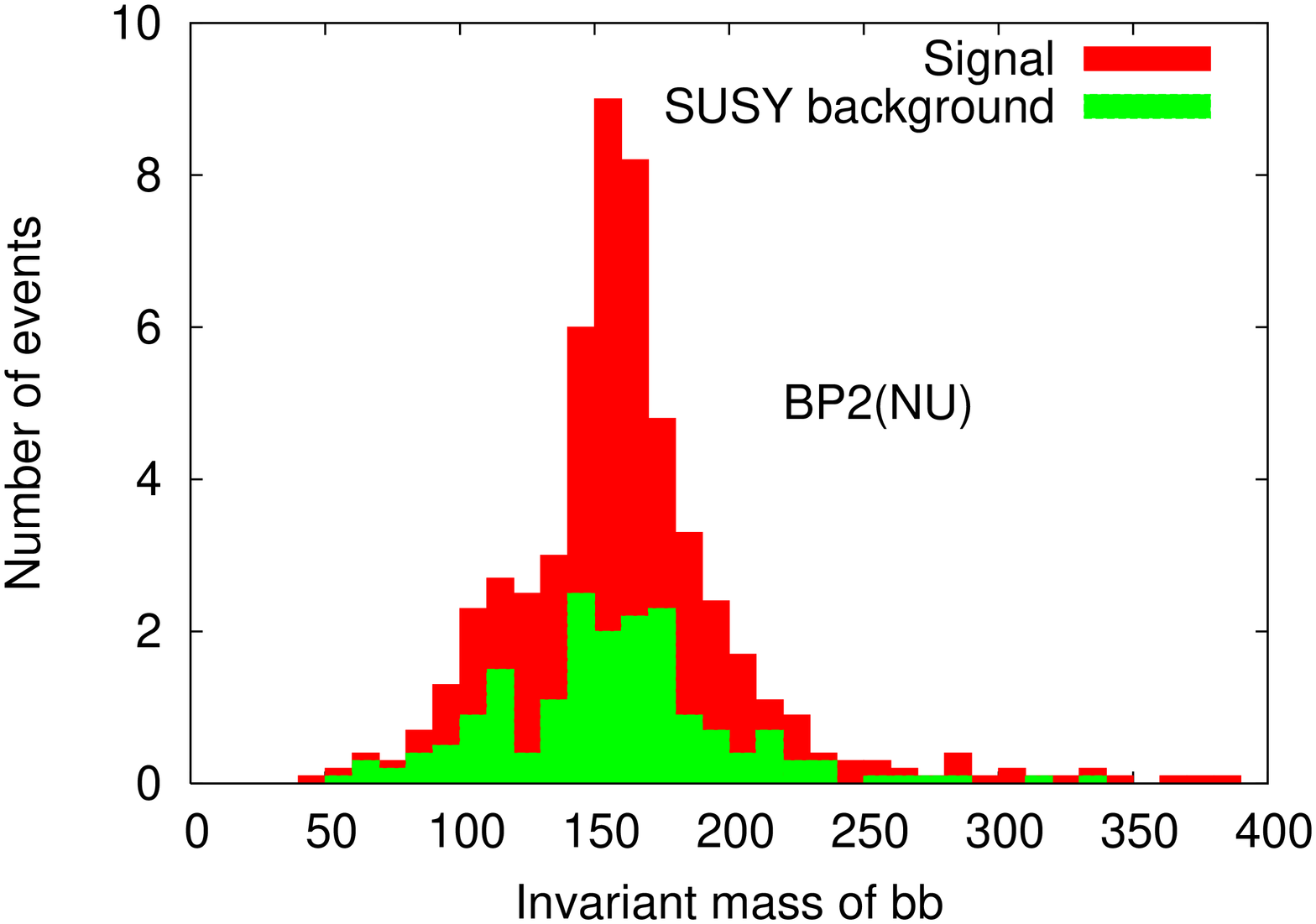,width=5.0cm,height=7.0cm,angle=-90.0}}
{\epsfig{file=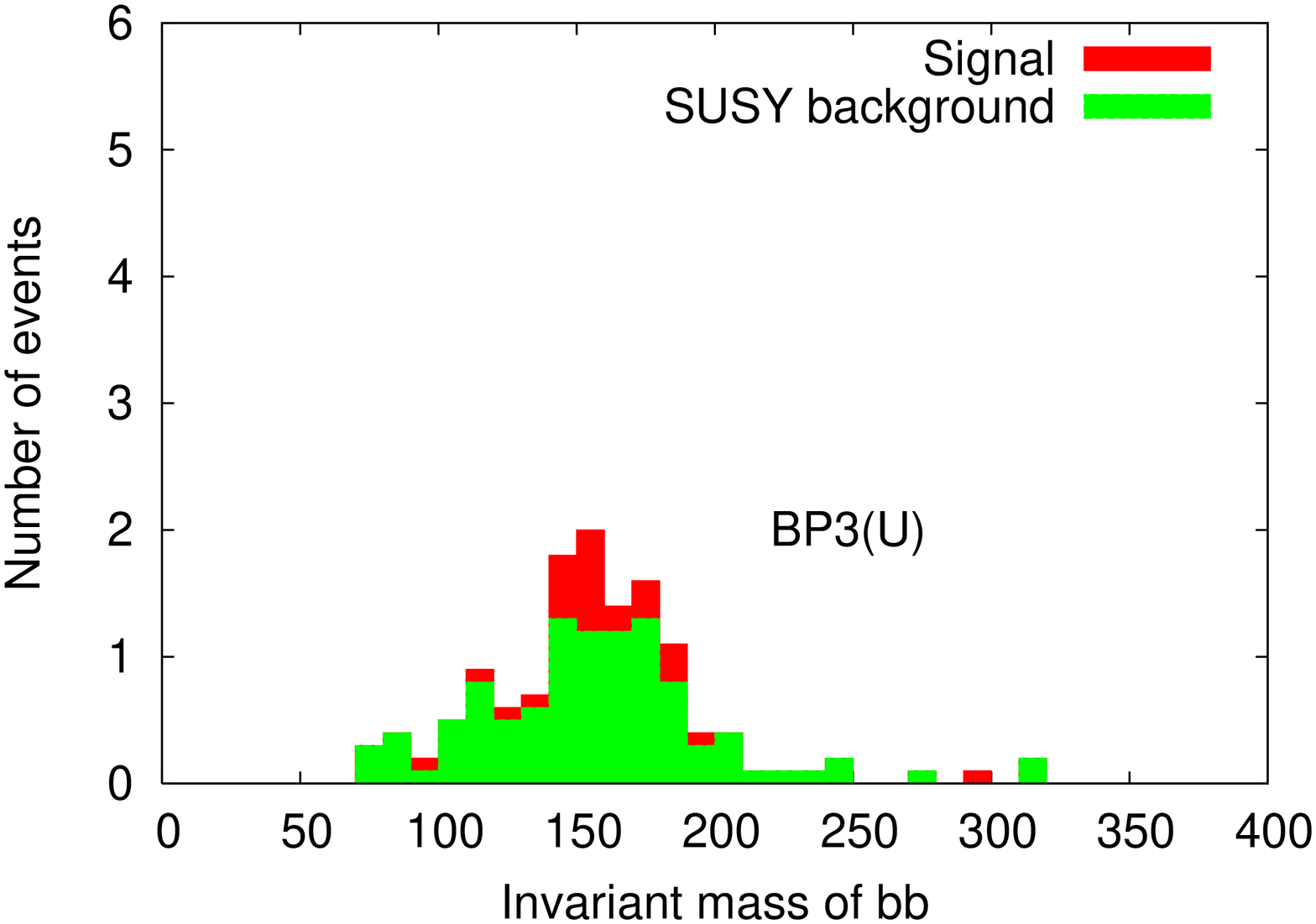,width=5.0 cm,height=7.0cm,angle=-90.0}}
\hskip -12pt 
{\epsfig{file=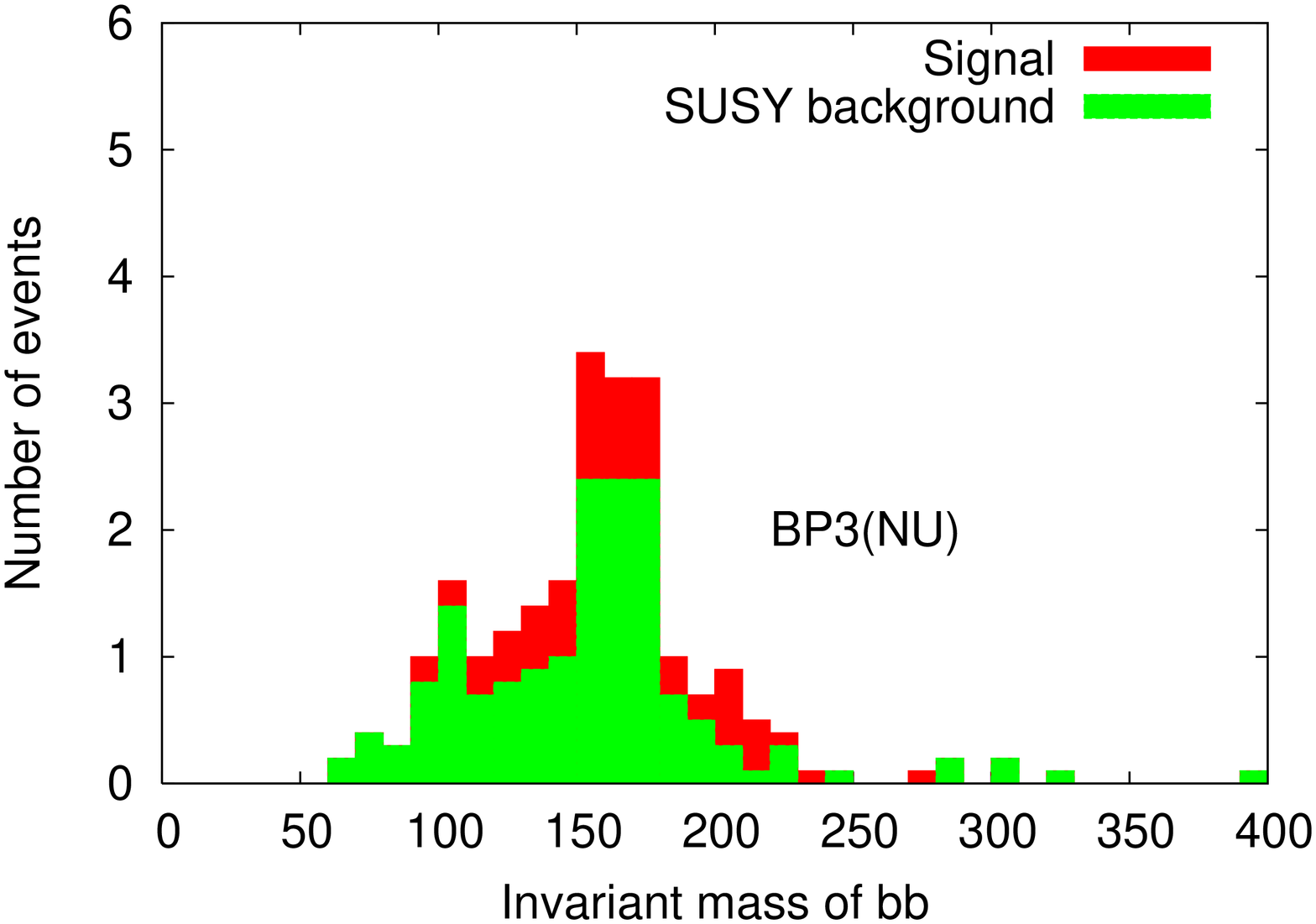,width=5.0cm,height=7.0cm,angle=-90.0}}
{\epsfig{file=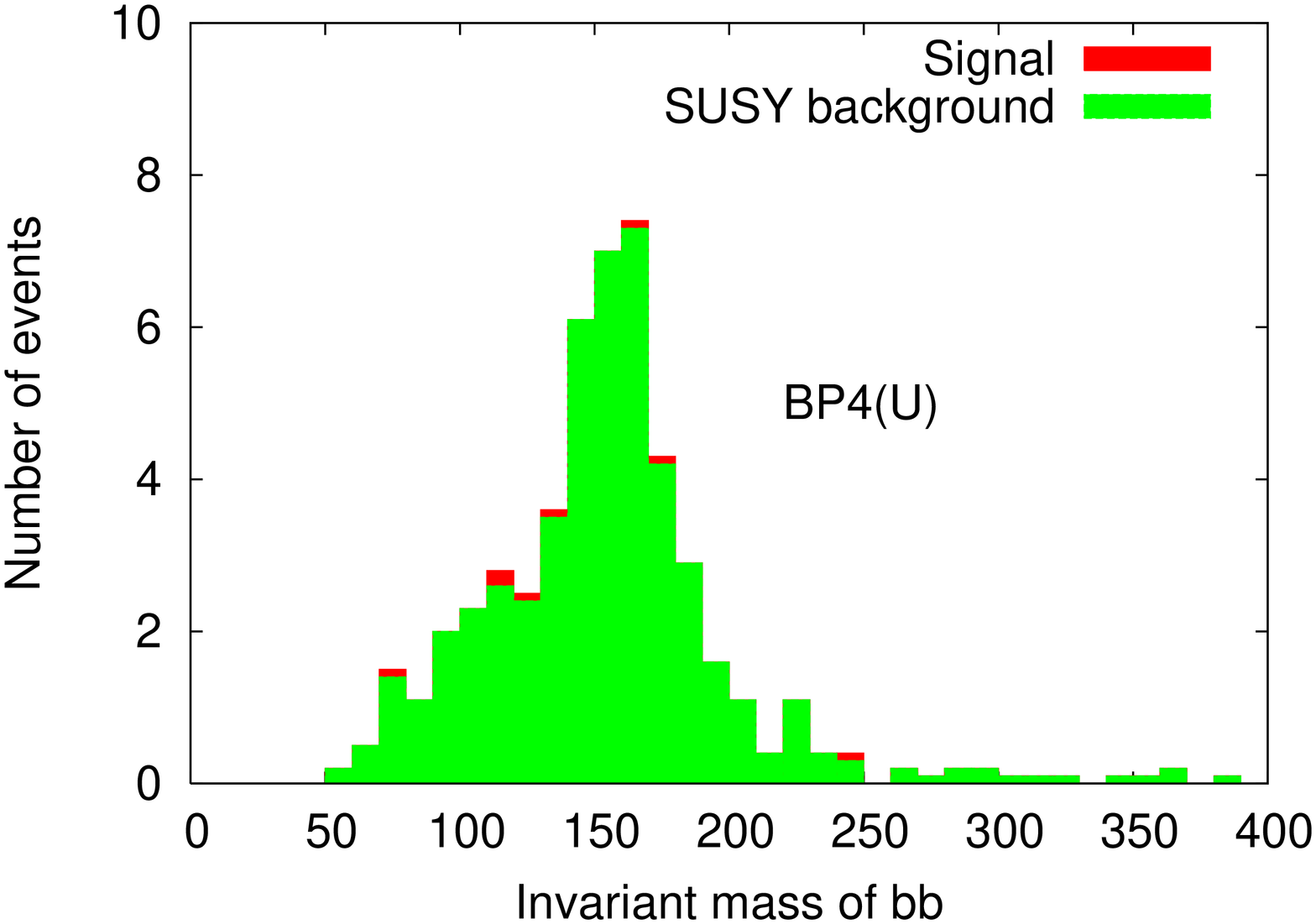,width=5.0 cm,height=7.0cm,angle=-90.0}}
\hskip -12pt 
{\epsfig{file=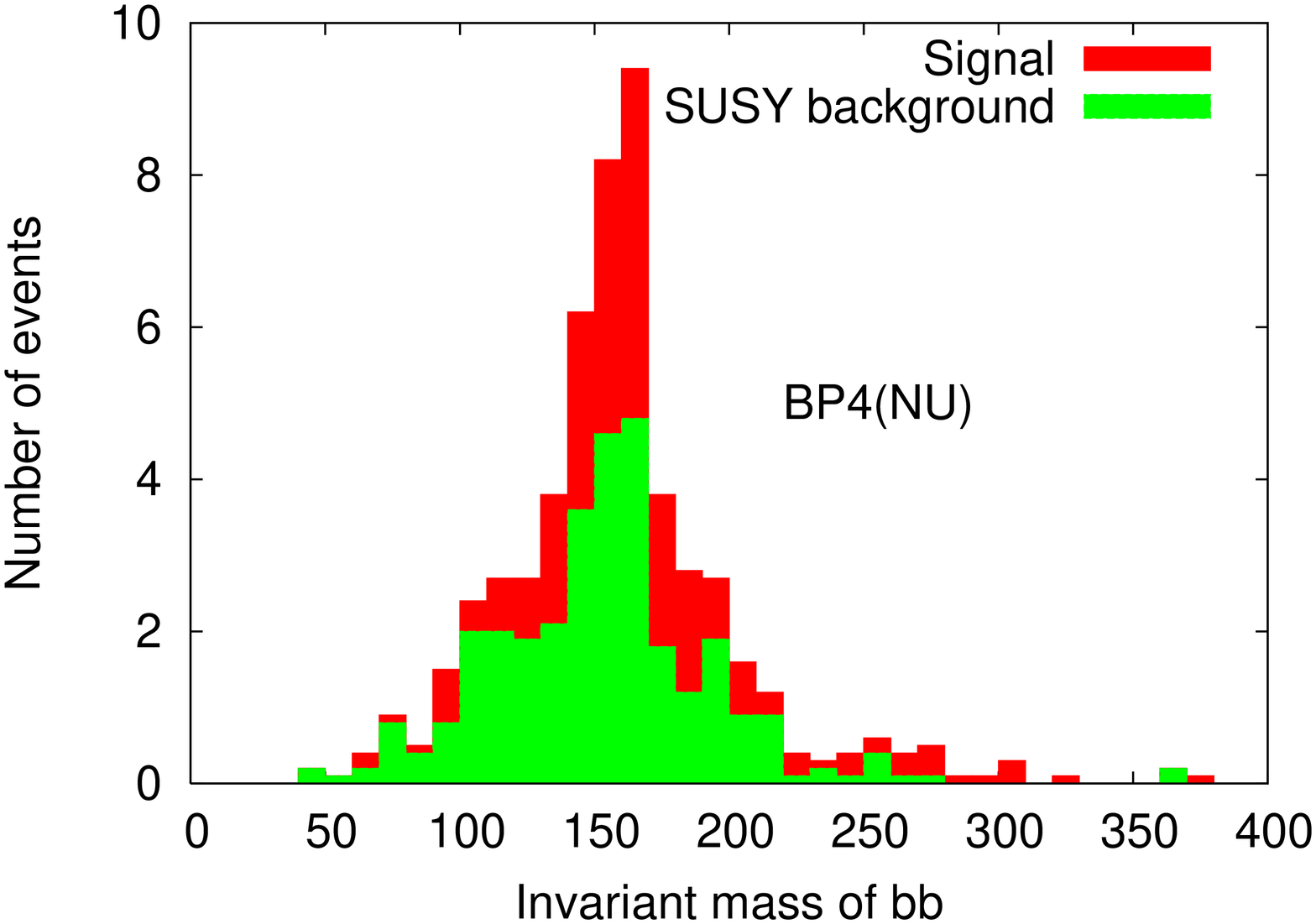,width=5.0cm,height=7.0cm,angle=-90.0}}
\caption{Invariant mass (in GeV) distribution for universal(left) and non-non-universal(right) for heavy neutral Higgs bosons and with $m_{H^{\pm}}=180$ GeV.} 
\end{center}
\label{fig11}
\vspace*{-1.0cm}
\end{figure}
\begin{figure}[hbtp]
\begin{center}
{\epsfig{file=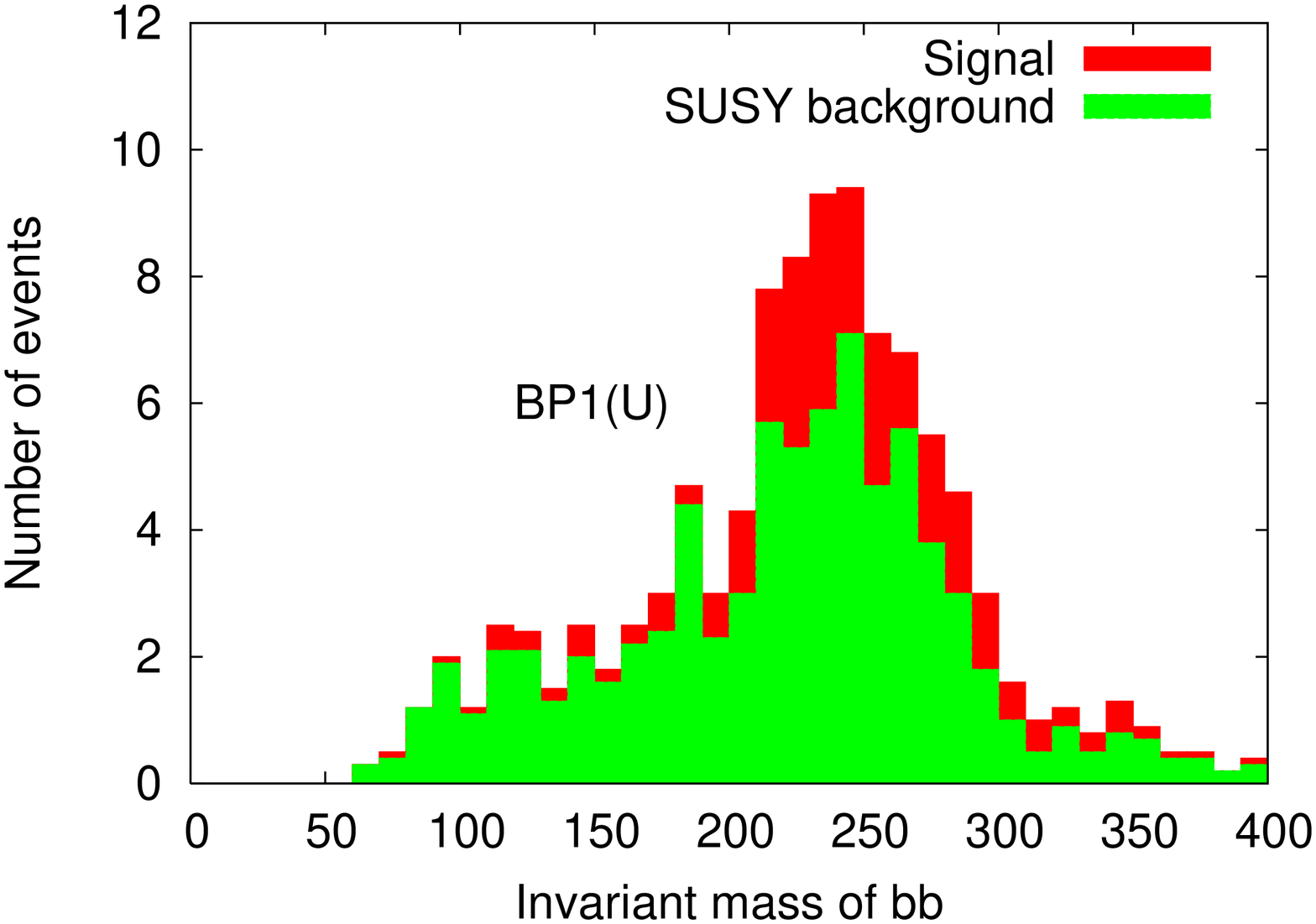,width=5.0 cm,height=7.0cm,angle=-90.0}}
\hskip -12pt 
{\epsfig{file=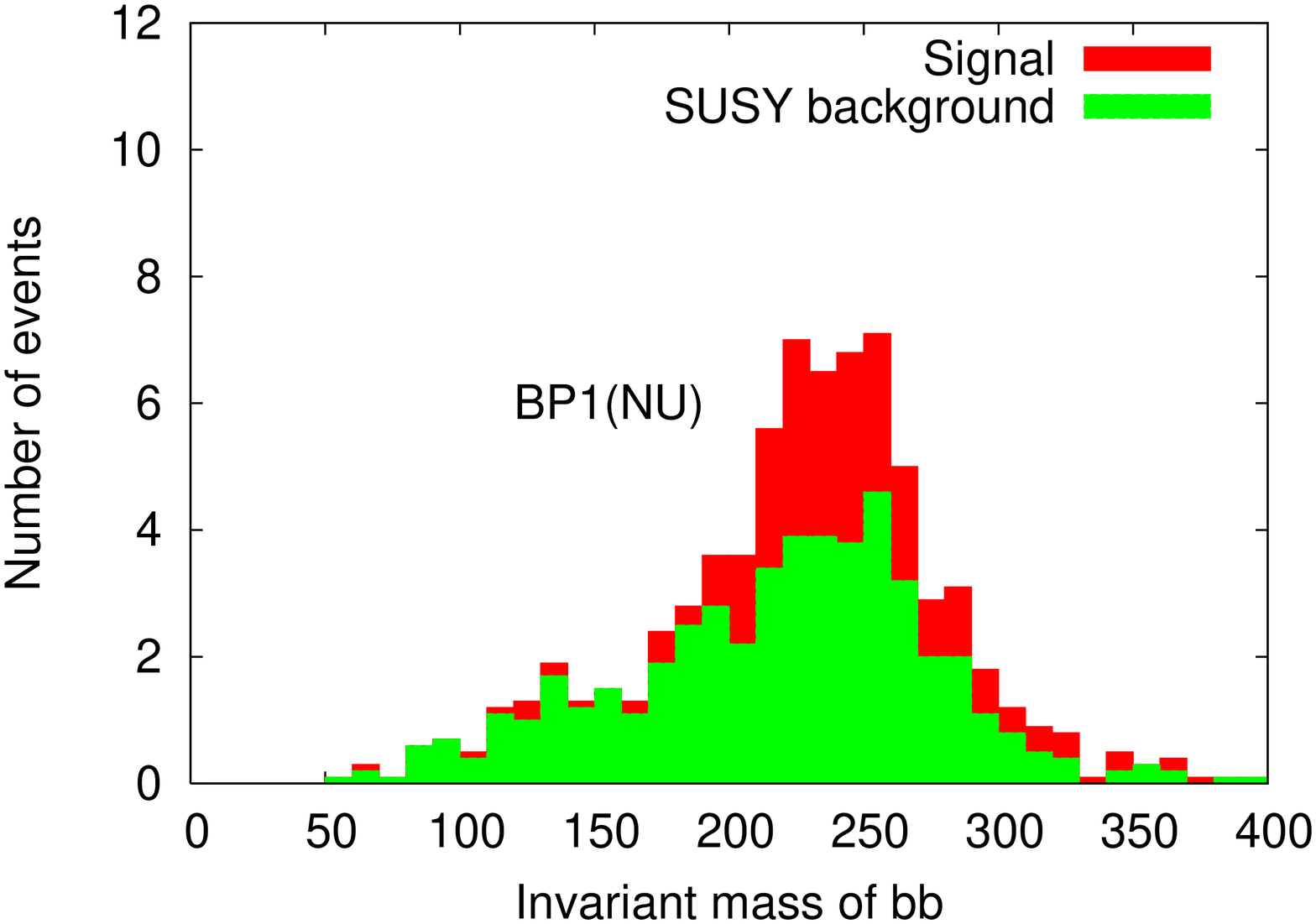,width=5.0cm,height=7.0cm,angle=-90.0}}
{\epsfig{file=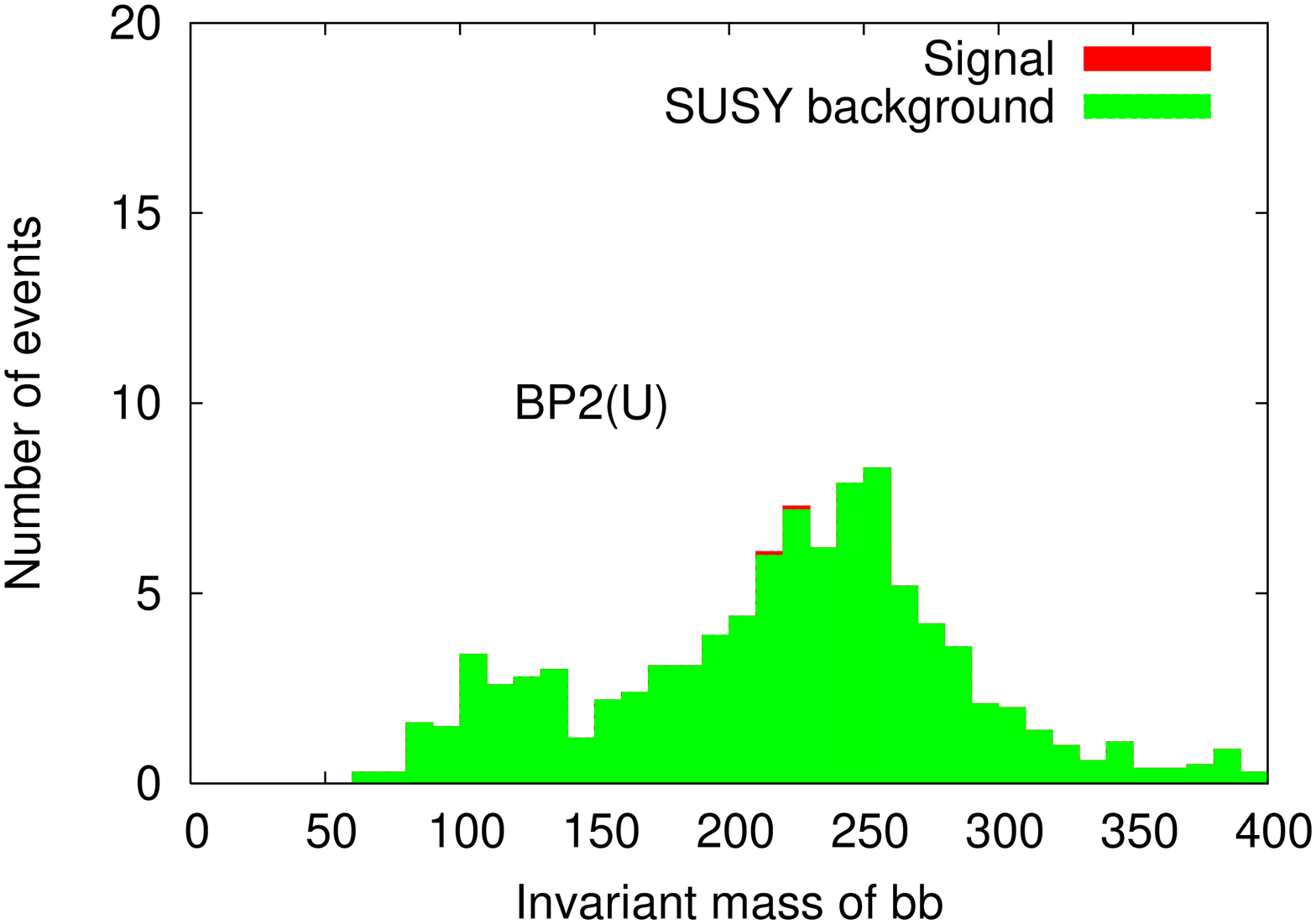,width=5.0 cm,height=7.0cm,angle=-90.0}}
\hskip -12pt 
{\epsfig{file=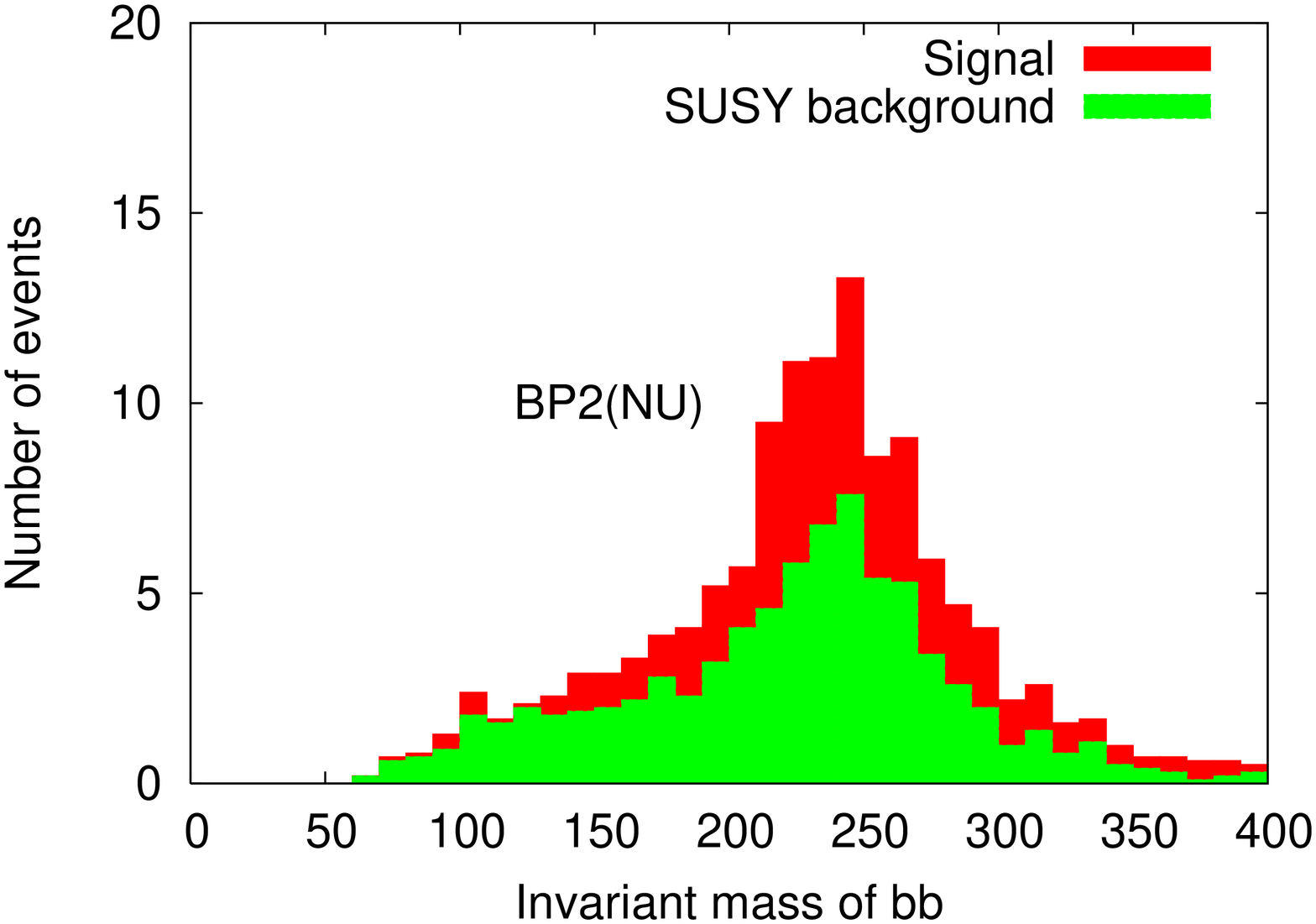,width=5.0cm,height=7.0cm,angle=-90.0}}
{\epsfig{file=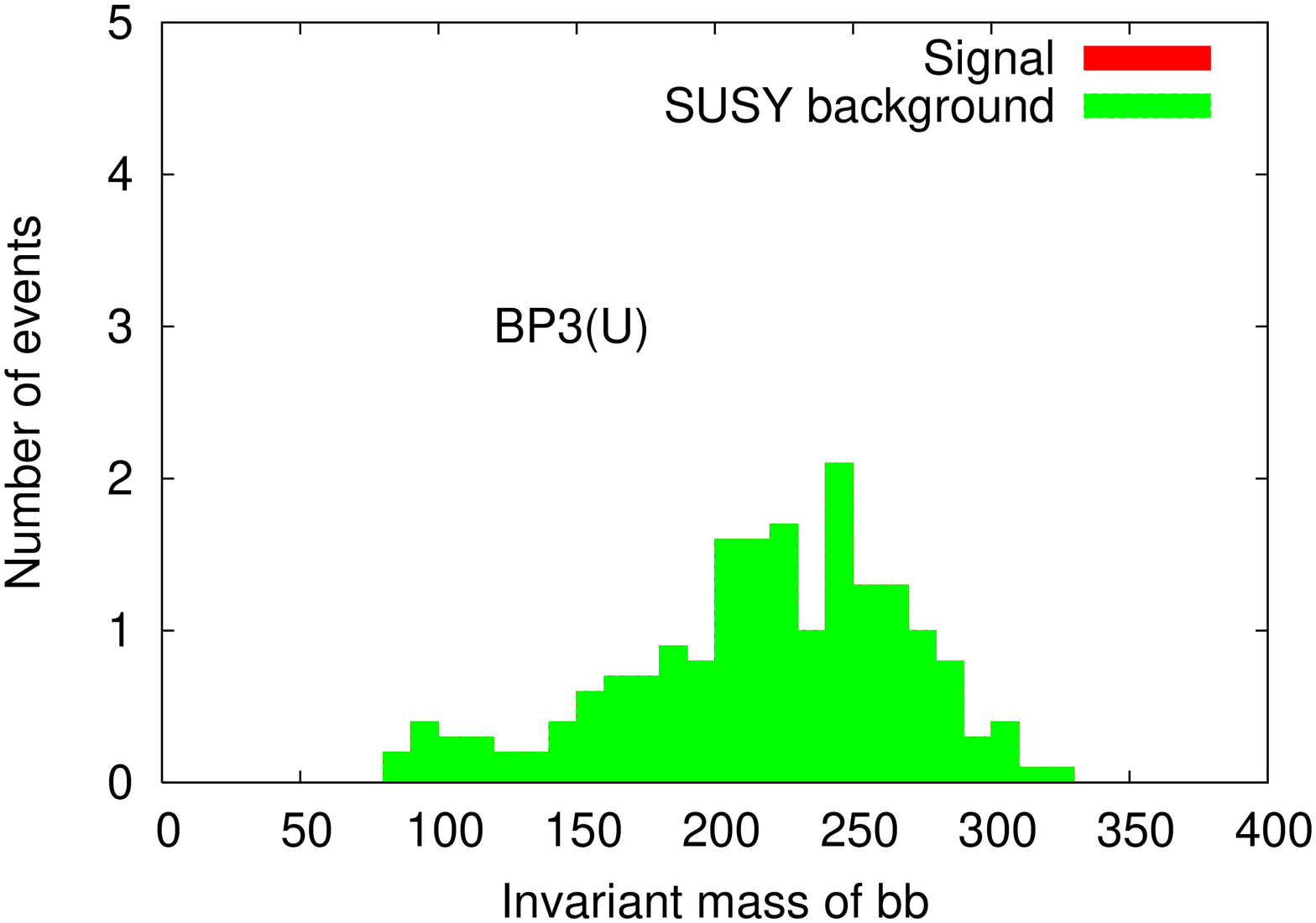,width=5.0 cm,height=7.0cm,angle=-90.0}}
\hskip -12pt 
{\epsfig{file=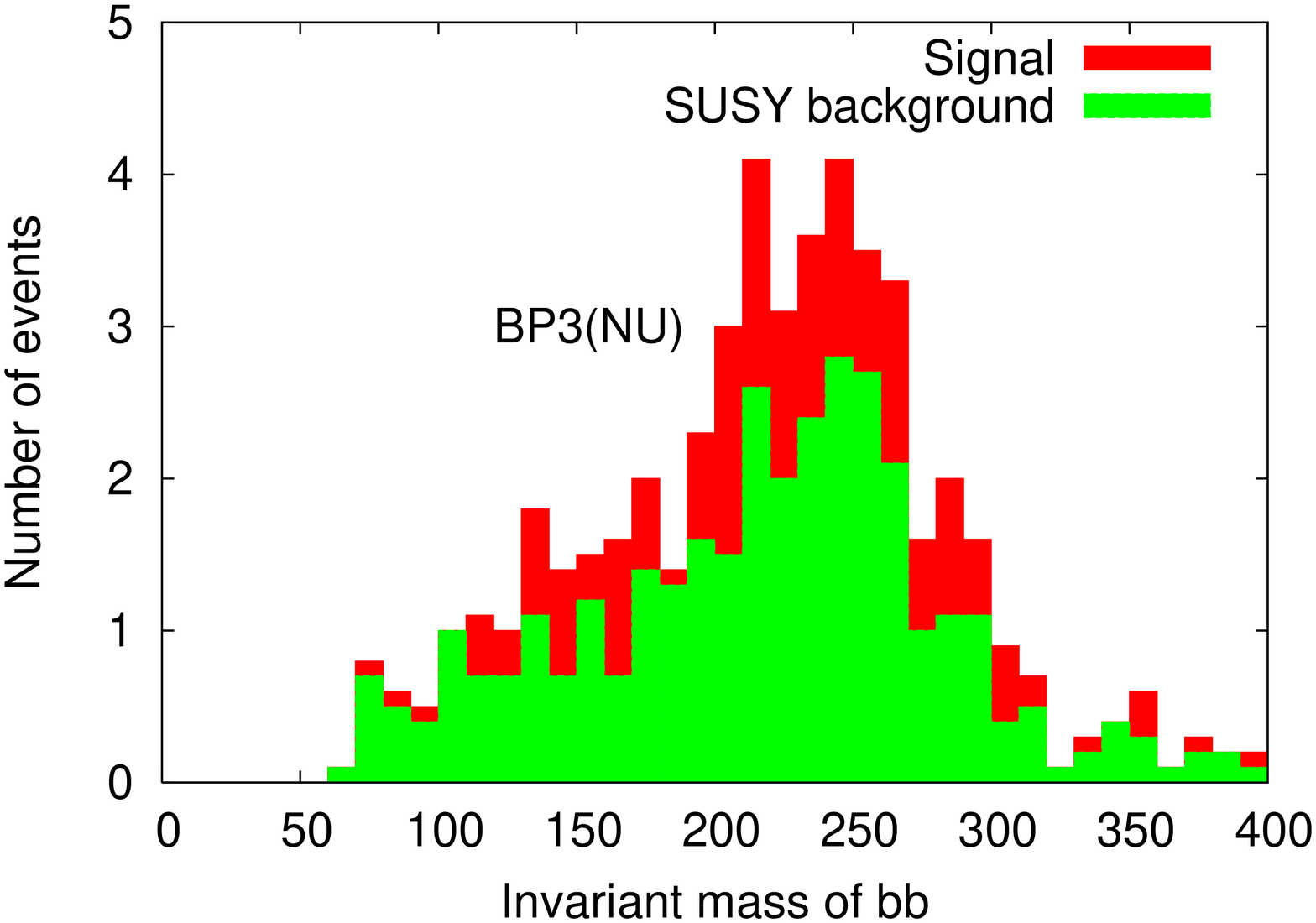,width=5.0cm,height=7.0cm,angle=-90.0}}
{\epsfig{file=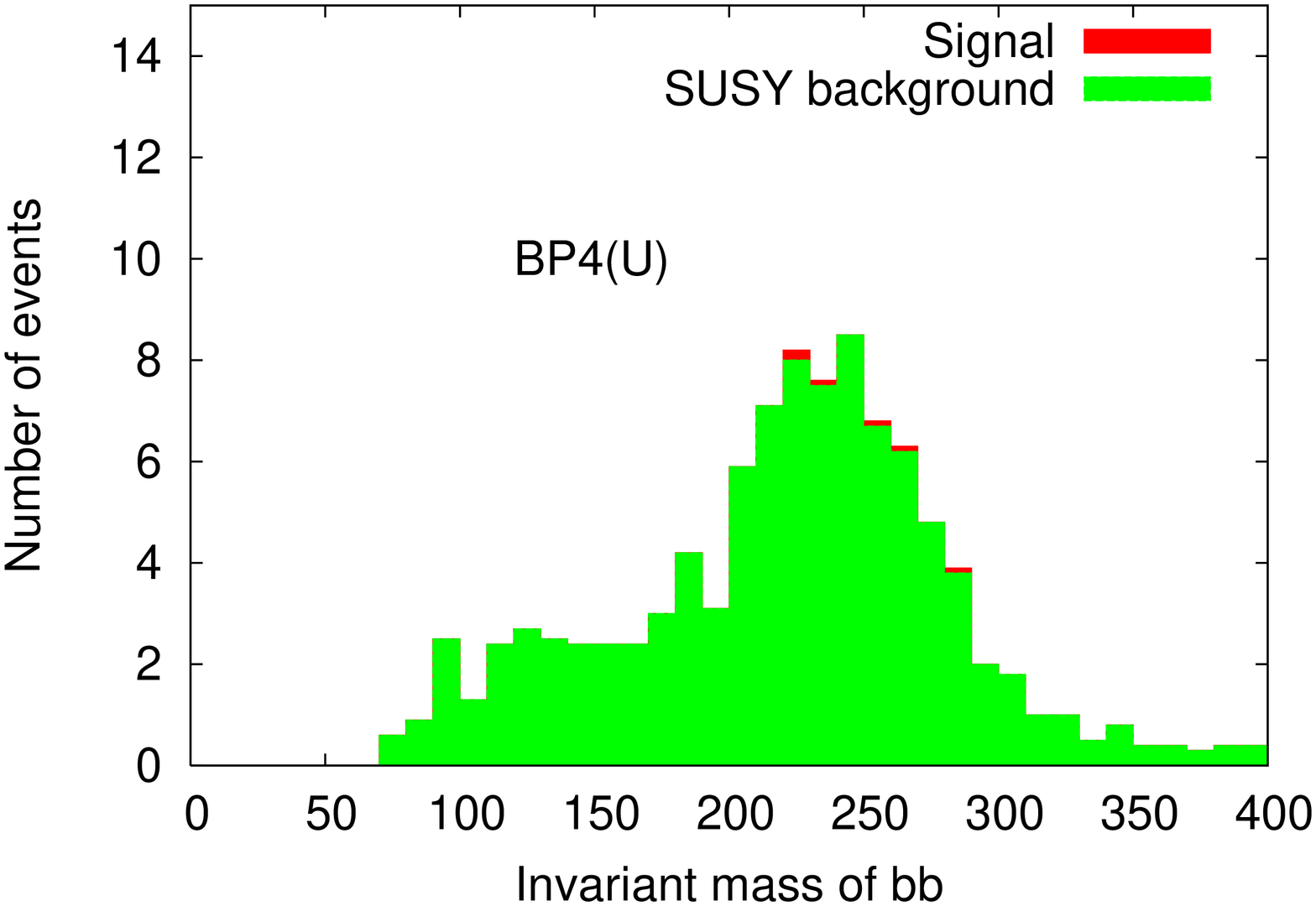,width=5.0 cm,height=7.0cm,angle=-90.0}}
\hskip -12pt 
{\epsfig{file=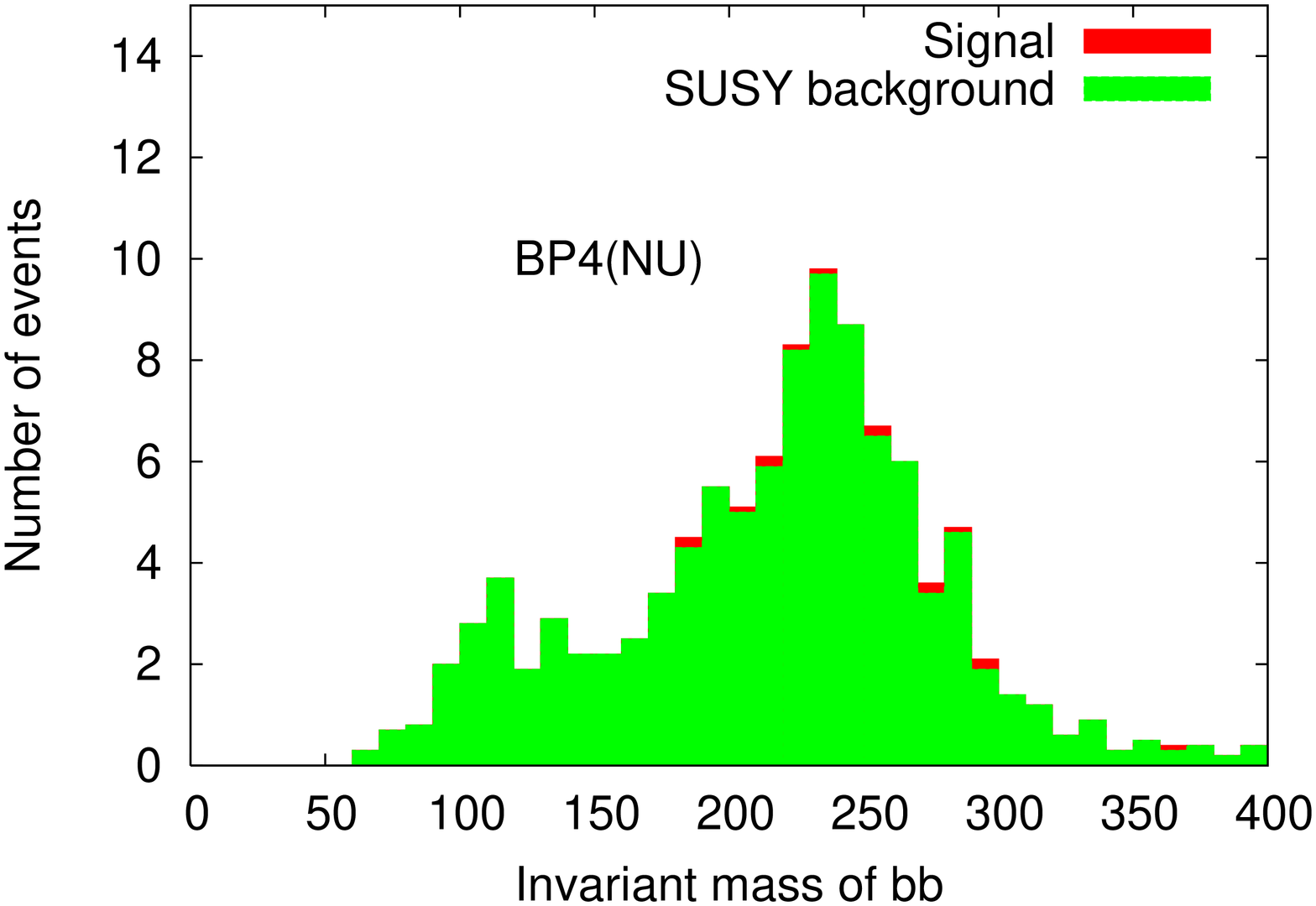,width=5.0cm,height=7.0cm,angle=-90.0}}
\caption{Invariant mass (in GeV) distribution for universal(left) and non-non-universal(right) for heavy neutral Higgs bosons and with $m_{H^{\pm}}=250$ GeV.} 
\end{center}
\label{fig11}
\vspace*{-1.0cm}
\end{figure}

\begin{table}[hbtp]
\begin{center}
\begin{tabular}{||c|c|c|c|c|c|c|c|c||} \hline\hline
&\multicolumn{4}{|c|}{ Universal}&\multicolumn{4}{|c|}{Non-universal}\\\hline
Process & BP1 &BP2&BP3&BP4 & BP1 &BP2&BP3&BP4  \\
\hline
\hline
Signal&30(10)&32(11)&3(2)&1(0)&6(5)&36(21)&6(4)&23(14)\\
\hline\hline
SUSY&21(19)&21(16)&11(6)&50(28)&21(11)&18(10)&17(9)&32(16)\\
Background&&&&&&&&\\
\hline\hline
$t\tbar$&\multicolumn{8}{|c|}{51}\\
Background&\multicolumn{8}{|c|}{(22)}\\
\hline\hline
\end{tabular}
\label{tab1}
\caption{Expected Number of events with an integrated luminosity of 10 fb$^{-1}$ for the case of heavy neutral Higgs bosons with $m_{H^{\pm}}= 180$ GeV. Numbers within the parenthesis are with an invariant mass cut of $140 \, \mathrm{GeV} \, \le m_{b_{j_1},b_{j_2}}\le 190$ GeV.}	
\end{center}
\end{table}
\begin{table}[hbtp]
\begin{center}
\begin{tabular}{||c|c|c|c|c|c|c|c|c||} \hline\hline
&\multicolumn{4}{|c|}{ Universal}&\multicolumn{4}{|c|}{Non-universal}\\\hline
Process & BP1 &BP2&BP3&BP4 & BP1 &BP2&BP3&BP4  \\
\hline
\hline
Signal&27(10)&2(2)&0(0)&1(1)&24(19)&53(36)&17(11)&2(1)\\
\hline\hline
SUSY&79(46)&99(55)&19(13)&100(61)&51(30)&79(48)&35(19)&104(60)\\
Background&&&&&&&&\\
\hline\hline
$t\tbar$&\multicolumn{8}{|c|}{91}\\
Background&\multicolumn{8}{|c|}{(5)}\\
\hline\hline
\end{tabular}
\label{tab1}
\caption{Expected Number of events with an integrated luminosity of 10 fb$^{-1}$ for the case of heavy neutral Higgs bosons with $m_{H^{\pm}}= 250$ GeV. Numbers within the parenthesis are with an invariant mass cut of $200 \, \mathrm{GeV} \, \le m_{b_{j_1},b_{j_2}}\le 300$ GeV.}	
\end{center}
\end{table}
\newpage

\section{Charged Higgs bosons under SUSY cascades}
In this section we discuss in detail the signal for the charged Higgs bosons under SUSY cascades. In the line of \cite{Bandyopadhyay:2008fp} and as motivated earlier in section 4 of the present work, we consider two specific cases with $m_{H^{\pm}}=180$ GeV and 250 GeV. To recapitulate, we note that for $m_{H^{\pm}}=180$ GeV, $H^{\pm}\to{\tau^{\pm} \nu_{\tau}}$ is the dominant decay mode while for $m_{H^{\pm}}=250$ GeV, $H^{\pm}\to{t \bar{b} \, (\bar{t}b)}$ becomes dominant. The corresponding branching fractions for $m_{H^{\pm}}=180, 250$ GeV are presented in Table 12.

\begin{table}[hbtp]
\begin{center}
\begin{tabular}{||c|c|c||} \hline\hline
$m_{H^{\pm}}$ in GeV &Br($H^{\pm}\to{\tau^{\pm} \nu_{\tau}})$ & Br($H^{\pm}\to{t \bar{b}}$) \\
\hline
\hline
180 & 0.87 & 0.13 \\
\hline\hline
250 & 0.24 & 0.74\\
\hline\hline
\end{tabular}
\label{tab1}
\caption{Dominant decay branching fractions of $H^{\pm}$ for $m_{H^{\pm}}$=180 and 250 GeV.}
\end{center}
\end{table}
Thus, we have two different kinds of signals for $H^{\pm}$ decaying into above two modes. In the following two subsections we discuss these two cases in some detail by taking into account the possible backgrounds.

\subsection{A Heavy Charged Higgs boson ($m_{H^{\pm}}\gg m_t +  m_b$)}
For a heavy charged Higgs boson (we take $m_{H^{\pm}}=250$ GeV), $H^{\pm}\to{t \bar{b}(\bar{t}b)}$ is the dominant decay mode. The resulting final state is thus,

$$H^{\pm}\to{t \bar{b}(\bar{t}b)}\to b W^{\pm}\bar{b} \to b\bar{b}q{\bar{q}}\quad .$$
However, the cascade decays of SUSY particles lead to increasing number of jets in the final state accompanied by large missing $p_T$ coming from the LSP.

In Figure 1, we already illustrated the jet multiplicity distributions of such a signal and that of its SM background which comes dominantly from $t\bar{t}$. Thus all our basic cuts of Table 7 can be directly used to our benefit.

We now attempt to reconstruct $H^{\pm}$ from the invariant mass of the set of particles, `$bbqq$' as 
indicated above. This is a multi-step process \cite{Ball:2007zza,:2008uu}. First, we reconstruct the $W^{\pm}$ from the invariant mass 
distribution of two `candidate' non-$b$-jets each of which has transverse momentum in the range 
$20 \, \mathrm{GeV}\leq p^{j_1,j_2}_{T} \leq 90 $ GeV. 
As before, this range of $p_T$-jets is  very characteristic of jets coming out of the decay of an 
on-shell $W^{\pm}$. Thus, it preselects the jets from $W^{\pm}$ rather efficiently (Figure 10).
\begin{figure}[hbtp]
\begin{center}
{\epsfig{file=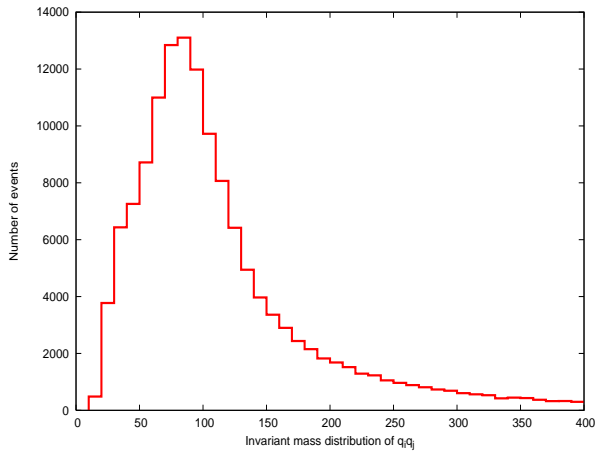,width=6.0cm,height=8.0cm,angle=-90.0}}
\caption{Invariant mass (in GeV) of $q_iq_i$ system, i.e., sytem of two non $b$-jets.} 
\end{center}
\label{fig11}
\end{figure}

Next, these two `candidate' non-$b$-jets  are required to satisfy a more stringent criterion of having
their invariant mass in the range of $\pm 10$ GeV about the $W$-mass. Such a pair of non-$b$-jets is then combined
with a suitable $b$-jet from the final state whose invariant mass gives a peak at around the top mass (Figure 11).
\begin{figure}[hbtp]
\begin{center}
{\epsfig{file=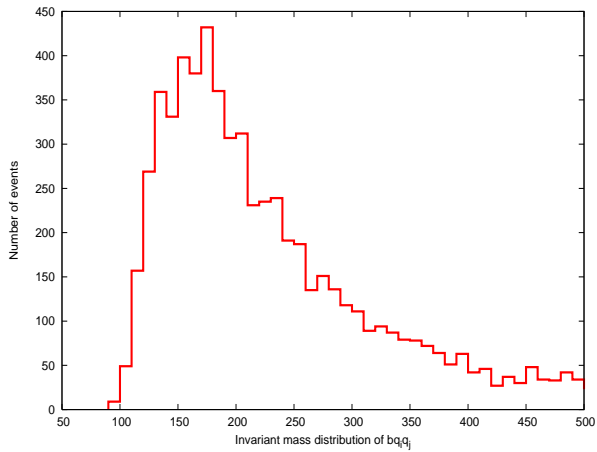,width=6.0cm,height=8.0cm,angle=-90.0}}
\caption{Invariant mass of (in GeV) $bq_iq_i$ system, i.e., system of two non $b$-jets with a $b$-jet.} 
\end{center}
\label{fig11}
\end{figure}

In the third step, with a similar approach, we subject a `candidate' set of $qqb$-system to have an invariant
mass in the range of 30 GeV on both side of the top quark mass.  

Note that in each successive step we are trying to preselect the candidates from the previous sample 
with more restriction before using them for the next phase of reconstruction. This effectively reduces the combinatoric effects.

As mentioned in the beginning of this subsection, again, the main background comes from the $t\bar{t}$ production. While the basic cuts introduced already take care of the SM $t\bar{t}$ background very efficiently, the SUSY background, which is combinatoric in nature, is dealt with additional cuts on jet-$p_T$ and invariant masses of different jet combinations.

In Figure 12 we illustrate the invariant mass of the $bbq_iq_j$ system. We follow the same convention for the signal and the background as we did for the neutral Higgs bosons (Figure 8 and Figure 9). It is apparent that while the signal distributions are more or less sharply peaked in the cases for the neutral Higgs bosons, we are left with extended tails. The reason can be traced back to the multi-stage reconstruction technique for which, each stage carries with it a significant spread in the concerned variable.   

Figure 12 indicates that out of several different benchmark points in both universal and non-universal scenarios, only in few cases the appearance of charged Higgs can be traced at a moderate level of significance with 10 fb$^{-1}$ of LHC data. These are the scenarios like BP1 (non-universal) and BP3 (non-universal). This is more evident from Table 13 for which we followed the same convention as in Table 10 and Table 11 for neutral Higgses. The Figure 12 also reveals a generic trend of backgrounds being more sever. This, in turn, can be traced back to the fact that the signal themselves are rather small in these cases. The reason behind this is that the mass of the charged Higgs boson chosen by us is too heavy (250 GeV, in this case) to be produced in a `little cascade' which, if open, would contribute more when compared to the contributions from the `big cascade'. In addition, the `big cascade' can be simultaneously closed for some cases. These features can   easily be read out from Table 4. Also note that in Table 13, the event counts for  $t\bar{t}$ are different from those in Table 10 and Table 11 (for the neutral Higgses). This is because the dedicated cuts in the case of charged Higgs boson are different from their counterparts in the neutral Higgs sector.

It should also be noted that for $m_{H^{\pm}}=250$ GeV, the branching fraction
for the charged Higgs boson into $H^{\pm}\to{\tau^{\pm} \nu_{\tau}}$ is about 
24\%, which is, though not so small by itself, a factor of 3 or more down compared 
to $H^\pm \to t\bar{b} \, (\bar{t} b)$. Also, we already discussed in the above 
paragraphs that in the bulk of the situations presented here 
($m_{H^\pm}=250$ GeV), the production cross-sections for the charged Higgs 
bosons under SUSY cascades are rather small. Hence, apriori, even without
 going into a detailed analysis, it is apparent that observing  the charged
 Higgs bosons in the $\tau$ mode stands less of a chance on the sole basis of 
effective yield (before any selection cuts are applied). On the other hand, if 
under certain circumstances, there is an appreciable rate for the charged Higgs
 boson  under SUSY cascades, then a dedicated study in the $\tau$ mode should be 
carried out. For this, one can, in principle, follow an analysis similar to what is done for a light charged Higgs boson as elaborated in the next section. 


\begin{table}[hbtp]
\begin{center}
\begin{tabular}{||c|c|c|c|c|c|c|c|c||} \hline\hline
&\multicolumn{4}{|c|}{ Universal}&\multicolumn{4}{|c|}{Non-universal}\\\hline
Process & BP1 &BP2&BP3&BP4 & BP1 &BP2&BP3&BP4  \\
\hline
\hline
Signal&140(68)&2(1)&0(0)&8(5)&128(68)&0(0)&207(122)&28(16)\\
\hline\hline
SUSY&431&517&110&700&299&469&168&605\\
Background&(231)&(269)&(58)&(401)&(140)&(220)&(85)&(328)\\
\hline\hline
$t\tbar$&\multicolumn{8}{|c|}{275}\\
Background&\multicolumn{8}{|c|}{(132)}\\
\hline\hline
\end{tabular}
\label{tab1}
\caption{Expected number of events of charged Higgs boson for $m_{H^{\pm}}=250$ GeV with  an integrated luminosity of 10 fb$^{-1}$. Numbers within the parenthesis are the corresponding  event counts with a further cut of  $200 \, \mathrm{GeV} \, \le m_{bbq_iq_j} \le 350$ GeV.}	
\end{center}
\end{table}

\begin{figure}[hbtp]
\vspace*{-2.2cm}
\begin{center}
\vspace*{-1.2cm}
{\epsfig{file=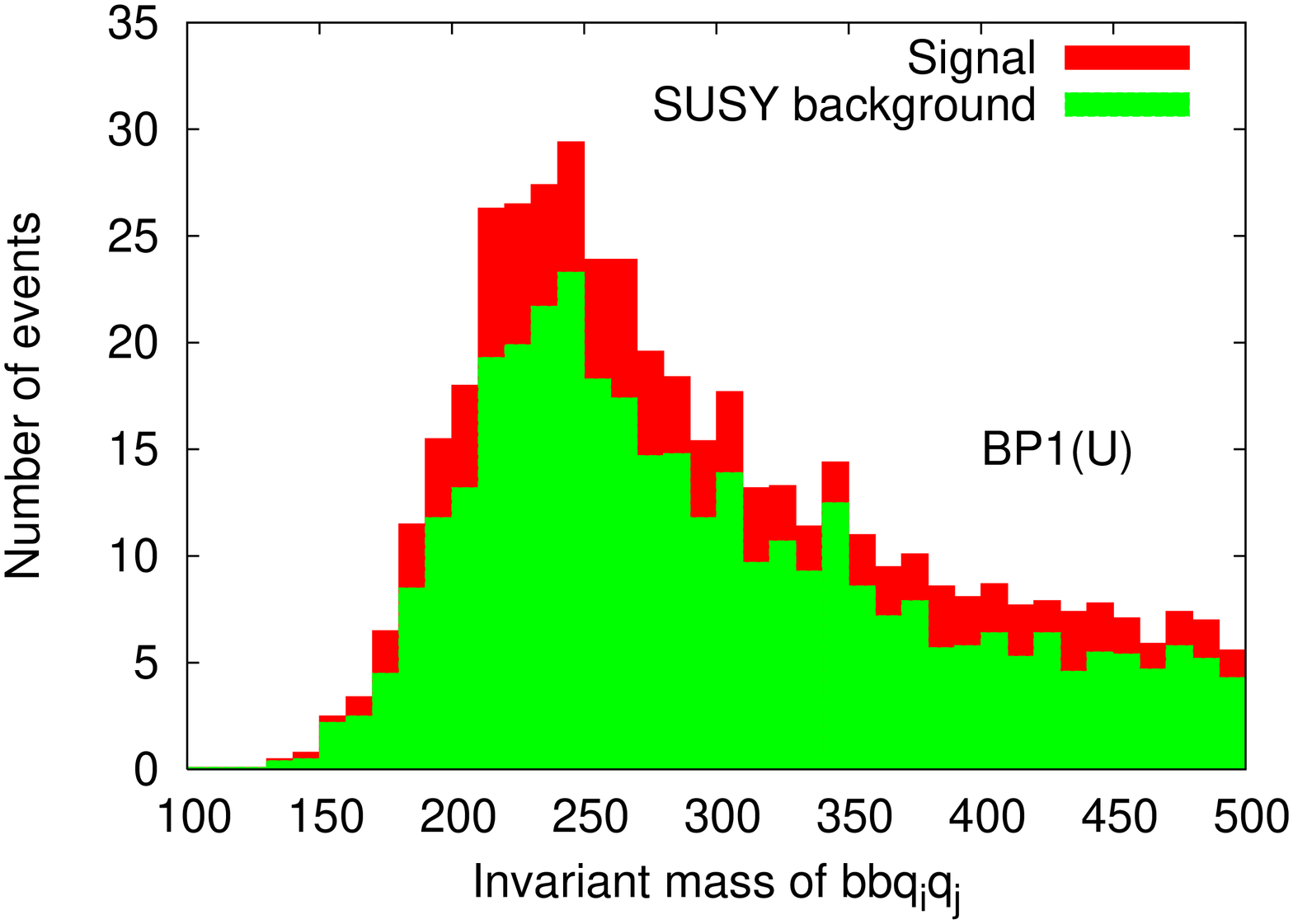,width=5.0 cm,height=7.0cm,angle=-90.0}}
\hskip -12pt 
{\epsfig{file=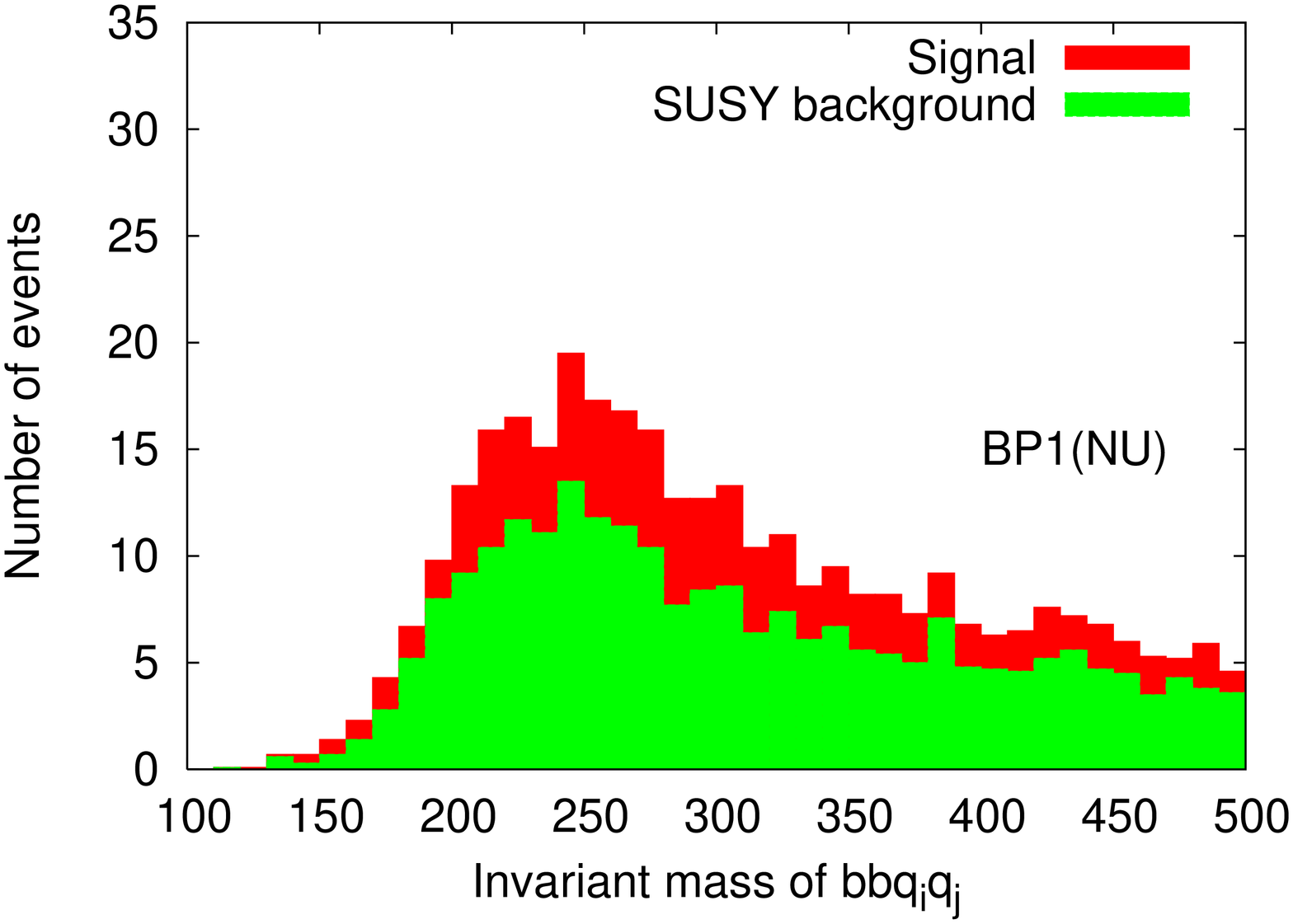,width=5.0cm,height=7.0cm,angle=-90.0}}
{\epsfig{file=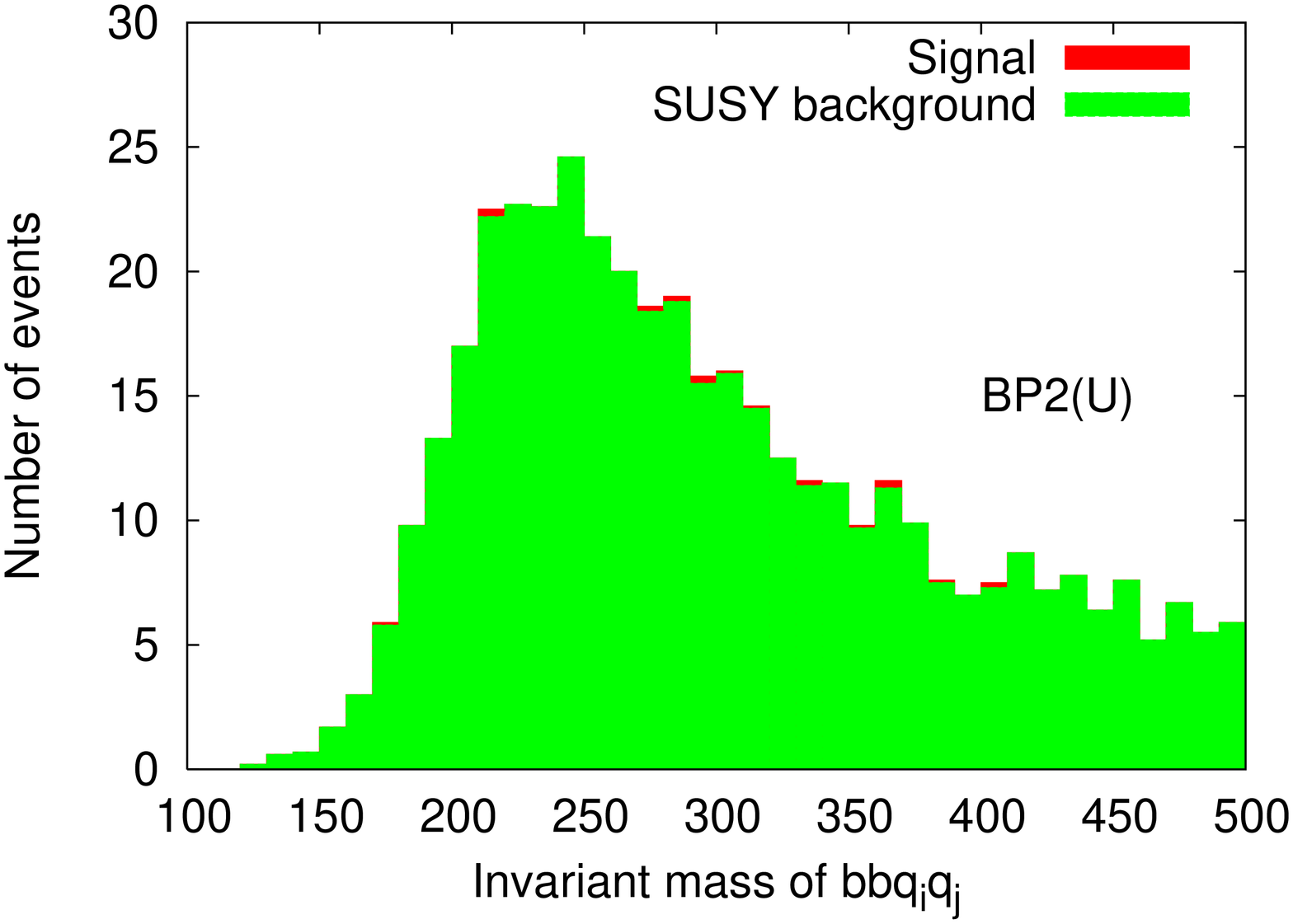,width=5.0 cm,height=7.0cm,angle=-90.0}}
\hskip -12pt 
{\epsfig{file=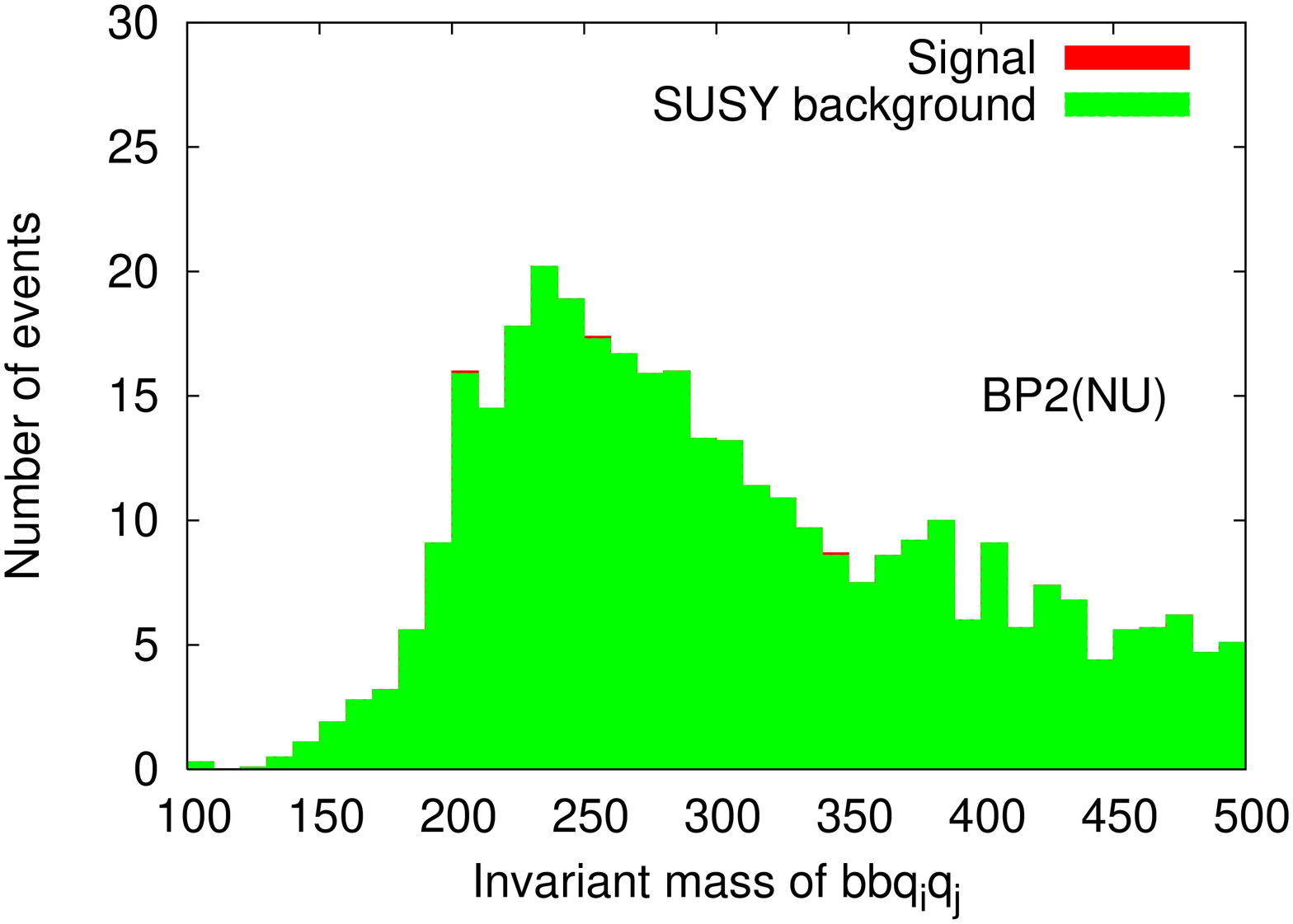,width=5.0cm,height=7.0cm,angle=-90.0}}
{\epsfig{file=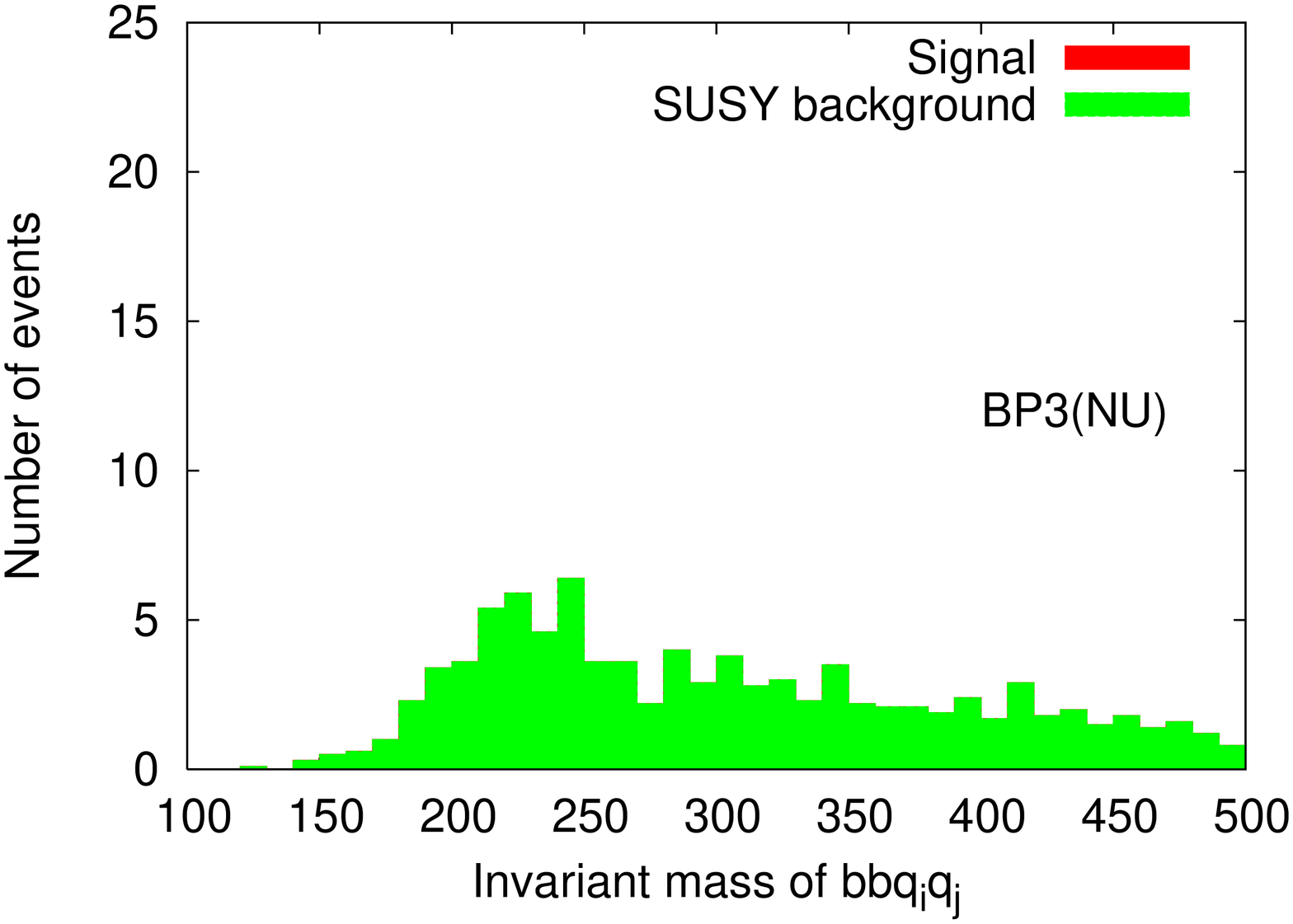,width=5.0 cm,height=7.0cm,angle=-90.0}}
\hskip -12pt 
{\epsfig{file=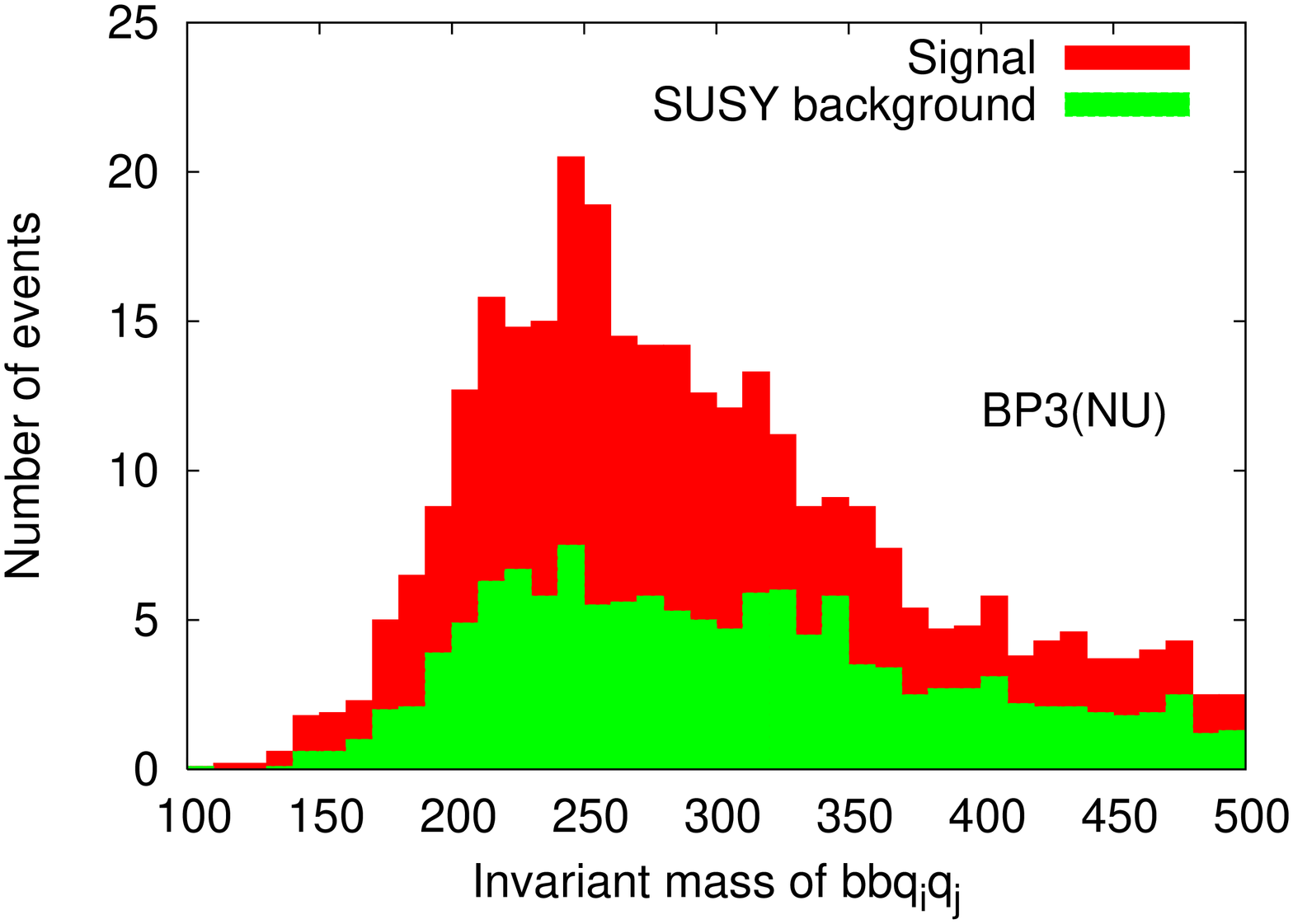,width=5.0cm,height=7.0cm,angle=-90.0}}
{\epsfig{file=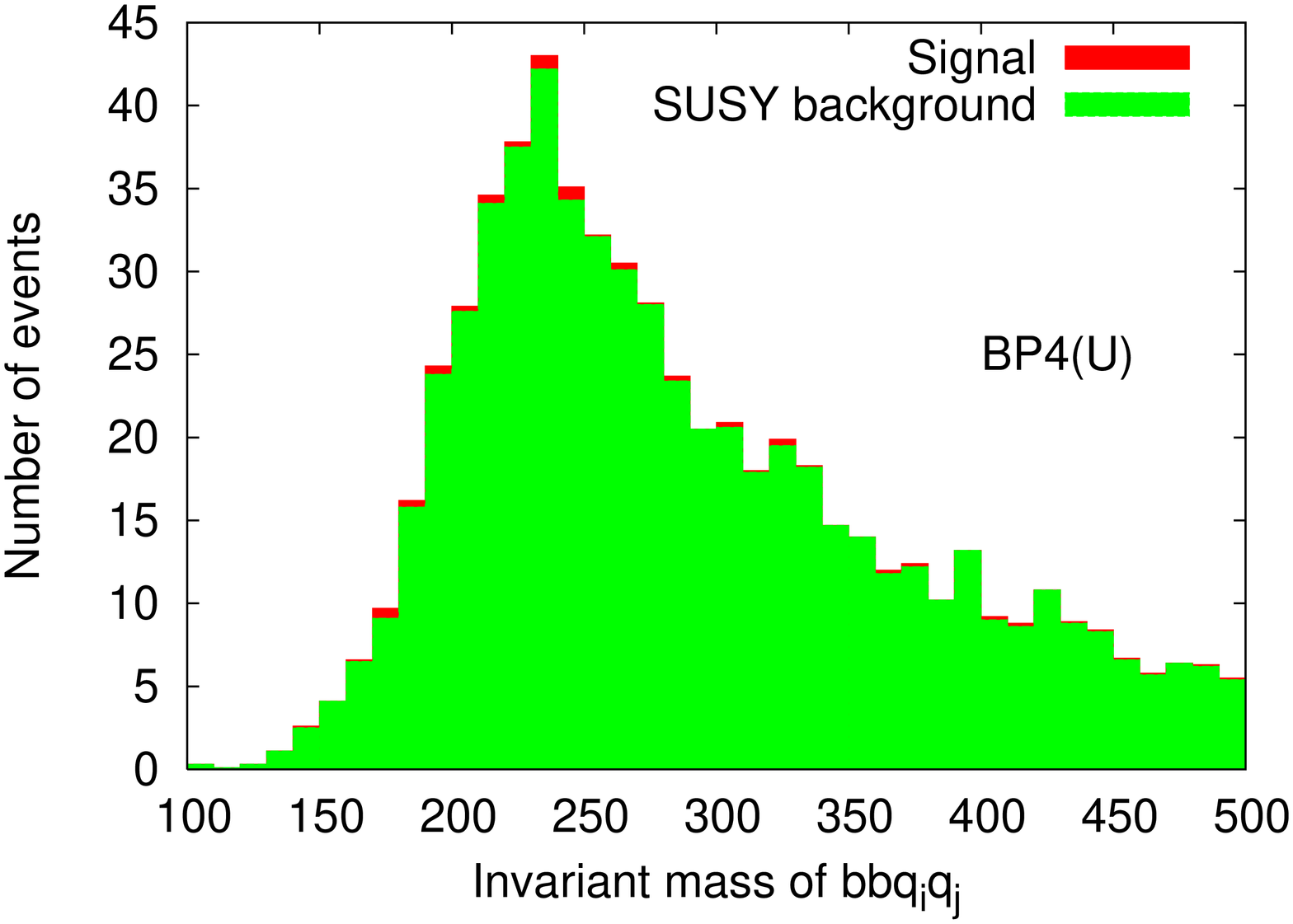,width=5.0 cm,height=7.0cm,angle=-90.0}}
\hskip -12pt 
{\epsfig{file=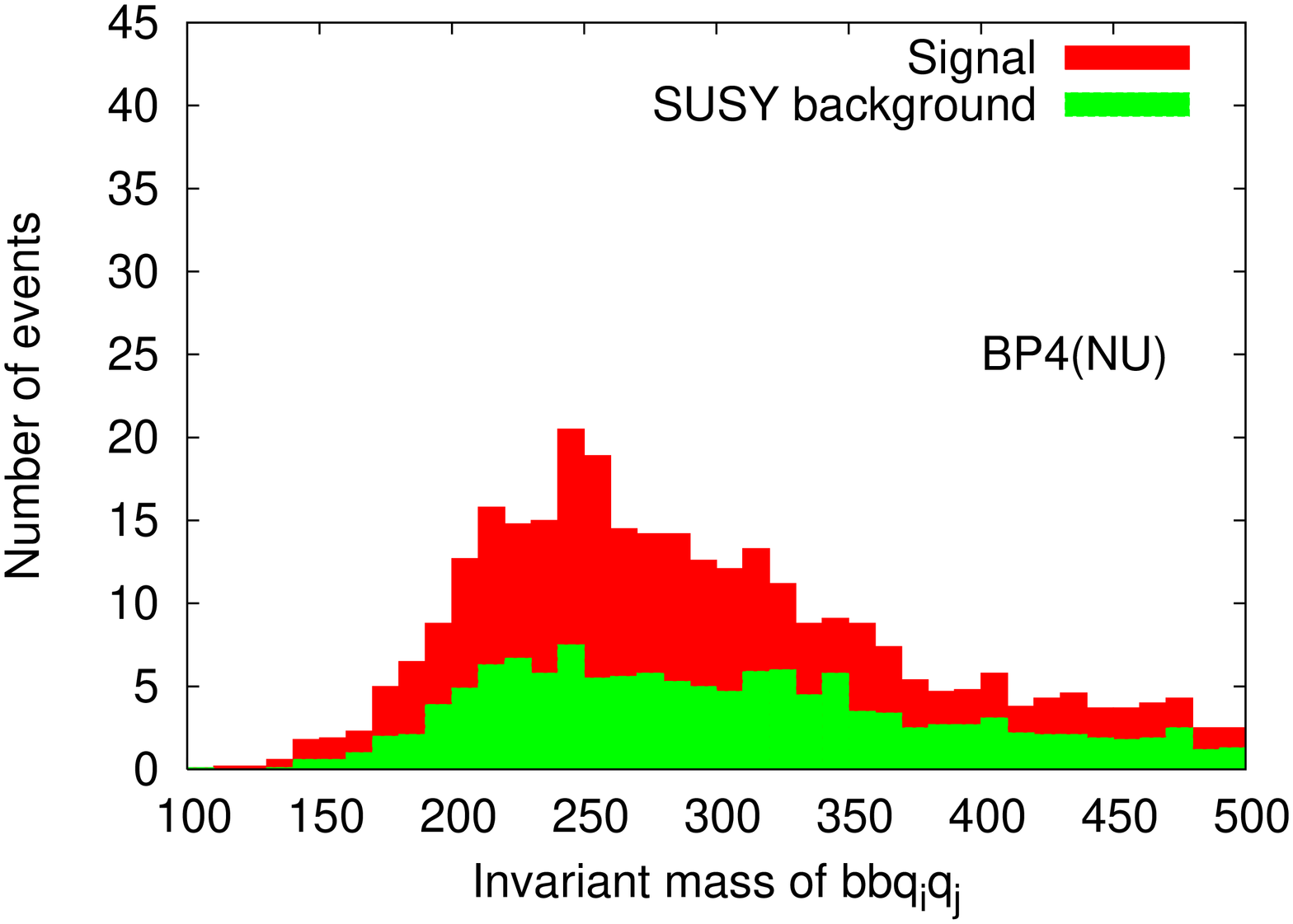,width=5.0cm,height=7.0cm,angle=-90.0}}
\caption{$bbq_iq_j$ invariant mass (in GeV) distribution for universal(left) and non-non-non-universal(right) for the charged Higgs bosons with $m_{H^{\pm}}=250$ GeV.} 
\end{center}
\label{fig11}
\end{figure}
\subsection{A Light Charged Higgs boson}
Production of a lighter Higgs boson under SUSY cascade is doubly blessed. 
First, for a given set of masses for charginos and neutralinos there is a chance that the `little cascades' may open up in addition to a generic presence of `big cascades'. Secondly, being lighter, there is an increased chance that such a Higgs boson can appear in decays of relatively lighter gauginos, which in turn would have larger rates. The argument holds for charged Higgs boson also. Table 5 already vindicates the observation by showing an increased cross-section for the charged Higgs boson for $m_{H^{\pm}}=180$ GeV as compared to the corresponding ones for 
$m_{H^{\pm}}=250$ GeV as presented in Table 6.

As far as the branching fractions are concerned a light charged Higgs boson with $m_{H^{\pm}}=180$ GeV predominantly decays to  $H^{+}\to \tau^{\pm}\nu_{\tau}$ with a branching fraction of $\sim 87\%$, while its probability of decaying into $t\bar{b} \, (\bar{t}b)$  is only 13\%. Hence, for such a light (or even lighter) charged Higgs boson the $\tau$ channel is expected to be the only viable mode for its discovery. In our present context, this would mean a final state comprising of $n$-jets plus missing energy and would involve $\tau$-leptons. As is well known, $\tau$-leptons are not at all stable and their signatures at colliders are subjects of intense and dedicated studies for quite some time now with substantial implication for the overall search strategies for the final states that involve them.

In the present study, we would be using the one-prong (one charged track) 
hadronic decays of the $\tau$-leptons which have a collective branching fraction of about 50\% of which almost 90\% is comprised of final states with $\pi^{\pm}, \rho$ and $a_1$ mesons. Thus, the signal for charged Higgs boson that we zero in on still remains to be multi-jets plus missing energy with a $\tau$-jet reconstructed with one charged track.

To establish a jet as a $\tau$-jet we take the following approach. We first check, for each jet coming out of PYCELL within $|\eta|\le 2.5$, if there is a partonic $\tau$ within a cone of $\Delta R\le 0.4 $ about the jet-axis. If there is one, then we further ensure that there is a single charged track within a cone of $\Delta R\le 0.1 $ of the same jet axis. This marks a narrow jet character of a $\tau$-jet. Of
course there is an efficiency associated to such kind of a geometric requirement which is a function of $p_T$ of the concerned jet and has been demonstrated  in the literature \cite{Ball:2007zza,Bagliesi:2007qx}. We closely reproduce the values of the efficiencies as indicated in references the $p_T$ range we adopted for the concerned jets. In particular, for $p^{jet}_T\ge 100$ GeV, the efficiency turns out to be $\sim 80\%$ which we used in our analysis. This requirement also ensure that there is no significant charge-activity in a cone of  $\Delta R\le 0.4 $ about the jet axis, i.e., we get an isolated $\tau$-jet with a probability which reflects the corresponding efficiency mentioned above.

The main SM background to $H^{\pm}\to \tau^{\pm}\nu_{\tau}$ is from $W^{\pm}$ production followed by $W^{\pm}\to \tau^{\pm}\nu_{\tau}$. As has been discussed in the literature this background can be efficiently reduced by exploiting the polarizations of the $\tau$-s which are different for the two cases \cite{Datta:2003iz, Roy:1991sf}. However, there are important backgrounds coming from 
$Z$, $\tilde{\tau}$ and other Higgs bosons which cannot be tackled by using
this property.\footnote{The reason behind this is that the
polarization property is directly applicable  only for particle-configurations
with definite polarization states of which $H^\pm \to \tau^\pm \nu$ and 
$W^\pm \to \tau^\pm \nu$ are good examples. For $\tilde{\tau}$, 
the polarization
of its daughter $\tau$-s depends crucially on the chiral-admixtures of the 
$\tilde{\tau}$ itslef which, in turn, is very much model-dependent. However, the
technique would be applicable unambiguously 
if one uses an event generator wherein the squared matrix-elements for different process are incorporated in the helicity basis, viz., HERWIG, SMadgraph etc., there the polarizations of particles are kept
track of down the cascades.} 
However, we have not employed this criteria in our present analysis. Instead we required a large enough $p_T$ ($\ge 100$) GeV for the $\tau$-jet which can efficiently reduce the background contamination. On the other hand, backgrounds may also arise from SUSY cascades via productions of $W,\, Z$ and other neutral Higgs bosons. Also $\tau$-leptons may appear in the decays of $\tilde{\tau}$ in the SUSY cascade. Now, $\tau$ coming out of all these 
particles can be efficiently eliminated by using a somewhat severe lower $p_T$ cut on the $\tau$-jet with $p^{\tau-jet}_T\ge 100$ GeV.  This works because $\tau$-leptons originating from a charged Higgs boson have a harder $p_T$ spectrum compared to those coming from the 
particles mentioned above. We also find that the basic set of cuts defined in Table 7 further improves the signal-significance in a cascade environment involving SUSY particles.

Among the $\tau$-s from the supersymmetric decays, mainly those from $\tilde{\tau}_1$ could become a significant background. As discussed in section 3,
sleptons as light as 400 GeV would crucially govern the nature of the 
SUSY cascades along with the charginos and neutralions. In the process, the $\tau$ which will be produced  down the cascasde could be hard 
 enough or miserably soft depending upon the mass splitting between the $\tilde{\tau}_{1}$ and the LSP. It is obvious that the hardness of the corresponding $\tau$ crucially depends on the available phase space. Typically, $\tilde{\tau}_{1}$ undergoes the decay  $\tilde{\tau}_{1} \to \tau \chi^{0}_1$. Now, in the  non-universal scenario we considered with $M_1=100$ GeV (thus, $m_{\ntrl1}\simeq 100$ GeV), this mass splitting is expected to be large and hence, the resulting $\tau$-s would be rather hard. In such a case $\tau$-s can survive the $p^{\tau-jet}_T$ cut imposed for their removal and start contributing to the background. However, the production rates for charged Higgs bosons receive an even larger boost because of the same reason. Thus, the  signal to background ratio generally increases in going from universal to the non-universal case.  On the other hand, for the universal scenario where  $m_{\chi^{0}_1}\sim M_2/2$,  the corresponding $\tau$ can be of lower $p_T$ depending on the phase space available and may fail to contribute as a dominant background.

In Table 14 we represent the event rates for the signal and the backgrounds in 
both universal and non-universal scenarios for the benchmark points defined 
earlier  (see Table 2) using the kinematic cuts as described earlier. Except for one 
benchmark point in the universal scenario (BP1(U)), all other 
cease to have a signal significances larger than 5. This is expected from the 
production rates of the charged Higgs boson as given in Table 5.

The absolute event rates for the signal are in general 
bigger than the corresponding cases for the neutral Higgs bosons. This is only a reflection of the fact that we did not apply any invariant -mass cut, which is anyway not feasible in this case since one of the decay products (the neutrino) is invisible.


\begin{table}[hbtp]
\begin{center}
\begin{tabular}{||c|c|c|c|c|c|c|c|c||} \hline\hline
&\multicolumn{4}{|c|}{ Universal}&\multicolumn{4}{|c|}{Non-universal}\\\hline
Process & BP1 &BP2&BP3&BP4 & BP1 &BP2&BP3&BP4  \\
\hline
\hline
Signal&125&1&23&7&137&590&139&147\\
\hline\hline
SUSY&338&154&157&155&267&103&268&332\\
Background&&&&&&&&\\
\hline\hline
$W^{\pm}W^{\mp}$&\multicolumn{8}{|c|}{}\\
$W^{\pm}Z$&\multicolumn{8}{|c|}{35}\\
$ZZ$&\multicolumn{8}{|c|}{}\\
Background&\multicolumn{8}{|c|}{}\\
\hline\hline
\end{tabular}
\label{tab1}
\caption{Final number of events of charged Higgs for $m_{H^{\pm}}=180$ GeV.}	
\end{center}
\end{table}

On the other hand, the decay mode $H^{\pm}\to t\bar{b} \, (\bar{t}b)$ is not at all promising as for as efficient reconstruction of $H^{\pm}$ is concerned. There are couple of reasons for that. First, the corresponding branching fraction is only 13\% as indicated in Table 12. Secondly, the technique of successive reconstruction fails to prove useful because of the small-mass splitting between the the charged Higgs and the top quark. Hence, $H^{\pm} \to \tau^{\pm} \nu_{\tau}$ mode is the only viable mode to look for such a light charged Higgs boson. 

\section{Squarks heavier than the gluino}
In this section we study the case of Higgs boson production under SUSY cascades in a
scenario with the mass-ordering of squarks and gluino reversed, i.e.,
scenario where the squarks are heavier than
the gluino. In this particular case, squarks not only decay to quarks and
charginos/neutralinos but also decay to gluino and corresponding quarks
with Br($\tilde{q}\to\tilde{g}) \sim 40\%$. In turn, the only possible decays of gluino decays to 
charginos and neutralinos in three-body modes, which then cascade
decay to Higgs bosons. For the present analysis we have taken $m_{\tilde{q}}=
900$ GeV and $m_{\tilde{g}}= 800$ GeV, which is just
opposite to what we have taken earlier.  

The situation turns out to be a bit complicated compared to the previous case, where both squarks and gluino
undergo two-body decay. $\tilde{g}\to q\tilde{q}$ followed by $\tilde{q}\to q\tilde{\chi}^{0}_{i}/q^{\prime}\tilde{\chi}^{\pm}_{i}$
and the charginos and neutralinos cascade down to Higgs bosons. In the case for  $\tilde{q} > \tilde{g}$, there would be some complicated
interplay od the two-body and the three-body decays of strongly
interacting particles  which lead to
the production of charginos and neutralinos. Below we have listed the  different types of contributions to the 
electro-weak `ino' production charginos and neutralions.

\begin{itemize}
\item Contributions from $\tilde{q}\tilde{q}$ involving two-body decays of quarks:
$$=\sigma(\tilde{q}\tilde{q})\times[Br(\tilde{q}\to
q\tilde{\chi}^{0}_{i}/q^{\prime}\tilde{\chi}^{\pm}_{i})]^2$$

\item Contributions from $\tilde{q}\tilde{q}$ via three-body decays of gluinos:
$$=\sigma(\tilde{q}\tilde{q})\times[Br(\tilde{q}\to q\tilde{g})\times
Br(\tilde{g} \to \tilde{\chi}^{0}_{i}/\tilde{\chi}^{\pm}_{i})]^2$$

\item Contributions from $\tilde{q}\tilde{q}$ involving direct two-body decays of squark to EW `ino's and three-body decays of gluinos to EW `inos':
$$=2\times \sigma(\tilde{q}\tilde{q})\times[Br(\tilde{q}\to \tilde{g})\times
Br(\tilde{g} \to \mathrm{gauginos}) \times Br(\tilde{q}\to
\tilde{\chi}^{0}_{i}/\tilde{\chi}^{\pm}_{i}) ]$$

\item Contributions from $\tilde{q}\tilde{g}$ via direct two-body decays of squarks to EW `ino's and three-body decyas of gluinos:
$$=\sigma(\tilde{q}\tilde{q})\times[Br(\tilde{q}\to \tilde{\chi}^{0}_{i}/\tilde{\chi}^{\pm}_{i})\times
Br(\tilde{g} \to \tilde{\chi}^{0}_{i}/\tilde{\chi}^{\pm}_{i})]$$

\item Contribution from $\tilde{q}\tilde{g}$ involving squark decays to glunios and three-body glinos decay to EW `ino's:
$$=\sigma(\tilde{g}\tilde{g})\times[Br(\tilde{q}\to\tilde{g})\times
Br(\tilde{g} \to \tilde{\chi}^{0}_{i}/\tilde{\chi}^{\pm}_{i})^2]$$

\item Contribution from $\tilde{g}\tilde{g}$ via three-body decays of gluinos:
$$=\sigma(\tilde{g}\tilde{g})\times[Br(\tilde{g} \to \tilde{\chi}^{0}_{i}/\tilde{\chi}^{\pm}_{i})^2]$$

\end{itemize} 

As the squarks are heavier ($\sim 900$ GeV) than the earlier case ($\sim 800$ GeV, as taken in section 3) the rates for strong production
 processes involving squarks drop down. On the other hand, as the gluino is relatively
light, it is produced more copiously. Overall, the
 total strong production drops down to 2 pb which
 was 3 pb in the previous case. 

Depending on the top of the cascade has a squark or a gluino the strong
production gets affected this later affects the gaugino
production via the possible decay schemes as shown above. 
 The extract of gauginos then decaying down
to Higgs bosons depends on the corresponding benchmark scenarios. Below
 (Table 15 \& 16) we list the effective production rates of different Higgs bosons as before. 


\begin{table}[hbtp]
\begin{center}
\begin{tabular}{||c||c|c|c|c||c|c|c|c||} \hline\hline
&\multicolumn{4}{|c||}{ Universal}&\multicolumn{4}{|c||}{Non-universal}\\\hline
Benchmark &\multicolumn{4}{|c||}{Effective
cross-section}&\multicolumn{4}{|c||}{Effective
cross-section}\\
Points &\multicolumn{4}{|c||}{(in fb)}
&\multicolumn{4}{|c||}{(in fb)}\\
\hline\hline
&$\sigma_{h}$&$\sigma_{H^+}$&$\sigma_A$&$\sigma_H$&$\sigma_{h}$&$\sigma_{H^\pm}$&$\sigma_A$&\multicolumn{1}{|c||}{$\sigma_{H}$}\\
\hline
BP1&214.2&90.9&48.4&69.5&64.7&60.5&45.3&\multicolumn{1}{|c||}{40.0}\\
\hline
BP2&646.1&0.8&4.6&294.5&345.6&1103.9&133.4&\multicolumn{1}{|c||}{466.9}\\
\hline
BP3&123.7&49.5&22.2&32.5&71.2&104.9&33.6&\multicolumn{1}{|c||}{31.5}\\
\hline
BP4&896.63&13.4&4.9&4.2&564.2&650.2&78.2&\multicolumn{1}{|c||}{293.5}\\
\hline
\end{tabular}
\label{tab1}
\caption{Effective production rates for the $h$, $H^{\pm}$, $A$ and $H$ for $m_H^{\pm}=180$ GeV and other parameters are described as in the Table 1.}
\end{center}
\end{table}

\begin{table}[hbtp]
\begin{center}
\begin{tabular}{||c||c|c|c|c||c|c|c|c||} \hline\hline
&\multicolumn{4}{|c||}{ Universal}&\multicolumn{4}{|c||}{Non-universal}\\\hline
Benchmark &\multicolumn{4}{|c||}{Effective
cross-section}&\multicolumn{4}{|c||}{Effective
cross-section}\\
Points &\multicolumn{4}{|c||}{(in fb)}
&\multicolumn{4}{|c||}{(in fb)}\\
\hline\hline
&$\sigma_{h}$&$\sigma_{H^+}$&$\sigma_A$&$\sigma_H$&$\sigma_{h}$&$\sigma_{H^\pm}$&$\sigma_A$&\multicolumn{1}{|c||}{$\sigma_{H}$}\\
\hline
BP1&206.7&76.5&40.2&38.4&66.6&75.7&39.7&\multicolumn{1}{|c||}{37.4}\\\hline
BP2&863.7&0.6&0.4&0.46&630.7&0.7&6.5&\multicolumn{1}{|c||}{246.6}\\\hline
BP3&97.9&0.04&0.0&0.04&69.2&76.5&20.1&\multicolumn{1}{|c||}{26.1}\\\hline
BP4&889.2&6.1&0.4&2.8&795.8&13.4&3.9&\multicolumn{1}{|c||}{4.0}\\\hline
\end{tabular}
\label{tab1}\caption{Effective production rates for the $h$, $H^{\pm}$, $A$ and $H$ for $m_H^{\pm}=250$ GeV and other parameters are described as in the Table 1.}

\end{center}
\end{table}

Because of the above mentioned reduction of strong production cross-section 
and the complicated interplay of the decay branching fractions down the subsequent SUSY cascades, both signal and the model background 
may get affected significantly. However, depending upon the scenarios, the suppression in the strong production cross-section can be effectively compensated for by enhancement of appropriat branching fractions in the later stages of the cascade.

To see what happens in the 
respective benchmark scenarios for ($m_{\tilde{q}} > m_{\tilde{g}}$) we carry out 
our analysis as before. Tables 17 and 18 give the
number of events for the signal and the model background (as defined earlier) for the light neutral CP-even
Higgs boson for an integrated luminosity of 10 fb$^{-1}$. From the tables it is clear that except for the scenarios BP1 and BP3
(for both universal and non-universal cases),  are almost
 similar as in the previous case, i.e., $m_{\tilde{q}} < m_{\tilde{g}}$.
  This is also consistent with the production rates given in Tables 15 and 16.
Only point to be noted that for BP1 (both in universal and non-universal cases), which is a higgsino type region, the model background gets an enhancement
 as the gluino decay to the lighter gauginos increases, which mainly contribute to the background. This is not so true for the other cases because of the kinematics involved in the respective scenarios.

The number of events with the heavy neutral Higgs bosons are given in
 Tables 19 and 20. This case is exactly similar to the one for lighter neutral Higgs boson
case, i.e., rates for only BP1 and BP3 differ from the corresponding ones for $m_{\tilde{q}} < m_{\tilde{g}}$,  which is quite expected from the effective production rates presented in Tables 15 and 16.

Again for the case of the lighter charged Higgs boson ($m_{H^\pm}$=180 GeV)
 the number of events decrease for the scenarios  BP1 and BP3 (both 
universal and non-universal cases)  when compared to the corresponding cases with 
$m_{\tilde{q}} < m_{\tilde{g}}$ pretty similar to that observed for the case
of the neutral Higgs boson. Similarly, for the heavier charged Higgs boson 
($m_{H^\pm}$=250 GeV) the situations get changed for BP1 (universal) and BP3 (non-universal),
 i.e, the number of signal events reduced
 as expected. On top of that, for BP1, the model-background  increases not only because
 of the higgsino-region, but also because of the combinatorial issues involved
in reconstructing several different masses (as explained in section 7.1).

Thus, in a nut-shell, the non-universal scenarios can in general be
 distinguished from a universal one irrespective of the relative hierarchy of 
the squarks and gluino masses, albeit, for $m_{\tilde{q}} > m_{\tilde{g}}$
and for scenarios like BP1 and BP3, a clear discrimination may turn out to
be statistically limitting for the reasons described above.

For BP1, the enhancement of the model background points to
the fact that for the higgsino-like region we are required to have some prior knowledge about the 
hierarchy of squark and gluino masses in order to estimate
the model background correctly which is expected to be crucial in searches involving cascade decays.

The case where the charged Higgs boson is lighter than the top mass
(i.e., $m_{H^{\pm}} < m_{t}$) can be analysed in the same way as it is
 done for $m_{H^\pm}$=180 GeV case. In the latter case, however,
the charged Higgs boson would almost entirely decay into $ \to \tau^{\pm}
\nu_{\tau}$ with Br[$H^{+} \to \tau \nu_{\tau}]\simeq 99\%$.

\begin{table}[hbtp]
\begin{center}
\begin{tabular}{||c||c|c|c|c||c|c|c|c||} \hline\hline
&\multicolumn{4}{|c||}{ Universal}&\multicolumn{4}{|c||}{Non-universal}\\\hline
Process & BP1 &BP2&BP3&BP4 & BP1 &BP2&BP3&BP4  \\
\hline
\hline
Signal&32(19)&86(54)&8(5)&128(80)&7(4)&42(25)&6(4)&84(51)\\
\hline\hline
SUSY&127(74)&35(22)&14(8)&17(10)&131(74)&75(45)&26(15)&58(36)\\
Background&&&&&&&&\\
\hline\hline
$t\tbar$&\multicolumn{8}{|c|}{91}\\
Background&\multicolumn{8}{|c|}{(50)}\\
\hline\hline
\end{tabular}
\label{tab1}
\caption{Number of events at an integrated luminosity of 10 fb$^{-1}$ for the case of lightest neutral Higgs boson with $m_{H^{\pm}}=180$ GeV. Numbers within the parenthesis are with an invariant mass cut  $95 \, \mathrm{GeV} \,\le m_{b_{j_1},b_{j_2}}\le 140$ GeV.}
\end{center}
\end{table}

\begin{table}[hbtp]
\begin{center}
\begin{tabular}{||c||c|c|c|c||c|c|c|c||} \hline\hline
&\multicolumn{4}{|c||}{ Universal}&\multicolumn{4}{|c||}{Non-universal}\\\hline
Process & BP1 &BP2&BP3&BP4 & BP1 &BP2&BP3&BP4  \\
\hline
\hline
Signal&32(20)&98(62)&5(3)&118(75)&8(5)&69(42)&7(5)&106(67)\\
\hline\hline
SUSY&130(72)&11(7)&10(6)&16(10)&140(78)&29(16)&34(20)&24(14)\\
Background&&&&&&&&\\
\hline\hline

$t\tbar$&\multicolumn{8}{|c|}{91}\\
Background&\multicolumn{8}{|c|}{(50)}\\
\hline\hline
\end{tabular}
\label{tab1}
\caption{Number of events at an integrated luminosity of 10 fb$^{-1}$ for the case of lightest neutral Higgs boson with $m_{H^{\pm}}= 250$ GeV. Numbers within the parenthesis are with an invariant mass cut  $95 \, \mathrm{GeV} \,\le m_{b_{j_1},b_{j_2}}\le 140$ GeV.}
\end{center}
\end{table}

\begin{table}[hbtp]
\begin{center}
\begin{tabular}{||c||c|c|c|c||c|c|c|c||} \hline\hline
&\multicolumn{4}{|c||}{ Universal}&\multicolumn{4}{|c||}{Non-universal}\\\hline
Process & BP1 &BP2&BP3&BP4 & BP1 &BP2&BP3&BP4  \\
\hline
\hline
Signal&8(4)&19(13)&1(0)&0(0)&5(3)&37(21)&0(0)&3(1)\\
\hline\hline
SUSY&70(37)&29(16)&6(4)&50(27)&64(36)&18(9)&1(1)&3(2)\\
Background&&&&&&&&\\
\hline\hline
$t\tbar$&\multicolumn{8}{|c|}{51}\\
Background&\multicolumn{8}{|c|}{(22)}\\
\hline\hline
\end{tabular}
\label{tab1}
\caption{Number of events at an integrated luminosity of 10 fb$^{-1}$ for the case of heavy neutral Higgs bosons with $m_{H^{\pm}}= 180$ GeV. Numbers within the parenthesis are with an invariant mass cut $140 \, \mathrm{GeV} \, \le m_{b_{j_1},b_{j_2}}\le 190$ GeV.}	
\end{center}
\end{table}
\begin{table}[hbtp]
\begin{center}
\begin{tabular}{||c||c|c|c|c||c|c|c|c||} \hline\hline
&\multicolumn{4}{|c||}{ Universal}&\multicolumn{4}{|c||}{Non-universal}\\\hline
Process & BP1 &BP2&BP3&BP4 & BP1 &BP2&BP3&BP4  \\
\hline
\hline
Signal&16(11)&0(0)&0(0)&0(0)&13(9)&49(36)&7(5)&1(1)\\
\hline\hline
SUSY&232(148)&106(63)&21(12)&105(66)&193(122)&74(44)&4(2)&12(7)\\
Background&&&&&&&&\\
\hline\hline
$t\tbar$&\multicolumn{8}{|c|}{91}\\
Background&\multicolumn{8}{|c|}{(5)}\\
\hline\hline
\end{tabular}
\label{tab1}
\caption{Number of events at an integrated luminosity of 10 fb$^{-1}$ for the case of heavy neutral Higgs bosons with $m_{H^{\pm}}= 250$ GeV. Numbers within the parenthesis are with an invariant mass cut $200 \, \mathrm{GeV} \, \le m_{b_{j_1},b_{j_2}}\le 300$ GeV.}	
\end{center}
\end{table}

\begin{table}[hbtp]
\begin{center}
\begin{tabular}{||c||c|c|c|c||c|c|c|c||} \hline\hline
&\multicolumn{4}{|c||}{ Universal}&\multicolumn{4}{|c||}{Non-universal}\\\hline
Process & BP1 &BP2&BP3&BP4 & BP1 &BP2&BP3&BP4  \\
\hline
\hline
Signal&36&1&9&5&41&610&45&331\\
\hline\hline
SUSY&200&150&70&145&170&92&121&127\\
Background&&&&&&&&\\
\hline\hline
$W^{\pm}W^{\mp}$&\multicolumn{8}{|c|}{}\\
$W^{\pm}Z$&\multicolumn{8}{|c|}{35}\\
$ZZ$&\multicolumn{8}{|c|}{}\\
Background&\multicolumn{8}{|c|}{}\\
\hline\hline
\end{tabular}
\label{tab1}
\caption{Number of events of charged Higgs for $m_{H^{\pm}}=180$ GeV.}	
\end{center}
\end{table}

\begin{table}[hbtp]
\begin{center}
\begin{tabular}{||c||c|c|c|c||c|c|c|c||} \hline\hline
&\multicolumn{4}{|c||}{ Universal}&\multicolumn{4}{|c||}{Non-universal}\\\hline
Process & BP1 &BP2&BP3&BP4 & BP1 &BP2&BP3&BP4  \\
\hline
\hline
Signal&95(51)&1(0)&0(0)&15(11)&131(51)&0(0)&77(43)&27(18)\\
\hline\hline
SUSY&2641&781&98&1177&2893&781&215&1281\\
Background&(963)&(385)&(47)&(610)&(1059)&(349)&(96)&(654)\\
\hline\hline
$t\tbar$&\multicolumn{8}{|c|}{275}\\
Background&\multicolumn{8}{|c|}{(132)}\\
\hline\hline
\end{tabular}
\label{tab1}
\caption{Number of events of charged Higgs boson for $m_{H^{\pm}}=250$ GeV for  an integrated luminosity of 10 fb$^{-1}$. Numbers within the parenthesis are the corresponding  event counts with a farther cut of,  $200 \, \mathrm{GeV} \, \le m_{bbq_iq_j} \le 350$ GeV.}	
\end{center}
\end{table}
\newpage
\section{Variation of $\tan{\beta}$ and the masses of squarks gluino}
In this section we comment on the effect of $\tan{\beta}$ and squark and gluino masses on the distinguishability
of universal and non-universal scenarios. To see the impact of $\tan{\beta}$ on the distinguishability
we check the production rates (as in section  4) for benchmark point 1 (BP1) with  two extreme $\tan{\beta}$
 values 5 and 50. From Table 23 it is clear that the `cross-over' behaviour is  retained for almost all 
$\tan{\beta}$ values from 5 to 50. Of course the absolute production rates get affected since the mass eigenvalues 
of the neutral Higgs bosons, the charginos and the neutralinos change as functions of  $\tan{\beta}$. However, we see 
that the relative behaviour of the cross-section remains the same. This is because the channel $\ntrl{3} \to h \ntrl{1}$
 is open only in the universal scenario and this decay is almost independent of $\tan{\beta}$.

\begin{table}[hbtp]
\begin{center}
\begin{tabular}{||c|c|c|c|c|c|c|c|c|c|c||} \hline\hline
&\multicolumn{4}{|c|}{BP1 Universal}&\multicolumn{4}{|c||}{BP1 Non-universal}\\\hline
$\tan{\beta}$ &\multicolumn{4}{|c|}{Effective
cross-section}&\multicolumn{4}{|c||}{Effective
cross-section}\\
&\multicolumn{4}{|c|}{(in fb)}&\multicolumn{4}{|c||}{(in fb)}
\\\hline\hline

&$\sigma_{h}$&$\sigma_{H^+}$&$\sigma_A$&$\sigma_H$&$\sigma_{h}$&$\sigma_{H^\pm}$&$\sigma_A$&\multicolumn{1}{|c||}{$\sigma_{H}$}\\
\hline
5&810.9&273.4&152.8&160.3&252.0&275.0&142.1&\multicolumn{1}{|c||}{114.9}\\\hline
50&648.60&334.1&176.4&348.2&192.9&323.9&175.6&\multicolumn{1}{|c||}{164.2}\\\hline
\end{tabular}
\label{tab1}
\caption{Effective production rates for the $h$, $H^{\pm}$, $A$ and $H$ for $m_H^{\pm}=180$ GeV for $\tan{\beta}$=5 and 50 . }
\end{center}
\end{table}
To see the dependence on squark and gluino masses, we increase squark, gluino and the slepton masses to about 
1200 GeV, 1500 GeV and 600 GeV (from 800 GeV, 900 GeV and 400 GeV) respectively. We keep the charged Higgss boson mass fixed to 180 GeV while changing the values of  $M_2$ to 800 GeV (from 600 GeV) and $\mu$ to 250 GeV (from 150 GeV). The rest of the parameters are the same as for BP1.  We call this point BP1$^\prime$. In this case the total strong production cross-section drops down to 0.19 pb from the earlier case (3 pb) as expected due to heavier squarks and glino. Here also,
 we see that the absolute value of the production rates change while the relative behaviour of the rates, especially 
the rates of the lightest neutral Higgs boson and that of the charged Higgs boson remain the  same as in BP1 (Table 5). 
As before, $\ntrl{3} \to h \ntrl{1}$  does not open up for the non-universal cases. From these observation,
 we can conclude that the `distinguishability' criteria is rather robust unless  we change $M_2$ in the non-universal case 
to such a value that could open up the $\ntrl{3}\to h \ntrl{1}$ channel.

\begin{table}[hbtp]
\begin{center}
\begin{tabular}{||c|c|c|c|c|c|c|c|c|c||} \hline\hline
\multicolumn{4}{|c|}{BP1$^\prime$ Universal}&\multicolumn{4}{|c||}{BP1$^\prime$ Non-universal}\\\hline
\multicolumn{4}{|c|}{Effective
cross-section}&\multicolumn{4}{|c||}{Effective
cross-section}\\
\multicolumn{4}{|c|}{(in fb)}&\multicolumn{4}{|c||}{(in fb)}
\\\hline\hline

$\sigma_{h}$&$\sigma_{H^+}$&$\sigma_A$&$\sigma_H$&$\sigma_{h}$&$\sigma_{H^\pm}$&$\sigma_A$&\multicolumn{1}{|c||}{$\sigma_{H}$}\\
\hline
61.4&25.3&14.2&18.4&19.6&24.8&13.7&\multicolumn{1}{|c||}{11.5}\\\hline
\end{tabular}
\label{tab1}
\caption{Effective production rates for the $h$, $H^{\pm}$, $A$ and $H$ for $m_H^{\pm}=180$ GeV.}
\end{center}
\end{table}

\section{Summary and Conclusions}

In the present work we study in detail how a possible
non-universal effect in soft SUSY breaking gaugino masses (as observed at
the weak scale) could
potentially affect the rates for the Higgs bosons under SUSY cascades of strongly interacting particles (like squarks and the 
gluino) at the LHC. The basic purpose of the present work is to find if the signature of gaugino-mass non-universality could 
indeed be deciphered at the LHC via simultaneous identification of different Higgs boson excitations under SUSY cascades and thereby studying their relative rates.

We study two sets of phenomenologically interesting mass-values for the SUSY Higgs
bosons. They are fixed by requiring two specific values for the mass of 
the charged Higgs boson, viz., $\mhpm=180,250$ GeV where the first value
is close to the mass of the top quark while the latter one is quite high.
 Thus, in these two cases the dominant decay modes of the charged Higgs boson
turn out to be different; $H^{+} \to \tau \nu_{\tau}$ in the first case while 
$H^{+} \to tb$  for the latter. These have significant implications for
their detectability in a cascade environment.
On the other hand, with the above inputs,
the masses of the 
heavier neutral Higgs bosons get fixed at values close to but somewhat less
than that of masses for the charged Higgs bosons in each case. 
However, as expected, the mass of the lightest Higgs boson is not much altered
for these two input mass values. 

The analyses are carried out at suitably chosen `benchmark' points in the
relevant SUSY parameter space. The points are so chosen that they represent
different corners in the $\mtwo-\mu$ plane where the relative rates for
the charged and the lightest neutral Higgs bosons have contrasting features. 
Thus, by studying the relative rates of the above Higgs bosons once they
are cleanly identified, one may attempt to refer back to the said plane to
locate the region in the $\mtwo-\mu$ plane we are in.

A detailed simulation is done  for both neutral and charged Higgs bosons
 with Pythia as the event-generator. Also, both SM (mainly from $t\bar{t}$ 
production) and SUSY backgrounds are studied by simulating events.

In case of the neutral Higgs bosons the generic signal we looked for is 
$n_{jets}\ge 5$ \emph{(with at least two} $b$-$jets)$ + missing transverse energy where the pair of $b$-jets 
has a reconstructed peak indicative of the mass of the neutral Higgs boson(s).
The conclusion is that the detection of the lightest neutral Higgs boson 
would be rather difficult in general in a non-universal scenario with
10 fb$^{-1}$ of data while for the universal case the signal significance
could be healthy ($5-10\sigma$) enough for our cases and for the same volume
of LHC data. For the non-universal case, the low significance is not at all
unexpected since it only reflects the fact that in the corresponding region of
the parameter space the rate for the lightest neutral Higgs boson is really low.
However, as discussed in section 6.1, this is by itself not a problem as
the rate for the charged Higgs boson is quite large and hence this region can 
be probed instead by looking at the charged Higgs boson. Also, it is noted
that for $\mhpm=250$ GeV, the signal significance for neutral Higgs boson could 
get enhanced in some cases. This is because although the mass of the lightest Higgs boson does not
change any appreciably when going from $\mhpm=180$ GeV to $\mhpm=250$ GeV,
the SUSY backgrounds, as described in section 5, definitely goes down.

As for the heavy neutral Higgses (corresponding to $\mhpm=250$ GeV), it is
clear that none of them can be efficiently probed with an integrated
luminosity of 10 fb$^{-1}$ and, at least, data equivalent to 30 fb$^{-1}$
is required for the purpose. However, one
can choose the invariant mass cut in such a way that the background from
$t\bar{t}$ can be reduced. In any case, this does not help much since the
signal rate is already very poor. Here also, a minimum of 30 fb$^{-1}$ is 
required for the purpose.

 An appropriate signal for the charged Higgs boson depends crucially 
on its mass. A heavy charged Higgs with mass $\mhpm=250$ GeV predominantly
decays in the $t\bar{b}(\bar{t} b)$ mode and thus a suitable signal final
state is again \emph{multijet (with at least two $b$-jets) plus missing
transverse energy}. The generic observation is that only for a few cases
under both universal and non-universal scenarios the charged Higgs boson
can be identified at a moderate significance with 10 fb$^{-1}$ of data.
This is because the mass of the charged Higgs boson (250 GeV) is already
too heavy such that their production via `little cascades' are already closed while in some 
cases, in addition, the `big cascades' could also be closed as well. It is also
observed, that the minimum accumulated luminosity required to probe such a 
heavy charged Higgs is roughly 30 fb$^{-1}$.

On the other hand, in case of a light charged Higgs boson with $\mhpm=180$ GeV,
the final state to look for would need to have a $\tau$-jet. This is again
dictated by the dominant branching fraction for $\hpm \to \tau^{\pm} \nu_\tau$
($\sim 87$\%). $\tau$-s are identified  in their one-prong decays. In most
of the cases we find that an integrated luminosity of 10 fb$^{-1}$ is enough
to have a signal significance of $5\sigma$ or more.

The bottom-line is that either the lightest neutral Higgs boson or a light
charged Higgs boson can be a potential indicator of the underlying scenario
in a complementary way while the heavier neutral Higgs bosons may only aid 
in this respect for an integrated luminosity
of a few tens of  fb$^{-1}$. 

The case of reverse hierarchy of squark and gluino masses (i.e. with $m_{\tilde{q}} > m_{\tilde{g}}$)
 also preserves  clear distinguishability between universal and non-universal scenarios  
except when being in the deep higgsino region or a mixed gaugino-higgsino region with both $M_2$ and $\mu$ at the higher side. In particular, for the latter case the distinguishability can be achieved with larger accumulated luminosity at the LHC. 
 The `distinguishability' is found to be robust in terms of varying $\tan{\beta}$ values as well as for higher values of squark
 and gluino masses.

It is also to be made clear 
that just by confining ourselves to a study of the Higgs bosons
under SUSY cascades a distinction between scenarios with universal and
non-universal scenarios can only be made by identifying at least one
heavy Higgs boson along with the lightest one. This is very much so because,
to start with, this constitutes the whole basis of starting we adopt to
contrast the two scenarios. Thus, with 10 fb$^{-1}$ of data we would
probably be aided only by the detection of the lightest neutral Higgs
boson and comparatively light charged Higgs boson of mass just
around the top quark mass.

\vskip 15pt
\noindent
{\bf Acknowledgments:} 
We thank AseshKrishna Datta and Biswarup Mukhopadhyaya for many useful 
discussions on various issues at different stages. We also thank Nabanita Bhattacharyya and Sujoy Poddar for sharing with us many  useful information and techniques. We also benefitted from discussions with Sudhir Gupta, Tao Han and Sourov Roy. 
I also thank Nishita Desai for valuable suggestions. Computational work was partially carried out on the cluster computing facility at Harish-Chandra Research Institute (http://cluster.mri.ernet.in). This work is partially supported by the Regional Center for Accelerator-based Particle Physics (RECAPP), Harish-Chandra Research Institute and funded by the Department of Atomic Energy, Government of India under the XIth 5-year Plan.

\end{document}